\documentstyle[12pt]{article}

\input epsf

\def\a{\alpha}
\def\b{\beta}
\def\c{\gamma}
\def\la{\langle}
\def\ra{\rangle}
\def\e{\epsilon}
\def\pa{\partial}

\newcommand{\be}{\begin{equation}}
\newcommand{\ee}{\end{equation}}
\newcommand{\ba}{\begin{eqnarray}}
\newcommand{\ea}{\end{eqnarray}}

\newcommand{\eg}{e.g.}
\newcommand{\ie}{i.e.}

\begin{document}

\rightline{DAMTP 94/12}
\begin{center}
{\large \bf Conserved Currents, Consistency Relations and Operator
Product Expansions  in the  Conformally Invariant \\ $O(N)$ Vector Model\\}
\vspace{1.5cm}
 {\large  Anastasios Petkou\footnote{Partially supported by PPARC.}\\}
\vspace{1cm}
 {Department of Applied Mathematics and Theoretical Physics\\
Silver Street\\
Cambridge CB3 9EW \\
England\\}
{e-mail: ap120@amtp.cam.ac.uk}
\end{center}

\vspace{1.5cm}

\begin{abstract}
We discuss conserved currents and  operator product expansions (OPE's)
in the context of a $O(N)$ invariant conformal field theory. Using
 OPE's we find explicit expressions for the first few terms in
suitable short-distance limits for various four-point functions
involving the fundamental $N$-component scalar field
$\phi^{\alpha}(x)$, $\alpha=1,2,..,N$. We propose an alternative
evaluation of these four-point functions based on graphical
expansions. Requiring consistency of the algebraic and graphical
treatments of the four-point functions we obtain the values of the dynamical
parameters
in either  a free theory of $N$ massless fields or  a
non-trivial conformally invariant $O(N)$ vector model in $2<d<4$, up to
next-to-leading order in a $1/N$ expansion. Our
approach  suggests  an interesting duality property of the critical
$O(N)$ invariant theory. Also, solving our consistency relations we obtain  the
next-to-leading order in $1/N$ correction for  $C_{T}$ which corresponds
to the normalisation of the
energy momentum tensor two-point function.

\end{abstract}

\vfill

\section*{Introduction}

There exists an enormous literature (see \cite{Fradkin,Cardy1,Ginsparg}
and references therein) devoted to the study of conformal field
theories (CFT's). Such discussions have been primarily motivated  by the
observation that scale invariance at the fixed points of the renormalisation
group also implies conformal invariance for
most quantum field theories \cite{Schroer}. An interesting possibility is that
CFT's  may be formulated in terms of  algebras of
local fields closed under operator product expansions which, if the
OPE coefficients are known, enable the evaluation of all correlation functions
in the theory
without using a Lagrangian \cite{Polyakov}.  Such
a formulation  has been successfully developed for a large class of
interesting CFT's
in two dimensions \cite{Belavin,Cardy1,Ginsparg} giving, among
other things, important information concerning the properties of
two-dimensional statistical mechanical systems. Although not many
explicitly constructed non-trivial CFT's are known in $d>2$,
statistical mechanical systems at criticality enjoying at least scale
 and in many cases conformal invariance exist in the real
world in more than two dimensions. Non-perturbative field theoretic methods for
studying these systems are  still rudimentary, however the investigation of
CFT's  provides a
viable framework  for  understanding  their properties  in $d>2$.

In \cite{anastasios} the subject of CFT's in $d>2$ was reconsidered
but  only free  field theories were discussed as
explicit examples in the context of that work. In the present work we explore
a more complicated CFT in $d>2$, the $O(N)$
invariant model
of a $N$-component scalar field $\phi^{\a}(x)$, $\a=1,2,..,N$. In this
model  the most singular contributions  in the OPE of $\phi^{\a}(x)$ with
itself are given by  a conserved vector
current, the energy momentum
tensor  and  a
scalar field $O(x)$ of low  dimension $\eta_{o}<d$.  As we also will see, this
model  naturally suggests  an
approximation scheme based on a $1/N$ expansion. Our main purpose in the
context of the present work is to provide evidence that
various  four-point functions involving  the
fundamental field $\phi^{\a}(x)$ can be consistently treated, at
least in  certain short-distance limits, on the basis of  conformally
invariant OPE's and also with the aid of  graphical expansions built using
internal
lines corresponding to the two-point functions of the  field $\phi^{\a}(x)$
and of another  scalar field $\tilde{O}(x)$ having dimension
$\tilde{\eta}_{o}$ with
$0<\tilde{\eta}_{o}<d$. The latter field can be identified either with the
scalar field $O(x)$ appearing in the OPE of $\phi^{\a}(x)$ with itself
or with its {\it{shadow field}} $O_{s}(x)$ of dimension $d-\eta_{o}$.

The paper is organised as follows. In section \ref{sec1} we derive Ward
identities associated with
a conserved vector current and the energy momentum tensor for Euclidean
CFT's in $d>2$. In section \ref{sec2} we focus on the discussion of the  $O(N)$
invariant field theory  at its critical point assuming conformal
invariance. We show
that  Ward identities relate the couplings in the three-point
functions of  $\phi^{\a}(x)$ with a conserved
vector current or the energy momentum tensor to the
dimension $\eta$ and the normalisation constant $C_{\phi}$ of the two-point
function  of $\phi^{\a}(x)$. Next,  making  a simple ansatz for the OPE of
$\phi^{\a}(x)$ with itself we obtain explicit
expressions for the first few terms in a suitable short-distance
limit of the four-point function $\langle\phi\phi\phi\phi\rangle$. In
section \ref{sec3} we show consistency of our results for the four-point
function with explicit
calculations in the context of the trivial CFT of $N$ massless free
fields in any dimension $d$.
Motivated by the form of the OPE ansatz we assume in section \ref{sec4} that
the four-point
function $\langle\phi\phi\phi\phi\rangle$ in a non-trivial
$O(N)$ invariant CFT in $2<d<4$ can alternatively be evaluated using a skeleton
graph
expansion with internal lines corresponding to  the two-point functions
of $\phi^{\a}(x)$ and of a scalar field $\tilde{O}(x)$ having
dimension $\tilde{\eta}_{o}$ with $0<\tilde{\eta}_{o}<d$. The first few graphs
in such an
expansion represent  the interaction of $\phi^{\a}(x)$ with
$\tilde{O}(x)$ via a unique vertex having coupling constant $g_{*}$.
Requiring agreement of the algebraic and  graphical treatments of the
four-point function we obtain  consistency relations for the
dynamical parameters of the theory which we
solve using a $1/N$ expansion. Our next-to-leading order in $1/N$
results agree with previous calculations \cite{Ruhl1,Vasiliev} in the context
of the $O(N)$ sigma model in
$2<d<4$. When $d\rightarrow 4$, our result for the next-to-leading
order in $1/N$ correction  for
the normalisation of the energy momentum tensor two-point function  $C_{T}$,
agrees  with the $\epsilon$-expansion
results \cite{Osborn,Cappelli}. Furthermore, the graphical expansion is shown
to possess a
{\it{shadow symmetry}} which, assuming that it holds to all orders,
suggests  an interesting duality property of the critical $O(N)$ vector model.
In section \ref{sec5} we present the graphical and
algebraic treatments  of the four-point function $\langle\phi\phi O
O\rangle$ involving  the dimension
$\eta_{o}<d$ field $O(x)$ which appears in the OPE of $\phi^{\a}(x)$ with
itself. The
consistency relations in this case provide a non-trivial check for the
associativity of the OPE in $d>2$. Our detailed calculations are in many
respects similar to those of Lang and R\"{u}hl \cite{Ruhl1,Ruhl2} and
in large part we agree on their results, however our perspective is rather
different. Discussions concerning  possible implications
of our results and concluding remarks are presented in section \ref{sec6}.
Four Appendices contain
technical details.

\section{Conserved Currents and Ward Identities in CFT} \label{sec1}

\setcounter{equation}{0}

We consider Euclidean CFT in dimensions $d>2$
which is invariant under the global action of a compact semisimple
finite-dimensional Lie group $G$. The
field  algebra  of such a theory consists of quasiprimary fields
$O_{i}^{\a}(x)$,  with spacetime index $i$
and internal index $\a$, together with their derivatives \cite{Polyakov}.
Quasiprimary
fields of dimension $\eta$
transform under a  finite-dimensional representation of the Euclidean
conformal group, (which is isomorphic to the pseudo-orthogonal group
$O(d+1,1)$), as \cite{anastasios}
\begin{equation}
O_{i}^{\a}(x)\rightarrow
O_{i}^{\prime\,\a}(x')=[\Omega(x)]^{\eta}D_{i}^{\,\,\,j}(R(x))O_{j}^{\a}(x),
\label{1.1.1}
\end{equation}
where $\Omega(x)^{-d}=| \pa x'/\pa x|$ is the Jacobian of the conformal
transformation $x_{\mu}\rightarrow
x'_{\mu}(x)$ and
$D_{i}^{\,\,\,j}(R)$ with $R_{\mu\nu}(x)=\Omega(x)\,\pa x'_{\mu} /\pa
x_{\nu}$,  $R_{\mu\nu}\in O(d)$, belongs to a representation of the rotation
group associated with that transformation. We only consider fields
transforming under
irreducible representations of $O(d)$ here. Correlation
functions of the fields $O_{i}^{\a}(x)$ are defined  to be statistical averages
with
Boltzmann weight $e^{-S[\phi]}$, formally
\begin{equation}
\la O_{i}^{\a}(x)\cdots\ra_{S}=\int ({\cal{D}}\phi )
\, O_{i}^{\a}(x)\cdots e^{-S[\phi]}. \label{1.1.2}
\end{equation}
The action (Hamiltonian) $S[\phi ]$ is a functional of some
fundamental fields $\phi_{i}^{\a}(x)$ and their derivatives.

Conserved currents play an important role in
CFT \cite{Mack} and in the present work we focus on the discussion of the
energy momentum
tensor $T_{\mu\nu}(x)$ and the  vector current
$J_{\mu}^{\kappa}(x)$, where $\kappa$ denotes indices in the adjoint
representation of $G$ - $T_{\mu\nu}(x)$ is a singlet under the action of $G$.
The conservation equations read
$\pa_{\mu}T_{\mu\nu}(x)=\pa_{\mu}J_{\mu}^{\kappa}(x)=0$. The energy
momentum tensor is here defined as  the response of the action $S[\phi]$ to
arbitrary infinitesimal  spacetime transformations $x_{\mu}\rightarrow
x_{\mu}+a_{\mu}(x)$ inducing suitable infinitesimal deformations
$\delta_{a}\phi(x)$ of the fields which reduce to the form (\ref{1.1.1}) for
$a_{\mu}(x)$ conformal transformations, namely
\begin{equation}
S\rightarrow S+\delta_{a} S\,\,\,\,\,,\,\,\,\,\, \delta_{a}S=\int
d^{d}\!x\Bigl(\pa_{\mu}a_{\nu}(x)\Bigl)T_{\mu\nu}(x). \label{1.1.3}
\end{equation}
This definition of the energy momentum tensor is arbitrary up to
$T_{\mu\nu}(x)\rightarrow
T_{\mu\nu}(x)+\pa_{\lambda}X_{\lambda\mu\nu}(x)$ with
$X_{\lambda\mu\nu}(x)=-X_{\mu\lambda\nu}(x)$. We assume that $T_{\mu\nu}(x)$
can be made
symmetric and traceless for a
suitable choice of $X_{\lambda\mu\nu}(x)$.

The conserved vector current $J_{\mu}^{\kappa}(x)$ is defined as the response
of the
action $S[\phi]$ to arbitrary infinitesimal group transformations
parametrised by a local field  $\epsilon^{\kappa}(x)$, namely
\begin{equation}
S\rightarrow S+\delta_{\epsilon}S
\,\,\,\,\,,\,\,\,\,\,\delta_{\epsilon}S=\int
d^{d}\!x\Bigl(\pa_{\mu}\epsilon^{\kappa}(x)\Bigl)J_{\mu}^{\kappa}(x).
\label{1.1.4}
\end{equation}
Then, the Ward identities follow from the requirement
that  correlation functions in the theory should satisfy
\begin{equation}
\la O_{i}^{\a}(x)\cdots\ra_{S}=\la
O_{i}^{\a}(x)+\delta_{a}O_{i}^{\a}(x)\cdots\ra_{S+\delta_{a} S}, \label{1.1.5}
\end{equation}
and similarly for $\delta_{\epsilon}O^{\a}_{i}(x)$,
$\delta_{\epsilon}S$. By virtue of  (\ref{1.1.2}) and (\ref{1.1.3}),
(\ref{1.1.4}) we  obtain from (\ref{1.1.5}) for $n$-point functions
\begin{eqnarray}
\sum_{k =1}^{n}\la
O_{i_{1}}^{\a_{1}}(x_{1})\cdots\delta_{a}O_{i_{k}}^{\a_{k}}(x_{k})\cdots\ra_{S}
 & = & \int
d^{d}\!x\Bigl(\pa_{\mu}a_{\nu}(x)\Bigl)\la
T_{\mu\nu}(x)\,O_{i_{1}}^{\a_{1}}(x_{1})\cdots\ra_{S}, \label{1.1.6}\\
\sum_{k=1}^{n}\la
O_{i_{1}}^{\a_{1}}(x_{1})\cdots\delta_{\epsilon}O_{i_{k}}^{\a_{k}}(x_{k})
\cdots\ra_{S} & = &
\int d^{d}\!x\Bigl(\pa_{\mu}\epsilon^{\kappa}(x)\Bigl)\la
J_{\mu}^{\kappa}(x)\,O_{i_{1}}^{\a_{1}}(x_{1})\cdots\ra_{S}.\label{1.1.7}
\end{eqnarray}
Given the particular representation of $G$ under
which the fields in the field algebra  transform,  the
deformations $\delta_{\epsilon}O_{i}^{\a}(x)$ are easily obtained.
However, when $a_{\mu}(x)$ corresponds to a conformal transformation \eg{}
{$\pa_{\mu}a_{\nu}(x)+\pa_{\nu}a_{\mu}(x)\propto\delta_{\mu\nu}$},  the
r.h.s. of (\ref{1.1.6}) vanishes due to the tracelessness of $T_{\mu\nu}(x)$.
For example, in the case of spacetime
dilatations with $a_{\mu}(x)\propto x_{\mu}$,  then $\delta_{a}$ becomes  the
configuration space analog of the
Callan-Symanzik operator with vanishing beta functions acting on the fields in
a massless theory at
a fixed point of the renormalisation group and we obtain
\begin{equation}
\sum_{k=1}^{n}\Bigl( x_{k}^{\mu}\frac{\pa}{\pa
x^{\mu}_{k}}+\eta_{k}\Bigl)\la O_{i_{1}}^{\a_{1}}(x_{1})\cdots
O_{i_{k}}^{\a_{k}}(x_{k})\cdots\ra_{S}=0.\label{1.1.8}
\end{equation}
More generally (\ref{1.1.6}) and (\ref{1.1.7}) determine Ward identities for
$\pa_{\mu}\langle T_{\mu\nu}(x)\cdots\rangle$ and  $\pa_{\mu}\langle
J_{\mu}^{\kappa}(x)\cdots\rangle$ or equivalently control the leading
terms in the OPE's of $T_{\mu\nu}(x)$ and $J_{\mu}^{\kappa}(x)$ with the
quasiprimary
fields. Note that for    $x\neq x_{i}$ , $i=1,...n$,  (\ref{1.1.6}) requires
${\pa_{\mu}\langle T_{\mu\nu}(x)\cdots\rangle =0}$. If the
transformations $a_{\mu}(x)$ ($\epsilon^{\kappa}(x)$)
satisfy the conformal invariance property
${\pa_{\mu}a_{\nu}(x)+\pa_{\nu}a_{\mu}(x)\propto\delta_{\mu\nu}}$ (are
$x$-independent group transformations) in the region
$|x-x_{i}|\leq r$ for $|x_{i}-x_{j}| >r$, $i\neq j$, $r>0$, then  we obtain
\begin{eqnarray}
\la \delta_{a}O_{i_{1}}^{\a^{1}}(x_{1})\cdots\ra_{S} & = &
-\int dS_{\mu}\,a_{\nu}(x)\,\la
T_{\mu\nu}(x) O_{i_{1}}^{\a^{1}}(x_{1})\cdots\ra_{S}, \label{1.1.9}\\
\la \delta_{\epsilon}O_{i_{1}}^{\a^{1}}(x_{1})\cdots\ra_{S} & =
& -\epsilon^{\kappa}\int dS_{\mu}\,\la
J_{\mu}^{\kappa}(x) O_{i_{1}}^{\a^{1}}(x_{1})\cdots\ra_{S},\label{1.1.10}
\end{eqnarray}
where $dS_{\mu}=d\Omega \,r_{\mu}\,r^{d-2}$ denotes integration on the surface
of the ball
$|x-x_{1}|=r$ and the angular
integration  is normalised as $\int d\Omega =S_{d}$ with
$S_{d}=2\pi^{d/2}/\Gamma(d/2)$ the surface of the
unit $d$-dimensional sphere. We find equations (\ref{1.1.9}) and (\ref{1.1.10})
to be a
useful form of the Ward identities expressing  conformal and group
symmetry of $n$-point correlation functions and we will use them for
explicit calculations in what follows. Note that the r.h.s. of (\ref{1.1.9})
and (\ref{1.1.10}) does not depend on $r$ as
$r\rightarrow 0$ because the difference of two surface integrals on
the balls with radii $r$ and $r'$, when $r>r'$, is a vanishing volume
integral. We may also remark  that with our presentation of the Ward identities
 $\langle T_{\mu\mu}(x)\cdots\rangle =0$ for all
$x$.\footnote{In \cite{anastasios}, alternative Ward identities were
derived leading to contributions to $\la
T_{\mu\mu}(x)O(x_{1})\cdots\ra$ proportional to $\delta^{d} (x-x_{1})\la
O(x_{1})\cdots\ra$ for
suitable  fields $O(x)$. Our approach is equivalent to the one
of  \cite{anastasios} up to a redefinition of $\la
T_{\mu\nu}(x)O(x_{1})\cdots\ra$ using sums of terms proportional to
$\delta^{d}(x-x_{1})\la O(x_{1})\cdots\ra$.}

\section{Conserved Currents and Operator Product  Expansions
 in the
Conformally Invariant   {$O(N)$ Vector  Model}}\label{sec2}

\setcounter{equation}{0}

\subsection{General Remarks}\label{sbsec21}

We consider the $O(N)$ invariant CFT having as fundamental field the
$N$-component vector, $O(d)$ scalar field $\phi^{\a}(x)$,
$\a=1,2,..,N$. In CFT the two-and three-point
correlation functions are fixed up to some arbitrary constants
\cite{anastasios}. The two-point function of
$\phi^{\a}(x)$ in this model is  diagonal in
the $O(N)$ indices and can be written as
\begin{equation}
\la\phi^{\a}(x_{1})\phi^{\b}(x_{2})\ra
=\Phi^{\a\b}(x_{12})=C_{\phi}\frac{1}{(x_{12}^{2})^{2\eta}}\delta^{\a\b}
\,\,\,\,\, , \,\,\,\,\,
x_{12}=x_{1}-x_{2},\label{2.1.1}
\end{equation}
with $C_{\phi}$ a normalisation constant and $\eta$ the dimension of
$\phi^{\a}(x)$. The three-point function of the
fundamental field in the conformally invariant $O(N)$ vector model
is zero. Conformal invariance alone does not fix the form of $n$-point
functions with
$n\geq 4$. One can show that conformal $n$-point functions will in general
depend on
$n(n-3)/2$ variables \cite{Ferrara1,Ginsparg}. The four-point function of
$\phi^{\a}(x)$
in the  $O(N)$ vector model can in general  be written  as
\begin{eqnarray}
\la\phi^{\a}(x_{1})\phi^{\b}(x_{2})\phi^{\c}(x_{3})\phi^{\delta}(x_{4})\ra
& \equiv &   \Phi^{\a\b\c\delta}(x_{1},x_{2},x_{3},x_{4}) \nonumber \\
 &  = &
\Phi(x_{1},x_{2},x_{3},x_{4})\delta^{\a\b}\delta^{\c\delta} \nonumber
\\
 & & {}+\Phi(x_{1},x_{3},x_{2},x_{4})\delta^{\a\c}\delta^{\b\delta}
\nonumber \\
 &  & {}+\Phi(x_{1},x_{4},x_{3},x_{2})\delta^{\a\delta}\delta^{\b\c},
\label{2.1.2}
\end{eqnarray}
where $\Phi(x_{1},x_{2},x_{3},x_{4})$ has the following crossing symmetry
properties
\begin{equation}
\Phi(x_{1},x_{2},x_{3},x_{4}) =
\Phi(x_{2},x_{1},x_{3},x_{4})=\Phi(x_{3},x_{4},x_{1},x_{2}).\label{2.1.3}
\end{equation}
By virtue of conformal invariance and (\ref{2.1.3}) we may then  write
\begin{equation}
\Phi(x_{1},x_{2},x_{3},x_{4})=H(\eta,x)F(u,v)\label{2.1.4}
\end{equation}
with
\begin{equation}
H(\eta,x)=\frac{1}{(x_{12}^{2}x_{34}^{2}x_{13}^{2}x_{24}^{2}x_{14}^{2}
x_{23}^{2})^{\frac{1}{3}\eta}}.\label{2.1.5}
\end{equation}
where  $F(u,v)=F(v,u)$ is an    arbitrary function of the two
independent invariant
ratios
\begin{equation}
u=\frac{x_{12}^{2}x_{34}^{2}}{x_{13}^{2}x_{24}^{2}}
\,\,\,\,\,\,\mbox{and}\,\,\,\,\,
v=\frac{x_{12}^{2}x_{34}^{2}}{x_{14}^{2}x_{23}^{2}}.\label{2.1.6}
\end{equation}
We may also
cast (\ref{2.1.2}) into a form convenient for what follows
\begin{eqnarray}
\Phi^{\a\b\c\delta}(x_{1},x_{2},x_{3},x_{4})& = &
H(\eta,x)F_{S}(u,v)\,\delta^{\a\b}\delta^{\c\delta} \nonumber \\
 & &
{}+H(\eta,x)F_{V}(u,v)\,\frac{1}{2}(\delta^{\a\c}\delta^{\b\delta}
-\delta^{\a\delta}\delta^{\b\c})
\nonumber \\
 &  &
{}+H(\eta,x)F_{T}(u,v)\,\frac{1}{2}(\delta^{\a\c}\delta^{\b\delta}
+\delta^{\a\delta}\delta^{\b\c}-\frac{2}{N}\delta^{\a\b}\delta^{\c\delta})
,\label{2.1.7}
\end{eqnarray}
where
\begin{eqnarray}
F_{S}(u,v) & = & F(u,v)
+\frac{1}{N}\left(F(\frac{1}{u},\frac{v}{u})+F(\frac{u}{v},\frac{1}{v})\right),
\nonumber \\
F_{V}(u,v) & = & F(\frac{1}{u},\frac{v}{u})-F(\frac{u}{v},\frac{1}{v}),
\nonumber \\
F_{T}(u,v) & = &
F(\frac{1}{u},\frac{v}{u})+F(\frac{u}{v},\frac{1}{v}).\label{2.1.8}
\end{eqnarray}
Correlation functions involving the group symmetry conserved current
$J_{\mu}^{\kappa}(x)$ and the energy momentum tensor
$T_{\mu\nu}(x)$ are of special interest in CFT. In the conformally
invariant  $O(N)$
vector model the conserved current of the
$O(N)$ symmetry has $O(d)$ spin-1, dimension $d-1$ and transforms
irreducibly under the action of the adjoint representation of $O(N)$, hence it
may be written as
$J_{\mu}^{\a\b}(x)=-J_{\mu}^{\b\a}(x)$. The energy momentum tensor
$T_{\mu\nu}(x)$ has $O(d)$ spin-2 and dimension $d$. Therefore, the
two-point functions of $T_{\mu\nu}(x)$ and $J_{\mu}^{\a\b}(x)$ can be
written as
\begin{eqnarray}
\la T_{\mu\nu}(x_{1})T_{\rho\sigma}(x_{2})\ra &  = &
C_{T}\frac{I_{\mu\nu ,\rho\sigma}(x_{12})}{x_{12}^{2d}}, \label{2.1.9}\\
\la J_{\mu}^{\a\b}(x_{1})J_{\nu}^{\c\delta}(x_{2})\ra & = &
C_{J}\frac{I_{\mu\nu}(x_{12})}{x_{12}^{2(d-1)}}\,(\delta^{\a\c}
\delta^{\b\delta}-\delta^{\a\delta}\delta^{\b\c}).\label{2.1.10}
\end{eqnarray}
The
functions $I_{\mu\nu ,\rho\sigma}(x)$ and $I_{\mu\nu}(x)$ are
determined from the conservation, symmetry and tracelessness properties
of the currents and have been explicitly
given  in  \cite{anastasios}
\begin{eqnarray}
I_{\mu\nu ,\rho\sigma}(x) & = & \frac{1}{2}\Bigl(
I_{\mu\rho}(x)I_{\nu\sigma}(x)+I_{\mu\sigma}(x)I_{\nu\rho}(x)\Bigl)
{}-{}\frac{1}{d}\delta_{\mu\nu}\delta_{\rho\sigma}, \label{2.1.11}\\
I_{\mu\nu}(x) & = &
\delta_{\mu\nu}-2\frac{x_{\mu}x_{\nu}}{x^{2}}.\label{2.1.12}
\end{eqnarray}
The quantities $C_{T}$ and $C_{J}$, which may be considered as the
normalisations of the conserved currents $T_{\mu\nu}(x)$ and
$J_{\mu}^{\a\b}(x)$, are fixed by the defining
relations (\ref{1.1.3}) and (\ref{1.1.4}) and they should be given in
terms of the dynamical variables of the theory {\ie} the
dimensions of the fields  and the couplings  \cite{Fradkin}. In the
following we
provide evidence  in favour of  this assertion.

It will also be useful for what follows to consider three-point
functions of the fundamental field $\phi^{\a}(x)$ involving one
$T_{\mu\nu}(x)$, one $J_{\mu}^{\a\b}(x)$ or
one   scalar field
$O(x)$
of dimension $\eta_{o}$. Conformal invariance fixes the form of
these  three-point functions up to a constant \cite{anastasios}
\begin{eqnarray}
\la \phi^{\a}(x_{1})\phi^{\b}(x_{2})O(x_{3})\ra & = & \frac{g_{\phi\phi
O}}{(x_{12}^{2})^{\eta
-\frac{1}{2}\eta_{o}}(x_{13}^{2}x_{23}^{2})^{\frac{1}{2}\eta_{o}}}
\delta^{\a\b},\label{2.1.13} \\
\la \phi^{\a}(x_{1})\phi^{\b}(x_{2})T_{\mu\nu}(x_{3})\ra & = &
\frac{-g_{\phi\phi T}}{(x_{12}^{2})^{\eta -\mu +1}(x_{13}^{2}x_{23}^{2})^{\mu
-1}}\left[
\Bigl(\!X_{12}\!\Bigl)_{\mu}\Bigl
(\!X_{12}\!\Bigl)_{\nu}\!-\frac{1}{d}\delta_{\mu\nu}
\Bigl(\!X_{12}\!\Bigl)^{2}\right]\delta^{\a\b},\label{2.1.14}\\
\la \phi^{\a}(x_{1})\phi^{\b}(x_{2})J_{\mu}^{\c\delta}(x_{3})\ra & = &
\frac{g_{\phi\phi J}}{(x_{12}^{2})^{\eta -\mu +1}(x_{13}^{2}x_{23}^{2})^{\mu
-1}}\Bigl( \!X_{12}\!\Bigl
)_{\mu}\,(\delta^{\a\c}\delta^{\b\delta}-\delta^{\a\delta}\delta^{\b\c})
,\label{2.1.15}
\end{eqnarray}
where
\begin{equation}
\Bigl(\!X_{12}\!\Bigl)_{\mu}=\frac{(x_{13})_{\mu}}{x_{13}^{2}}
-\frac{(x_{23})_{\mu}}{x_{23}^{2}}.\label{2.1.16}
\end{equation}
The couplings  $g_{\phi\phi O}$, $g_{\phi\phi T}$ and
$g_{\phi\phi J}$ above  depend on the dynamics of the
particular CFT model. However, in the case of (\ref{2.1.14}) and (\ref{2.1.15})
the Ward
identities (\ref{1.1.9}) and (\ref{1.1.10}) relate $g_{\phi\phi T}$ and
$g_{\phi\phi J}$ to the dimension
$\eta$ and the normalisation  constant $C_{\phi}$ of  the two-point
function of $\phi^{\a}(x)$. To see this we use the following infinitesimal
field deformations
\begin{eqnarray}
a_{\mu}(x)=\lambda x_{\mu}  \Rightarrow  \delta_{a}\phi^{\a}(x) & = & \lambda
(\eta+x_{\mu} \pa_{\mu} )\phi^{\a}(x), \label{2.1.17}\\
\delta_{\e}\phi^{\a}(x) & = & \e^{\a\b}\phi^{\b}(x),\label{2.1.18}
\end{eqnarray}
where $\lambda$ and $\e^{\a\b}=-\epsilon^{\b\a}$ are arbitrary
infinitesimal parameters. By virtue of the antisymmetry of
$\epsilon^{\a\b}$ we may now take  in (\ref{1.1.10})
$\epsilon^{\kappa}J_{\mu}^{\kappa}(x)\rightarrow\frac{1}{2}\epsilon^{\a\b}
J_{\mu}^{\a\b}(x)$
and then from (\ref{1.1.9}) and (\ref{1.1.10}) we obtain
\begin{eqnarray}
\la\delta_{a}\phi^{\a}(x_{1})\phi^{\b}(x_{2})\ra & = & -\lambda\int
d\Omega \,r^{d-2}\,r_{\nu}r_{\rho}\,\la
T_{\nu\rho}(x)\phi^{\a}(x_{1})\phi^{\b}(x_{2})\ra,\label{2.1.19}\\
\la\delta_{\epsilon}\phi^{\a}(x_{1})\phi^{\b}(x_{2})\ra & = &
-{\textstyle{\frac{1}{2}}}\epsilon^{\c\delta}\int d\Omega
\,r^{d-2}\,r_{\nu}\,\la
J_{\nu}^{\c\delta}(x)\phi^{\a}(x_{1})\phi^{\b}(x_{2})\ra,\label{2.1.20}
\end{eqnarray}
where $r=|x-x_{1}|$. Performing the integrals in (\ref{2.1.19}) and
(\ref{2.1.20})  in the
limit  $r\rightarrow 0$ we
obtain
\begin{eqnarray}
g_{\phi\phi T} & = & \frac{1}{S_{d}}\frac{d\eta}{d-1}C_{\phi}, \label{2.1.21}\\
g_{\phi\phi J}& = & \frac{1}{S_{d}}C_{\phi}. \label{2.1.22}
\end{eqnarray}
It is clear
that in
the place of  $\phi^{\a}(x)$ in (\ref{2.1.19}) and (\ref{2.1.20}) we could have
had any $O(d)$ scalar
field. We conclude that the
couplings (such as  $g_{\phi\phi T}$ and
$g_{\phi\phi J}$ above) of all $O(d)$ scalar fields
with the conserved currents $T_{\mu\nu}(x)$ and $J_{\mu}^{\kappa}(x)$
in any CFT can be found solely from the
knowledge of the complete two-point functions of those  scalar fields.

\subsection{Operator Product Expansions  and  the Four-Point Function
in the Conformally Invariant  $O(N)$ Vector Model}\label{sbsec22}

Within non-Lagrangian approaches to quantum field theory and in
particular to CFT the  operator product expansion is
a basic starting point. In the context of the present work we begin our
discussion with the assumption of the following OPE
\begin{equation}
\phi^{\a}(x_{1})\phi^{\b}(x_{2})=C_{\phi}\frac{1}{(x_{12}^{2})^{\eta}}
\delta^{\a\b}+\sum_{O_{n}}C^{\phi\phi
O_{n}}(x_{12},\pa_{2})\,
O^{\a\b}_{n}(x_{2}),\label{2.2.1}
\end{equation}
where we have suppressed $O(d)$ indices and the index  $n$ here denotes
all different quasi-primary  fields which  may appear in
the OPE above. For the purposes
of the present work we follow \cite{Ferrara1} and  assume that the fields
$O_{n}^{\a\b}(x)$ are realised by symmetric $O(d)$ tensors
$O_{\mu_{1}\mu_{2}..\mu_{k}}^{\a\b}(x)$ of rank $k$ and dimension
$\eta_{k}$. Moreover, these fields can be $O(N)$ singlets, symmetric traceless
$O(N)$
tensors or antisymmetric  $O(N)$ tensors. Inserting (\ref{2.2.1}) into
(\ref{2.1.2}) we obtain formally
\begin{eqnarray}
\Phi^{\a\b\c\delta}(x_{1},x_{2},x_{3},x_{4}) & = &
C_{\phi}^{2}\frac{1}{(x_{12}^{2}x_{34}^{2})^{\eta}}\delta^{\a\b}
\delta^{\c\delta}
\nonumber \\
 &  & +\sum_{O_{n}}C^{\phi\phi O_{n}}(x_{12},\pa_{2})C^{\phi\phi
O_{n}}(x_{34},\pa_{4})\la
O_{n}^{\a\b}(x_{2})O_{n}^{\c\delta}(x_{4})\ra,\label{2.2.2}
\end{eqnarray}
where we have used the selection rule for quasiprimary fields
\cite{Ferrara1,anastasios}
\begin{equation}
\la
O_{\mu_{1}\mu_{2}..\mu_{k}}^{\a\b}(x_{1})\,\,O_{\nu_{1}\nu_{2}..
\nu_{l}}^{\c\delta}(x_{2})\ra
\neq 0
\,\,\,\,\,\mbox{only
if}\,\,\,\,\,k=l\,\,\,\,\,\mbox{and}\,\,\,\,\,\eta_{k}=\eta_{l},\label{2.2.3}
\end{equation}
and we have also assumed that if $k=l$ and $\eta_{k}=\eta_{l}$ the two-point
functions (\ref{2.2.3}) are diagonal in  the $O(N)$ representation. On
the r.h.s. of (\ref{2.2.2}),  the first
term represents the contribution from the
unit field in the field algebra and the sum   involves the
two-point functions of all fields in the OPE (\ref{2.2.1}). Clearly, the OPE
(\ref{2.2.1}) and the
resulting expression (\ref{2.2.2}) determine the form of
the functions $F_{S}(u,v)$, $F_{V}(u,v)$ and $F_{T}(u,v)$ in (\ref{2.1.7}).
Next, we make the
following ansatz for the OPE of the fundamental field $\phi^{\a}(x)$ with
itself
\begin{eqnarray}
\phi^{\a}(x_{1})\phi^{\b}(x_{2})  & = &
C_{\phi}\frac{1}{x_{12}^{2\eta}}\delta^{\a\b}
+C^{\phi\phi O}(x_{12},\pa_{2})O(x_{2})\delta^{\a\b}
\nonumber \\
 &   & {}+\frac{g_{\phi\phi
J}}{C_{J}}\,\frac{(x_{12})_{\mu}}{(x_{12}^{2})^{\eta -\mu
+1}}J_{\mu}^{\a\b}(x_{2}) +\cdots \nonumber \\
&  & {}-\frac{g_{\phi\phi
T}}{C_{T}}\,\frac{(x_{12})_{\mu}(x_{12})_{\nu}}{(x_{12}^{2})^{\eta
-\mu +1}}T_{\mu\nu}(x_{2})\delta^{\a\b}+\cdots . \label{2.2.4}
\end{eqnarray}
Namely, the most singular terms as
$x_{12}^{2}\rightarrow 0$ in the OPE
(\ref{2.2.1}) are assumed to be, apart from the contribution of the unit field,
 the  coefficients of the
conserved vector current $J_{\mu}^{\a\b}(x)$, of the energy
momentum tensor $T_{\mu\nu}(x)$ and also of
some   scalar field $O(x)$ of
dimension $\eta_{o}$ with  $0< \eta_{o}<d$ whose two-point function
in normalised as
\begin{equation}
\la O(x_{1})\, O(x_{2})\ra = \Phi_{O}(x_{12})=
C_{O}\frac{1}{x_{12}^{2\eta_{o}}}.\label{2.2.6}
\end{equation}
Other possible fields neglected in  (\ref{2.2.4}) include all
symmetric traceless rank-2 $O(N)$ tensors. The dots in the second
(third) row of (\ref{2.2.4})  stand for derivatives of $J_{\mu}^{\a\b}(x)$
$(T_{\mu\nu}(x))$ having for coefficients less singular
functions of $x_{12}^{2}$ as $x_{12}^{2}\rightarrow 0$ and  also for other
$O(N)$ singlets and antisymmetric tensors. The leading
OPE coefficients of   $J_{\mu}^{\a\b}(x)$ and
$T_{\mu\nu}(x)$ are determined from the conservation, symmetry and
tracelessness properties of
these currents. The
normalisation coefficients $g_{\phi\phi T}/C_{T}$ and
$g_{\phi\phi J}/C_{J}$ are  determined by requiring  consistency of
(\ref{2.2.4}) with (\ref{2.1.14}) and (\ref{2.1.15}).

The OPE coefficient $C^{\phi \phi O}(x_{12},\pa_{2})$ is determined
\cite{Ferrara1} by
requiring consistency of (\ref{2.2.4}) with the three-point function
(\ref{2.1.13}). If we let
\begin{equation}
C_{O}C^{\phi\phi
O}(x_{12},\pa_{2}) = g_{\phi\phi
O}\frac{1}{(x_{12}^{2})^{\eta-\frac{1}{2}\eta_{o}}}
C^{\eta_{o}}(x_{12},\pa_{2}),
\end {equation}
then we require
\begin{equation}\nonumber \\
 C^{\eta_{o}}(x_{12},\pa_{2})\frac{1}{x_{23}^{2\eta_{o}}} =
\frac{1}{(x_{13}^{2}x_{23}^{2})^{\frac{1}{2}\eta_{o}}}.\label{2.2.5}
\end{equation}
To determine  $C^{\eta_{o}}(x_{12},\pa_{2})$ we use  \cite{Gradshteyn}
\begin{equation}
\frac{1}{(ab)^{\rho}}=\frac{1}{B(\rho,\rho)}\int_{0}^{1}dt[t(1-t)]^{\rho
-1}\frac{1}{[at+(1-t)b]^{2\rho}},\label{2.2.7}
\end{equation}
and we find from  (\ref{2.2.5})
\begin{eqnarray}
C^{\eta_{o}}(x_{12},\pa_{2})\frac{1}{x_{23}^{2\eta_{o}}} & =  &
\frac{1}{B(\frac{1}{2}\eta_{o},\frac{1}{2}\eta_{o})}\int_{0}^{1}
dt[t(1-t)]^{\frac{1}{2}\eta_{o}-1}\frac{1}{[t(x_{23}+x_{12})^{2}
+(1-t)x_{23}^{2}]^{\eta_{o}}}
 \nonumber \\
 & =  &
\frac{1}{B(\frac{1}{2}\eta_{o},\frac{1}{2}\eta_{o})}\int_{0}^{1}
dt[t(1-t)]^{\frac{1}{2}\eta_{o}-1}\sum_{m=0}^{\infty}
\frac{(\eta_{o})_{m}}{m!}\frac{[-x_{12}^{2}t(1-t)]^{m}}{[(x_{23}
+tx_{12})^{2}]^{\eta_{o}+m}},\label{2.2.8}
\nonumber \\
\end{eqnarray}
where $(a)_{m}=\Gamma(a+m)/\Gamma(a)$ is the Pochhammer
symbol. Then, using
\begin{equation}
\left(\pa^{2}\right)^{m}\!\frac{1}{x^{2a}}=4^{m}\,(a)_{m}\,(a+1-\mu)_{m}
\,\frac{1}{(x^{2})^{a+m}},\label{2.2.9}
\end{equation}
we obtain from  (\ref{2.2.8})
\begin{eqnarray}
C^{\eta_{o}}(x_{12},\pa_{2})\frac{1}{x_{23}^{2\eta_{o}}}
=\frac{1}{B(\frac{1}{2}\eta_{o},\frac{1}{2}\eta_{o})} & & \int_{0}^{1}
dt[t(1-t)]^{\frac{1}{2}\eta_{o}-1}\,\sum_{m=0}^{\infty}\frac{1}{m!}
\frac{1}{(\eta_{o}+1-\mu)_{m}}
\nonumber \\
 &  & \times
\left[-{\textstyle{\frac{1}{4}}}x_{12}^{2}t(1-t)\right]^{m}\pa_{2}^{2m}
\frac{1}{(x_{23}+tx_{12})^{2\eta_{o}}}.\label{2.2.10}
\end{eqnarray}
Comparing (\ref{2.2.5}) with (\ref{2.2.10}) we finally obtain
\begin{eqnarray}
 C^{\eta_{o}}(x_{12},\pa_{2}) & = &
\frac{1}{B(\frac{1}{2}\eta_{o},\frac{1}{2}\eta_{o})}\int_{0}^{1}
dt[t(1-t)]^{\frac{1}{2}\eta_{o}-1}\sum_{m=0}^{\infty}\frac{1}{m!}
\frac{1}{(\eta_{o}+1-\mu)_{m}}
\nonumber \\
 & &
{}\times\left[-{\textstyle{\frac{1}{4}}}x_{12}^{2}t(1-t)\right]^{m}
\pa_{2}^{2m}e^{tx_{12}\pa_{2}},
\nonumber \\
& = &
1+\frac{1}{2}(x_{12})_{\mu}\pa_{2,\mu}+\frac{\eta_{o}+2}{8(\eta_{o}+1)}
(x_{12})_{\mu}(x_{12})_{\nu}\pa_{2,\mu}\pa_{2,\nu}
{}.
\nonumber \\
& &
\hspace{2cm}{}-\frac{\eta_{o}}{16(\eta_{o}+1)(\eta_{o}+1-\mu)}
x_{12}^{2}\pa_{2}^{2}+O(x_{12}^{3},\pa_{2}^{3}).\label{2.2.11}
\end{eqnarray}
Only the first few most singular terms as $x_{12}^{2}\rightarrow 0$ of
the differential operator $C^{\eta_{o}}(x_{12},\pa_{2})$ are essential for our
purposes and
are explicitly  displayed on the r.h.s. of (\ref{2.2.11}).
The contribution to the four-point function (\ref{2.2.2}) from such a
scalar field in the OPE can   be
found using
\begin{equation}
C^{\eta_{o}}(x_{12},\pa_{2})\,C^{\eta_{o}}(x_{34},\pa_{4})
\frac{1}{x_{24}^{2\eta_{o}}}=\frac{1}{(x_{13}^{2}x_{24}^{2})^{\frac{1}{2}
\eta_{o}}}{\cal{H}}_{\eta_{o}}(u,v),\label{2.2.12}
\end{equation}
where
\begin{equation}
{\cal{H}}_{\eta_{o}}(u,v) =\left(\frac{v}{u}\right)^{\frac{1}{2}\eta_{o}}
\sum_{n=0}^{\infty}\frac{v^{n}}{n!}\frac{({\textstyle{\frac{1}{2}}}
\eta_{o})_{n}^{4}}{(\eta_{o})_{2n}(\eta_{o}
+1-\mu)_{n}}{}_{2}F_{1}({\textstyle{\frac{1}{2}}}\eta_{o}
+n,{\textstyle{\frac{1}{2}}}\eta_{o}+n;\eta_{o}+2n;1-\frac{v}{u}).
\label{2.2.13}
\end{equation}
which is proved in Appendix \ref{apdx1}. Recalling
the following property of the hypergeometric function \cite{Gradshteyn}
\begin{equation}
_{2}F_{1}(a,b;c;z)=(1-z)^{-a}\,_{2}F_{1}(a,c-b;c;\frac{z}{z-1}\,),
\label{2.2.14}
\end{equation}
we easily verify that (\ref{2.2.12}) is symmetric in
$(x_{1},x_{2})\leftrightarrow (x_{3},x_{4})$ or
$u\leftrightarrow v$ as
it should be. We emphasise once more that under our basis assumption, $O(x)$ is
the only scalar
field with dimension $<d$ contributing to the OPE (\ref{2.2.4}).

The leading contribution of the energy momentum tensor $T_{\mu\nu}(x)$ to the
four-point function in the limit
$x_{12}^{2}$, $x_{34}^{2}\rightarrow 0$ is also seen from (\ref{2.1.9}) and
(\ref{2.2.4}) to
be
\begin{eqnarray}
  &  & \frac{g_{\phi\phi
T}^{2}}{C_{T}}
\frac{1}{(x_{12}^{2}x_{34}^{2})^{\eta-\mu+1}}\frac{1}{(x_{24}^{2})^{2\mu}}
(x_{12})_{\mu}(x_{12})_{\nu}(x_{34})_{\rho}(x_{34})_{\sigma}
\,I_{\mu\nu,\rho\sigma}(x_{24})\,\delta^{\a\b}\delta^{\c\delta}\,
\nonumber \\
\sim & & \frac{g^{2}_{\phi\phi
T}}{C_{T}}  H(\eta,x)  (uv)^{-\frac{1}{3}\eta+\frac{1}{2}(\mu -1)}
\nonumber \\
 &  & {}\times\Bigg(\left[\frac{(x_{12}\cdot
x_{34})}{x_{24}^{2}}-2\frac{(x_{12}\cdot x_{24})(x_{34}\cdot
x_{24})}{x_{24}^{4}}\right]^{2}-\frac{1}{d}\frac{x_{12}^{2}x_{34}^{2}}
{x_{24}^{4}}+\cdots\Bigg)\,\delta^{\a\b}\delta^{\c\delta}.\label{2.2.15}
\end{eqnarray}
Similarly, the leading contribution of the conserved current
$J_{\mu}^{\kappa}(x)$ to the four-point function in the limit
$x_{12}^{2}$, $x_{34}^{2}\rightarrow 0$ can be found from (\ref{2.1.10}) and
(\ref{2.2.4}) to
be
\begin{eqnarray}
 & & \frac{g_{\phi\phi
J}^{2}}{C_{J}}
\frac{1}{(x_{12}^{2}x_{34}^{2})^{\eta-\mu+1}}\frac{1}{(x_{24}^{2})^{2\mu-1}}
(x_{12})_{\mu}(x_{24})_{\nu}\,I_{\mu\nu}(x_{24})\,
(\delta^{\a\c}\delta^{\b\delta}-\delta^{\a\delta}\delta^{\b\c})
\nonumber \\
  \sim  & & \frac{g_{\phi\phi
J}^{2}}{C_{J}}  H(\eta,x)(uv)^{-\frac{1}{3}\eta +\frac{1}{2}(\mu -1)}
\nonumber \\
 &  & {}\times\Bigg(\frac{(x_{12}\cdot
x_{34})}{x_{24}^{2}}-2\frac{(x_{12}\cdot\!
x_{24})(x_{34}\cdot x_{24})}{x_{24}^{4}}+\cdots
\Bigg)(\delta^{\a\c}\delta^{\b\delta}-\delta^{\a\delta}\delta^{\b\c}).
\label{2.2.16}
\end{eqnarray}
The dots in (\ref{2.2.15}) and (\ref{2.2.16})  stand for less singular terms as
$x_{12}^{2}\rightarrow 0$ and $x_{34}^{2}\rightarrow 0$
{\it{independently}}. Equations (\ref{2.2.12}), (\ref{2.2.15})  and
(\ref{2.2.16})  suggest that it is convenient for what follows to consider
$H^{-1}(\eta ,x)\Phi^{\a\b\c\delta}(x_{1},x_{2},x_{3},x_{4})$ as a function of
the two new independent variables
\begin{eqnarray}
Y=1-\frac{v}{u} & = & 2\frac{1}{x_{24}^{2}}\left[(x_{12}\cdot
x_{34})-2\frac{(x_{12}\cdot x_{24})(x_{34}\cdot
x_{24})}{x_{24}^{2}}\right]+\cdots, \label{2.2.17}\\
W=(uv)^{\frac{1}{2}} & = &
\frac{x_{12}^{2}x_{34}^{2}}{x_{24}^{4}}+\cdots,\label{2.2.18}
\end{eqnarray}
where the dots stand for terms which tend to zero faster as
$x_{12}^{2}\rightarrow 0$ and $x_{34}^{2}\rightarrow 0$. Note that as
$x_{12}^{2},x_{34}^{2}\rightarrow 0$, then $Y,W\rightarrow 0$. Also
note that $Y^{2}$ and $W$ are of the same order in the above limit.
Comparing (\ref{2.1.7}) with (\ref{2.2.12}), (\ref{2.2.15}), (\ref{2.2.16}) and
expanding
${\cal{H}}_{\eta_{o}}(u,v)$  for  $u,v\rightarrow 0$ with
$v/u\rightarrow 1$, we obtain\footnote{The crossing symmetry properties
(\ref{2.1.3}) imply
among other things that $F_{S}(u,v)$, $\left(F_{V}(u,v)\right)$, is symmetric
(antisymmetric)
in $u\leftrightarrow v$. These
functions can be expanded as infinite power series in
$(1-v/u)$. Truncating such infinite series to a finite order
makes the crossing symmetry properties of the resulting expressions
less transparent. Alternatively, we might consider the expansion in
the variable $\tilde{Y}=\Bigl[
\left(u/v\right)^{1/2}-\left(v/u\right)^{1/2}\Bigl]=Y+O(Y^{3})$.}
\begin{eqnarray}
F_{S}(u,v) & \equiv & {\cal{F}}_{S}(Y,W) \nonumber \\
 & = & C_{\phi}^{2}W^{-\frac{2}{3}\eta} \nonumber \\
&  & {}+\frac{g_{\phi\phi
O}^{2}}{C_{O}}\,W^{\frac{1}{2}\eta_{o}-\frac{2}{3}\eta}\,
\Bigg(1-\frac{\eta_{o}^{2}}{32(\eta_{o}
+1)}Y^{2}+\frac{\eta_{o}^{3}}{16(\eta_{o}
+1)(\eta_{o}+1-\mu)}\,W\,\Bigg)\nonumber \\
&   & {}+\frac{g_{\phi\phi T}^{2}}{C_{T}}\,W^{\mu
-1-\frac{2}{3}\eta}\,\Bigg(\frac{1}{4}\,Y^{2}-\frac{1}{d}\,W\Bigg){}
+\cdots,\label{2.2.19}
 \\
F_{V}(u,v) & \equiv &  {\cal{F}}_{V}(Y,W) \nonumber \\
 & = & \frac{1}{2}\frac{g_{\phi\phi
J}^{2}}{C_{J}}\,W^{\mu -1-\frac{2}{3}\eta}\,Y+\cdots, \label{2.2.20}
\end{eqnarray}
where the dots stand for less singular terms in the limit
$Y , W\rightarrow 0$.  The discussion of $F_{T}(u,v)$ is outside the
scope of the present work. The field
dimensions and the couplings appearing in (\ref{2.2.19}) and (\ref{2.2.20}) are
the
dynamical parameters of the theory and can only be
determined within the context of explicit CFT models.

\section{The Four-Point Function in the Free Field Theory}\label{sec3}

\setcounter{equation}{0}

Consider the trivial theory of $N$ massless
free scalar fields $\phi^{\a}(x)$ with the normalisation of the
fundamental
two-point function being  as in (\ref{2.1.1}). In this case an explicit
expression for the
four-point function (\ref{2.1.2}) can be found using Wick's theorem with
elementary contraction (\ref{2.1.1}). It is then easy to see that
\begin{equation}
F_{f}(u,v)=C_{\phi}^{2}\,(uv)^{-\frac{1}{3}\eta},\label{3.1.1}
\end{equation}
where the subscript $f$ stands for ``free field theory''. We may also represent
graphically the above
result as   shown in Fig. \ref{fg1}  where the solid lines stand for the
two-point function (\ref{2.1.1}) (without
the $O(N)$ indices).
\begin{figure}[t]

\setlength{\unitlength}{0.01100in}%
\begingroup\makeatletter
\def\x#1#2#3#4#5#6#7\relax{\def\x{#1#2#3#4#5#6}}%
\expandafter\x\fmtname xxxxxx\relax \def\y{splain}%
\ifx\x\y   
\gdef\SetFigFont#1#2#3{%
  \ifnum #1<17\tiny\else \ifnum #1<20\small\else
  \ifnum #1<24\normalsize\else \ifnum #1<29\large\else
  \ifnum #1<34\Large\else \ifnum #1<41\LARGE\else
     \huge\fi\fi\fi\fi\fi\fi
  \csname #3\endcsname}%
\else
\gdef\SetFigFont#1#2#3{\begingroup
  \count@#1\relax \ifnum 25<\count@\count@25\fi
  \def\x{\endgroup\@setsize\SetFigFont{#2pt}}%
  \expandafter\x
    \csname \romannumeral\the\count@ pt\expandafter\endcsname
    \csname @\romannumeral\the\count@ pt\endcsname
  \csname #3\endcsname}%
\fi
\endgroup
\begin{picture}(190,71)(-30,530)
\thicklines
\put(320,590){\line( 1, 0){ 60}}
\put(320,545){\line( 1, 0){ 60}}

\put(0,565){\makebox(0,0)[lb]{\smash{\SetFigFont{10}{14.4}{rm}
$\Phi_{f}(x_{1},x_{2},x_{3},x_{4})=
C_{\phi}^{2}{\displaystyle{\frac{1}{(x_{12}^{2}x_{34}^{2})^{\eta}}}}
=H(\eta,x)F_{f}(u,v)=$}}}
\put(343,608){\makebox(0,0)[lb]{\smash{\SetFigFont{10}{14.4}{rm}$
{\cal{G}}_{0}$}}}

\put(310,595){\makebox(0,0)[lb]{\smash{\SetFigFont{10}{14.4}{rm}$x_{1}$}}}
\put(380,595){\makebox(0,0)[lb]{\smash{\SetFigFont{10}{14.4}{rm}$x_{2}$}}}
\put(310,535){\makebox(0,0)[lb]{\smash{\SetFigFont{10}{14.4}{rm}$x_{3}$}}}
\put(380,535){\makebox(0,0)[lb]{\smash{\SetFigFont{10}{14.4}{rm}$x_{4}$}}}
\end{picture}

\caption{The Graphical Expansion for
$\Phi_{f}(x_{1},x_{2},x_{3},x_{4})$}\label{fg1}

\end{figure}
It is then easy to obtain
\begin{eqnarray}
F_{S,f}(u,v) & = &
F_{f}(u,v)+\frac{1}{N}\left(F_{f}(\frac{1}{u},\frac{v}{u})
+F_{f}(\frac{1}{v},\frac{u}{v})\right)
\nonumber \\
 &  \equiv &{\cal{F}}_{S,f}(Y,W)=
C_{\phi}^{2}\,W^{-\frac{2}{3}\eta}+\frac{1}{N}C_{\phi}^{2}
\,W^{\frac{1}{3}\eta}\Bigg(2+\frac{1}{4}\eta^{2}
Y^{2}+\cdots\Bigg),\label{3.1.2}
\end{eqnarray}
and
\begin{eqnarray}
F_{V,f}(u,v) & = &
F_{f}(\frac{1}{u},\frac{v}{u})-F_{f}(\frac{1}{v},\frac{u}{v})
\nonumber \\
 &  \equiv &{\cal{F}}_{V,f}(Y,W)=
\eta\,C_{\phi}^{2}\,W^{\frac{1}{3}\eta}\,Y+\cdots.\label{3.1.3}
\end{eqnarray}
Equations (\ref{3.1.2}) and (\ref{3.1.3}) have to be compared with
(\ref{2.2.19}) and (\ref{2.2.20})
respectively. Consistency of (\ref{3.1.2}) and (\ref{2.2.19}) requires
\begin{eqnarray}
\mu -1 -{\textstyle{\frac{2}{3}\eta}} =  {\textstyle{\frac{1}{3}}}\eta
&  \Rightarrow & \eta =\mu -1, \label{3.1.4}\\
{\textstyle{\frac{1}{2}}}\eta_{o}-{\textstyle{\frac{2}{3}}}\eta =
{\textstyle{\frac{1}{3}}}\eta & \Rightarrow & \eta_{o}=2\eta, \label{3.1.5}\\
g_{\phi\phi O}^{2} & = & \frac{2}{N}C_{O}C_{\phi}^{2}. \label{3.1.6}
\end{eqnarray}
Comparing the coefficient of $Y^{2}$ in (\ref{3.1.2}) and (\ref{2.2.19}) we
obtain by virtue
of (\ref{3.1.4})-(\ref{3.1.6}) using   also (\ref{2.1.21}) for $g_{\phi\phi T}$
\begin{eqnarray}
\frac{\eta^{2}}{N}C_{\phi}^{2} & = &
-\frac{g_{\phi\phi
O}^{2}}{C_{O}}\frac{\eta_{o}^{2}}{8(\eta_{o}+1)}
+\frac{d^{2}\eta^{2}}{(d-1)^{2}S_{d}^{2}C_{T}}C_{\phi}^{2},
\nonumber \\
 & \Rightarrow & C_{T}=N\frac{d}{(d-1)S_{d}^{2}}.\label{3.1.7}
\end{eqnarray}
Using the results above  we also see that
\begin{equation}
\frac{g_{\phi\phi
O}^{2}}{C_{O}}\frac{\eta_{o}^{3}}{16(\eta_{o}+1)(\eta_{o}+1-\mu)}
-\frac{g_{\phi\phi
T}^{2}}{C_{T}}\frac{1}{d}=0.\label{3.1.8}
\end{equation}
as required for consistency of the coefficients of $W$ in (\ref{3.1.2}) and
(\ref{2.2.19}).

Comparing (\ref{3.1.3}) and (\ref{2.2.20})
we  obtain by virtue of the results above and also  (\ref{2.1.22}) for
$g_{\phi\phi J}$
\begin{equation}
C_{J}=\frac{2}{(d-2)S_{d}^{2}}.\label{3.1.9}
\end{equation}
The values for the dimensions $\eta$ and $\eta_{o}$, for the
coupling $g_{\phi\phi O}$ and for the normalisations of the conserved
currents $C_{T}$ and $C_{J}$ obtained above, are in agreement with the
results given
\eg{} in \cite{anastasios} for the theory on $N$ massless free scalar
fields in any dimension $d$.

\section{The Four-Point Function in   a Non-Trivial CFT in
${2<d<4}$}\label{sec4}
\setcounter{equation}{0}

\subsection{The Graphical Expansion}\label{sbsec41}

Although the theory considered in the previous section is
trivial, it suggests a possible treatment for  a
non-trivial conformally invariant $O(N)$ vector model. In particular,
we showed  that consistency of the (trivial) graphical
representation in Fig. \ref{fg1} with the OPE (\ref{2.2.4}) and the resulting
expressions
(\ref{2.2.19}) and (\ref{2.2.20}) for the four-point function, determines the
values of the dimensions $\eta$, $\eta_{o}$ and the other dynamical
parameters to be the corresponding ones for  a free
theory of $N$ massless scalar fields.

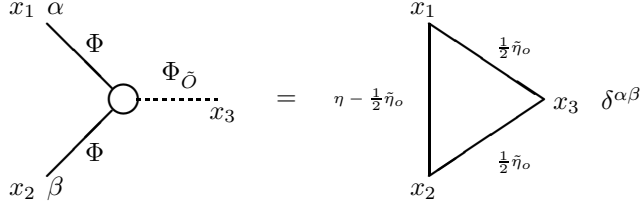
\begin{figure}[t]

\setlength{\unitlength}{0.01000in}%
\begingroup\makeatletter
\def\x#1#2#3#4#5#6#7\relax{\def\x{#1#2#3#4#5#6}}%
\expandafter\x\fmtname xxxxxx\relax \def\y{splain}%
\ifx\x\y   
\gdef\SetFigFont#1#2#3{%
  \ifnum #1<17\tiny\else \ifnum #1<20\small\else
  \ifnum #1<24\normalsize\else \ifnum #1<29\large\else
  \ifnum #1<34\Large\else \ifnum #1<41\LARGE\else
     \huge\fi\fi\fi\fi\fi\fi
  \csname #3\endcsname}%
\else
\gdef\SetFigFont#1#2#3{\begingroup
  \count@#1\relax \ifnum 25<\count@\count@25\fi
  \def\x{\endgroup\@setsize\SetFigFont{#2pt}}%
  \expandafter\x
    \csname \romannumeral\the\count@ pt\expandafter\endcsname
    \csname @\romannumeral\the\count@ pt\endcsname
  \csname #3\endcsname}%
\fi
\endgroup
\begin{picture}(275,101)(100,530)
\thicklines
\put(280,580){\circle{15}}
\put(240,620){\line( 1,-1){ 34}}
\put(240,540){\line( 1, 1){ 34}}
\multiput(287,580)(4.5,0.0){10}{\line( 1, 0){ 2}}
\put(440,620){\line( 3,-2){ 60}}
\put(500,580){\line(-3,-2){ 60}}
\put(440,540){\line( 0, 1){ 80}}
\put(360,577){\makebox(0,0)[lb]{\smash{\SetFigFont{10}{14.4}{rm}$=$}}}
\put(220,625){\makebox(0,0)[lb]{\smash{\SetFigFont{10}{14.4}{rm}$x_{1}$}}}
\put(220,530){\makebox(0,0)[lb]{\smash{\SetFigFont{10}{14.4}{rm}$x_{2}$}}}
\put(325,570){\makebox(0,0)[lb]{\smash{\SetFigFont{10}{14.4}{rm}$x_{3}$}}}
\put(430,625){\makebox(0,0)[lb]{\smash{\SetFigFont{10}{14.4}{rm}$x_{1}$}}}
\put(505,575){\makebox(0,0)[lb]{\smash{\SetFigFont{10}{14.4}{rm}$x_{3}$}}}
\put(430,530){\makebox(0,0)[lb]{\smash{\SetFigFont{10}{14.4}{rm}$x_{2}$}}}
\put(390,578){\makebox(0,0)[lb]{\smash{\SetFigFont{6}{14.4}{rm}$\eta
-\frac{1}{2}\tilde{\eta}_{o}$}}}
\put(475,605){\makebox(0,0)[lb]{\smash{\SetFigFont{6}{14.4}{rm}$
\frac{1}{2}\tilde{\eta}_{o}$}}}
\put(475,545){\makebox(0,0)[lb]{\smash{\SetFigFont{6}{14.4}{rm}$
\frac{1}{2}\tilde{\eta}_{o}$}}}
\put(260,605){\makebox(0,0)[lb]{\smash{\SetFigFont{10}{14.4}{rm}$\Phi$}}}
\put(260,548){\makebox(0,0)[lb]{\smash{\SetFigFont{10}{14.4}{rm}$\Phi$}}}
\put(300,590){\makebox(0,0)[lb]{\smash{\SetFigFont{10}{14.4}{rm}$
\Phi_{\tilde{O}}$}}}
\put(240,625){\makebox(0,0)[lb]{\smash{\SetFigFont{10}{14.4}{rm}$\a$}}}
\put(240,530){\makebox(0,0)[lb]{\smash{\SetFigFont{10}{14.4}{rm}$\b$}}}
\put(530,575){\makebox(0,0)[lb]{\smash{\SetFigFont{10}{14.4}{rm}$
\delta^{\a\b}$}}}
\end{picture}

\caption{The Conformally Invariant Three-Point Function  $\la\phi\phi
\tilde{O}\ra$.}\label{fg2}

\end{figure}

We propose that a graphical expansion for a non-trivial
$\Phi(x_{1},x_{2},x_{3},x_{4})$ in (\ref{2.1.2})
may  be obtained by introducing a
conformally invariant vertex in the theory. This vertex  is assumed to
describe the interaction of $\phi^{\a}(x)$ with  an
arbitrary scalar field $\tilde{O}(x)$, which is a $O(N)$ singlet
and has   dimension $\tilde{\eta}_{o}$
with $0<\tilde{\eta}_{o}<d$, whose two-point function is normalised as
\begin{equation}
\la\tilde{O}(x_{1})\tilde{O}(x_{2})\ra\equiv\Phi_{\tilde{O}}(x_{12})
=C_{\tilde{O}}\frac{1}{x_{12}^{2\tilde{\eta}_{o}}},\label{4.1.1}
\end{equation}
and it is represented diagrammatically as a dashed line. Then, the  full
three-point function of $\tilde{O}(x)$ with $\phi^{\a}(x)$ is
analogously to (\ref{2.1.13})
\begin{equation}
\la\phi^{\a}(x_{1})\phi^{\b}(x_{2})\tilde{O}(x_{3})\ra
=g_{*}\frac{1}{(x_{12}^{2})^{\eta-\frac{1}{2}\tilde{\eta}_{o}}
(x_{13}^{2}x_{23}^{2})^{\frac{1}{2}\tilde{\eta}_{o}}}\delta^{\a\b}.
\label{4.1.2}
\end{equation}
Diagrammatically this can be represented as shown in Fig. \ref{fg2}.
The coupling constant $g_{*}$ has to be determined from the
dynamics of the theory. Next, we  assume that the amplitudes for
$n$-point functions of $\phi^{\a}(x)$ with $n\geq 4$ in our non-trivial CFT are
constructed in terms of skeleton graphs, with no self-energy or vertex
insertions, with internal lines corresponding to the two-point
functions of $\phi^{\a}(x)$ and
$\tilde{O}(x)$. Symmetry factors are determined as in the usual
Feynman perturbation expansion. Graphical expansions in field theory are
usually
connected with a Lagrangian formalism. In the present work however, we
consider a formulation for a non-trivial CFT based on a  graphical
expansion without explicit reference to an underlying Lagrangian.

For use in the graphical expansion
it is necessary to require  amputation of the three-point function
(\ref{4.1.2}) for any leg linked to an internal line of a graph. For example,
the inverse kernel for the two-point function (\ref{4.1.1}) is
\begin{equation}
\Phi^{-1}_{\tilde{O}}(x_{12})=\frac{1}{C_{\tilde{O}}}\rho(\tilde{\eta}_{o})
\frac{1}{(x_{12}^{2})^{d-\tilde{\eta}_{o}}},\label{4.1.3}
\end{equation}
where
\begin{equation}
\rho(\tilde{\eta}_{o})=\frac{1}{\pi^{d}}\frac{\Gamma(d-\tilde{\eta}_{o})
\Gamma(\tilde{\eta}_{o})}{\Gamma(\tilde{\eta}_{o}-\mu)\Gamma(\mu
-\tilde{\eta}_{o})},\label{4.1.4}
\end{equation}
and it is represented diagrammatically by a dotted line. We may then
amputate the three-point function (\ref{4.1.2}) on the $\tilde{O}$ leg
obtaining the following vertex
\begin{eqnarray}
V_{\phi\phi\tilde{O}}^{\a\b}(x_{1},x_{2},x_{3}) & = & \int
d^{d}x\la\phi^{\a}(x_{1})\phi^{\b}(x_{2})\tilde{O}(x)\ra
\Phi^{-1}_{\tilde{O}}(x-x_{3})
\nonumber \\
 &  = &
\frac{g_{*}}{C_{\tilde{O}}}\rho({\tilde{\eta}}_{o})
U({\textstyle{\frac{1}{2}}}\tilde{\eta}_{o},{\textstyle{\frac{1}{2}}}
\tilde{\eta}_{o},d-\tilde{\eta}_{o})\frac{1}{(x_{12}^{2})^{\eta-\mu
+\frac{1}{2}\tilde{\eta}_{o}}(x_{13}^{2}x_{23}^{2})^{\mu-\frac{1}{2}
\tilde{\eta}_{o}}}\delta^{\a\b}.\label{4.1.5}
\end{eqnarray}
This amputation can be diagrammatically represented as shown in Fig. \ref{fg3}.
\begin{figure}

\setlength{\unitlength}{0.0100in}%
\begingroup\makeatletter
\def\x#1#2#3#4#5#6#7\relax{\def\x{#1#2#3#4#5#6}}%
\expandafter\x\fmtname xxxxxx\relax \def\y{splain}%
\ifx\x\y   
\gdef\SetFigFont#1#2#3{%
  \ifnum #1<17\tiny\else \ifnum #1<20\small\else
  \ifnum #1<24\normalsize\else \ifnum #1<29\large\else
  \ifnum #1<34\Large\else \ifnum #1<41\LARGE\else
     \huge\fi\fi\fi\fi\fi\fi
  \csname #3\endcsname}%
\else
\gdef\SetFigFont#1#2#3{\begingroup
  \count@#1\relax \ifnum 25<\count@\count@25\fi
  \def\x{\endgroup\@setsize\SetFigFont{#2pt}}%
  \expandafter\x
    \csname \romannumeral\the\count@ pt\expandafter\endcsname
    \csname @\romannumeral\the\count@ pt\endcsname
  \csname #3\endcsname}%
\fi
\endgroup
\begin{picture}(275,101)(100,530)
\thicklines

\put(180,620){\line( 3,-2){ 60}}
\put(180,540){\line( 0, 1){ 80}}
\put(240,580){\line(-3,-2){ 60}}
\multiput(240,580)( 3.0, 0.0){15}{\circle*{2}}
\put(580,620){\line( 3,-2){ 60}}
\put(640,580){\line(-3,-2){ 60}}
\put(580,540){\line( 0, 1){ 80}}
\put(310,577){\makebox(0,0)[lb]{\smash{\SetFigFont{10}{14.4}{rm}$=$}}}

\put(334,577){\makebox(0,0)[lb]{\smash{\SetFigFont{10}{14.4}{rm}$
\frac{1}{C_{\tilde{O}}}\rho(\tilde{\eta}_{o})\,U({\textstyle{\frac{1}{2}}}
\tilde{\eta}_{o},{\textstyle{\frac{1}{2}}}\tilde{\eta}_{o},
d-\tilde{\eta}_{o})$}}}

\put(160,625){\makebox(0,0)[lb]{\smash{\SetFigFont{10}{14.4}{rm}$x_{1}$}}}
\put(160,530){\makebox(0,0)[lb]{\smash{\SetFigFont{10}{14.4}{rm}$x_{2}$}}}
\put(280,570){\makebox(0,0)[lb]{\smash{\SetFigFont{10}{14.4}{rm}$x_{3}$}}}
\put(240,565){\makebox(0,0)[lb]{\smash{\SetFigFont{10}{14.4}{rm}$x$}}}
\put(570,625){\makebox(0,0)[lb]{\smash{\SetFigFont{10}{14.4}{rm}$x_{1}$}}}
\put(645,575){\makebox(0,0)[lb]{\smash{\SetFigFont{10}{14.4}{rm}$x_{3}$}}}
\put(570,530){\makebox(0,0)[lb]{\smash{\SetFigFont{10}{14.4}{rm}$x_{2}$}}}
\put(515,578){\makebox(0,0)[lb]{\smash{\SetFigFont{6}{14.4}{rm}$\eta
-\mu+\frac{1}{2}\tilde{\eta}_{o}$}}}
\put(615,605){\makebox(0,0)[lb]{\smash{\SetFigFont{6}{14.4}{rm}$
\mu-\frac{1}{2}\tilde{\eta}_{o}$}}}
\put(615,545){\makebox(0,0)[lb]{\smash{\SetFigFont{6}{14.4}{rm}$
\mu-\frac{1}{2}\tilde{\eta}_{o}$}}}
\put(215,605){\makebox(0,0)[lb]{\smash{\SetFigFont{6}{14.4}{rm}$
\frac{1}{2}\tilde{\eta}_{o}$}}}
\put(215,545){\makebox(0,0)[lb]{\smash{\SetFigFont{6}{14.4}{rm}$
\frac{1}{2}\tilde{\eta}_{o}$}}}
\put(130,578){\makebox(0,0)[lb]{\smash{\SetFigFont{6}{14.4}{rm}$\eta
-\frac{1}{2}\tilde{\eta}_{o}$}}}
\put(250,590){\makebox(0,0)[lb]{\smash{\SetFigFont{10}{14.4}{rm}$
\Phi^{-1}_{\tilde{O}}$}}}
\put(180,625){\makebox(0,0)[lb]{\smash{\SetFigFont{10}{14.4}{rm}$\a$}}}
\put(180,530){\makebox(0,0)[lb]{\smash{\SetFigFont{10}{14.4}{rm}$\b$}}}
\put(670,575){\makebox(0,0)[lb]{\smash{\SetFigFont{10}{14.4}{rm}$
\delta^{\a\b}$}}}
\end{picture}

\caption{The Three-Point Function $\la\phi\phi\tilde{O}\ra$ with
amputation on the $\tilde{O}$ leg.}\label{fg3}

\end{figure}
In obtaining (\ref{4.1.5}) we have used  the D'EPP formula \cite{DEPP}
\begin{eqnarray}
&  & \int
d^{d}x\frac{1}{(x_{1}-x)^{2a_{1}}(x_{2}-x)^{2a_{2}}(x_{3}-x)^{2a_{3}}}
\nonumber \\
 &  & {}= U(a_{1},a_{2},a_{3})\frac{1}{(x_{12}^{2})^{\mu
-a_{3}}(x_{13}^{2})^{\mu -a_{2}}(x_{23}^{2})^{\mu
-a_{1}}},\label{4.1.6}
\end{eqnarray}
which is valid for
$a_{1}+a_{2}+a_{3}=d$, with
\begin{equation}
U(a_{1},a_{2},a_{3})=\pi^{\mu}\frac{\Gamma(\mu
-a_{1})\Gamma(\mu -a_{2})\Gamma(\mu
-a_{3})}{\Gamma(a_{1})\Gamma(a_{2})\Gamma(a_{3})}.\label{4.1.7}
\end{equation}

We denote by $\Phi^{(\tilde{\eta}_{o})}(x_{1},x_{2},x_{3},x_{4})$ the
amplitude of interest in our graphical treatment  of the four-point
function (\ref{2.1.2}). The first few
graphs in the skeleton expansion for this amplitude in increasing
order according to the number of vertices are displayed in Fig. \ref{fg4} where
\begin{figure}[b]

\setlength{\unitlength}{0.01100in}%
\begingroup\makeatletter
\def\x#1#2#3#4#5#6#7\relax{\def\x{#1#2#3#4#5#6}}%
\expandafter\x\fmtname xxxxxx\relax \def\y{splain}%
\ifx\x\y   
\gdef\SetFigFont#1#2#3{%
  \ifnum #1<17\tiny\else \ifnum #1<20\small\else
  \ifnum #1<24\normalsize\else \ifnum #1<29\large\else
  \ifnum #1<34\Large\else \ifnum #1<41\LARGE\else
     \huge\fi\fi\fi\fi\fi\fi
  \csname #3\endcsname}%
\else
\gdef\SetFigFont#1#2#3{\begingroup
  \count@#1\relax \ifnum 25<\count@\count@25\fi
  \def\x{\endgroup\@setsize\SetFigFont{#2pt}}%
  \expandafter\x
    \csname \romannumeral\the\count@ pt\expandafter\endcsname
    \csname @\romannumeral\the\count@ pt\endcsname
  \csname #3\endcsname}%
\fi
\endgroup
\begin{picture}(444,120)(40,570)
\thicklines
\put(165,665){\line( 1, 0){ 60}}
\put(165,615){\line( 1, 0){ 60}}
\put(285,615){\line( 1, 0){ 60}}
\put(405,665){\line( 1, 0){ 60}}
\put(405,615){\line( 1, 0){ 60}}
\put(285,665){\line( 1, 0){ 60}}
\put(315,665){\circle*{8}}
\put(420,665){\circle*{8}}
\put(450,665){\circle*{8}}
\put(315,615){\circle*{8}}
\put(420,615){\circle*{8}}
\put(450,615){\circle*{8}}

\multiput(315,665)(0.00000,-4.00000){13}{\line( 0,-1){  2.000}}
\multiput(420,665)(0.00000,-4.00000){13}{\line( 0,-1){  2.000}}
\multiput(450,665)(0.00000,-4.00000){13}{\line( 0,-1){  2.000}}
\thinlines
\put(405,700){\line(-1, 0){ 15}}
\put(390,700){\line( 0,-1){120}}
\put(390,580){\line( 1, 0){ 15}}
\put(550,700){\line( 1, 0){ 15}}
\put(565,700){\line( 0,-1){120}}
\put(565,580){\line(-1, 0){ 15}}
\thicklines
\put(160,670){\makebox(0,0)[lb]{\smash{\SetFigFont{10}{14.4}{rm}$x_{1}$}}}
\put(220,670){\makebox(0,0)[lb]{\smash{\SetFigFont{10}{14.4}{rm}$x_{2}$}}}
\put(160,605){\makebox(0,0)[lb]{\smash{\SetFigFont{10}{14.4}{rm}$x_{3}$}}}
\put(220,605){\makebox(0,0)[lb]{\smash{\SetFigFont{10}{14.4}{rm}$x_{4}$}}}
\put(280,605){\makebox(0,0)[lb]{\smash{\SetFigFont{10}{14.4}{rm}$x_{3}$}}}
\put(280,670){\makebox(0,0)[lb]{\smash{\SetFigFont{10}{14.4}{rm}$x_{1}$}}}
\put(340,670){\makebox(0,0)[lb]{\smash{\SetFigFont{10}{14.4}{rm}$x_{2}$}}}
\put(340,605){\makebox(0,0)[lb]{\smash{\SetFigFont{10}{14.4}{rm}$x_{4}$}}}
\put(400,670){\makebox(0,0)[lb]{\smash{\SetFigFont{10}{14.4}{rm}$x_{1}$}}}
\put(460,670){\makebox(0,0)[lb]{\smash{\SetFigFont{10}{14.4}{rm}$x_{2}$}}}
\put(400,605){\makebox(0,0)[lb]{\smash{\SetFigFont{10}{14.4}{rm}$x_{3}$}}}
\put(460,605){\makebox(0,0)[lb]{\smash{\SetFigFont{10}{14.4}{rm}$x_{4}$}}}
\put(40,637){\makebox(0,0)[lb]{\smash{\SetFigFont{10}{14.4}{rm}$
\Phi^{(\tilde{\eta}_{o})}(x_{1},x_{2},x_{3},x_{4})$ = }}}
\put(250,637){\makebox(0,0)[lb]{\smash{\SetFigFont{10}{14.4}{rm}$+$}}}
\put(350,637){\makebox(0,0)[lb]{\smash{\SetFigFont{10}{14.4}{rm}$+$}}}

\put(475,637){\makebox(0,0)[lb]{\smash{\SetFigFont{10}{14.4}{rm}$+\,\,\,\,\,
(x_{3}\leftrightarrow x_{4}$)}}}
\put(575,637){\makebox(0,0)[lb]{\smash{\SetFigFont{10}{14.4}{rm}$+\cdots$}}}
\put(190,685){\makebox(0,0)[lb]{\smash{\SetFigFont{10}{14.4}{rm}${
\cal{G}}_{0}$}}}
\put(310,685){\makebox(0,0)[lb]{\smash{\SetFigFont{10}{14.4}{rm}${
\cal{G}}_{1}^{(\tilde{\eta}_{o})}$}}}
\put(430,685){\makebox(0,0)[lb]{\smash{\SetFigFont{10}{14.4}{rm}${
\cal{G}}_{2}^{(\tilde{\eta}_{o})}$}}}

\end{picture}

\caption{The Skeleton Graph Expansion for
$\Phi^{(\tilde{\eta}_{o})}(x_{1},x_{2},x_{3},x_{4})$}\label{fg4}

\end{figure}
the dots stand for graphs with more than
four vertices. The solid
lines stand for the two-point function (\ref{2.1.1}) and the dashed lines stand
for the two-point function (\ref{4.1.1}). The vertices (dark blobs) formed by
two solid and one dashed lines stand for the three-point function (\ref{4.1.2})
with amputation on all  legs. Integration over the
coordinates of all internal vertices is understood. The superscript
$(\tilde{\eta}_{o})$ denotes skeleton graphs built using dashed lines
corresponding  to the two-point
function of the scalar field $\tilde{O}(x)$ of  dimension
$\tilde{\eta}_{o}$.

Note that the amplitude
$\Phi^{(\tilde{\eta}_{o})}(x_{1},x_{2},x_{3},x_{4})$ as given by the
skeleton expansion of Fig. \ref{fg4}  satisfies the crossing symmetry
properties
(\ref{2.1.3}). The graphs displayed
in Fig. \ref{fg4} have the interesting  property
\begin{eqnarray}
\Phi^{(d-\tilde{\eta}_{o})}(x_{1},x_{2},x_{3},x_{4})& = &
{\cal{G}}_{0}(x_{1},x_{2},x_{3},x_{4})+C(\tilde{\eta}_{o})
{\cal{G}}_{1}^{(\tilde{\eta}_{o})}(x_{1},x_{2},x_{3},x_{4})\nonumber
\\
 &  &
\!\!\!\!\!{}+\left[C(\tilde{\eta}_{o})\right]^{2}\left({\cal{G}}_{2}^{
(\tilde{\eta}_{o})}(x_{1},x_{2},x_{3},x_{4})+{\cal{G}}_{2}^
{(\tilde{\eta}_{o})}(x_{1},x_{2},x_{4},x_{3})\right)+\cdot\cdot .
\label{4.1.8}
\end{eqnarray}
The superscript $(d-\tilde{\eta}_{o})$ on the l.h.s. of (\ref{4.1.8}) denotes
the same  graphical expansion as previously but  using for the
internal dashed
lines the two-point function (\ref{4.1.1}) with dimension
$d-\tilde{\eta}_{o}$ and the same normalisation $C_{\tilde{O}}$, while
the dark blobs now correspond to the interaction of $\phi^{\a}(x)$
with a scalar field $\tilde{O}_{s}(x)$ of dimension
$d-\tilde{\eta}_{o}$ via a unique vertex having the same coupling
constant $g_{*}$. The coefficient
$C(\tilde{\eta}_{o})$ is given by
\begin{eqnarray}
C(d-\tilde{\eta}_{o}) = C^{-1}(\tilde{\eta}_{o}) & = &
\frac{\rho(\tilde{\eta}_{o})U({\textstyle{\frac{1}{2}}}\tilde{\eta}_{o},
{\textstyle{\frac{1}{2}}}\tilde{\eta}_{o},d-\tilde{\eta}_{o})}
{\rho(d-\tilde{\eta}_{o})U(\mu
-{\textstyle{\frac{1}{2}}}\tilde{\eta}_{o},\mu
-{\textstyle{\frac{1}{2}}}\tilde{\eta}_{o},\tilde{\eta}_{o})} \nonumber \\
 & = &
\frac{\Gamma(\tilde{\eta}_{o})\Gamma(\tilde{\eta}_{o}-\mu)\Gamma^{4}
(\mu-\frac{1}{2}\tilde{\eta}_{o})}{\Gamma(2\mu-\tilde{\eta}_{o})
\Gamma(\mu-\tilde{\eta}_{o})\Gamma^{4}(\frac{1}{2}\tilde{\eta}_{o})}.
\label{4.1.9}
\end{eqnarray}
The
{\it{shadow symmetry}} properties of the skeleton graphs
${\cal{G}}_{1}^{(\tilde{\eta}_{o})}$
and ${\cal{G}}_{2}^{(\tilde{\eta}_{o})}$ required for obtaining (\ref{4.1.8})
are
proved in  Appendices \ref{apdx2} and \ref{apdx3}. An obvious conjecture is
that
each graph in the full  skeleton expansion \footnote{Assuming that
only vertices formed by two solid and one dashed lines are present.} for
$\Phi^{(d-\tilde{\eta}_{o})}(x_{1},x_{2},x_{3},x_{4})$ is proportional
to the corresponding graph in the skeleton expansion for
$\Phi^{(\tilde{\eta}_{o})}(x_{1},x_{2},x_{3},x_{4})$ with
proportionality constant $\left[C(\tilde{\eta}_{o})\right]^{n}$ where $2n$ is
the
number of vertices formed by two solid and one dashed lines in the graph. In
the following we will see that assuming this symmetry property of the skeleton
graph expansion holds  to
all orders we arrive at an interesting duality
of the $O(N)$ theory.

\subsection{The Consistency Relations}\label{sbsec42}

The crucial consistency requirement regarding the present work is that
amplitudes constructed according to graphical expansions such as the
one  in  Fig. \ref{fg4}
correspond to a CFT having operator content in agreement with the OPE
ansatz (\ref{2.2.4}) and are therefore compatible with amplitudes obtained by
straightforward application of this ansatz on $n$-point functions.
Without further input at this point we have no intrinsic means in estimating
the magnitude of
$g_{*}$ and hence we cannot hope to obtain  a weak coupling
expansion.  However, on account of the  $O(N)$ symmetry we
subsequently see that the assumption   $g_{*}^{2}=O(1/N)$ leads naturally to a
well defined perturbation expansion in
$1/N$ for the theory. Therefore, from now on we consider the theory
for large $N$.

An important point concerning the consistency relations is that
one has to conduct the calculation of the amplitudes corresponding to
skeleton graphs like the ones in Fig. \ref{fg4}, in such a way that the
resulting
expressions can be compared with the OPE results (\ref{2.2.19}) and
(\ref{2.2.20}). In the context of this subsection this means that one
must ensure that these amplitudes
are explicitly expressed as  functions of $Y$ and $W$ which allow
for the transparent and unambiguous evaluation of the limit as
$Y\rightarrow 0$, $W\rightarrow 0$ {\it{independently}}. Our
subsequent calculations will clarify further this point.

Then, by virtue of the results in the Appendices we can write
\begin{eqnarray}
{\cal{F}}_{S}^{(\tilde{\eta}_{o})}(Y,W) & =  &
C_{\phi}^{2}W^{-\frac{2}{3}\eta}+\frac{1}{N}C_{\phi}^{2}
W^{\frac{1}{3}\eta}\left[(1-Y)^{-\frac{1}{2}\eta}+(1-Y)^{\frac{1}{2}\eta}
\right]
\nonumber \\
&  &
{}+\frac{g_{*}^{2}}{C_{\tilde{O}}}W^{-\frac{2}{3}\eta}
\Bigl[{\cal{H}}_{\tilde{\eta}_{o}}(u,v)+C(d-\tilde{\eta}_{o})
{\cal{H}}_{d-\tilde{\eta}_{o}}(u,v)\Bigl]
\nonumber \\
&  &
{}+\frac{1}{N}\frac{g_{*}^{2}}{C_{\tilde{O}}}{\cal{F}}_{S,1}^
{(\tilde{\eta}_{o})}(Y,W)+\left(\frac{g_{*}^{2}}{C_{\tilde{O}}}\right)^{2}
{\cal{F}}_{S,2}^{(\tilde{\eta}_{o})}(Y,W)+O(\frac{1}{N^{3}}),\label{4.2.1}
\end{eqnarray}
where
\begin{eqnarray}
\frac{g_{*}^{2}}{C_{\tilde{O}}}{\cal{F}}_{S,1}^{(\tilde{\eta}_{o})}(Y,W)
& = &
H^{-1}(\eta,x)\left[{\cal{G}}_{1}^{(\tilde{\eta}_{o})}(x_{1},x_{3},
x_{2},x_{4})+{\cal{G}}_{1}^{(\tilde{\eta}_{o})}
(x_{1},x_{4},x_{3},x_{2})\right], \\
\left(\frac{g_{*}^{2}}{C_{\tilde{O}}}\right)^{2}{\cal{F}}_{S,2}^
{(\tilde{\eta}_{o})}(Y,W)
& =  &
H^{-1}(\eta,x)\left[{\cal{G}}_{2}^{(\tilde{\eta}_{o})}(x_{1},x_{2},
x_{3},x_{4})+{\cal{G}}_{2}^{(\tilde{\eta}_{o})}
(x_{1},x_{2},x_{4},x_{3})\right].\label{4.2.2}
\end{eqnarray}
We also find
\begin{equation}
{\cal{F}}_{V}^{(\tilde{\eta}_{o})}(Y,W)  =
C_{\phi}^{2}W^{\frac{1}{3}\eta}\Bigl[(1-Y)^{-\frac{1}{2}\eta}-(1-Y)^
{\frac{1}{2}\eta}\Bigl]+\frac{g_{*}^{2}}{C_{\tilde{O}}}{\cal{F}}_{V,1}^
{(\tilde{\eta}_{o})}(Y,W)+O(\frac{1}{N^{2}}),\label{4.2.4}
\end{equation}
where
\begin{equation}
\frac{g_{*}^{2}}{C_{\tilde{O}}}{\cal{F}}_{V,1}^{(\tilde{\eta}_{o})}(Y,W)
=H^{-1}(\eta,x)\left[{\cal{G}}_{1}^{(\tilde{\eta}_{o})}(x_{1},x_{3},x_{2},
x_{4})-{\cal{G}}_{1}^{(\tilde{\eta}_{o})}(x_{1},x_{4},x_{3},x_{2})\right].
\label{4.2.5}
\end{equation}

To $O(1/N)$ we just need to consider the first two lines on the
r.h.s. of (\ref{4.2.1}). From the results for  free fields in section
\ref{sec3} we see
that the first line on the r.h.s. of (\ref{4.2.1}) is compatible with an OPE
for
a field $\phi^{\a}(x)$ of dimension $\eta =\mu -1$  with itself,
including  the energy momentum tensor and a scalar field of dimension
$2\eta=d-2$. Next, we note that for $2<d<4$ and $0<\tilde{\eta}_{o}<d$ it is
easy to show that  $C(d-\tilde{\eta}_{o})<0$. Therefore, to
leading order in $1/N$ if we take $\tilde{\eta}_{o}=2$ we must
impose
\begin{equation}
\frac{g_{*}^{2}}{C_{\tilde{O}}}C(d-\tilde{\eta}_{o})+\frac{2}{N}
C_{\phi}^{2}=0,\label{4.2.6}
\end{equation}
to ensure that only one scalar field appears in the the OPE (\ref{2.2.4}) in
accordance with our basic requirement. By virtue of the required agreement  of
${\cal{F}}_{S}^{(\tilde{\eta}_{o})}(Y,W)$ in (\ref{4.2.1})  with
the operator product expansion result ${\cal{F}}_{S}(Y,W)$ in (\ref{2.2.19}),
we
may  therefore identify the field $O(x)$ in the  (\ref{2.2.4}) with the field
$\tilde{O}(x)$ which is associated  with the graphical expansion, so
we set
$\eta_{o}=\tilde{\eta}_{o}$ and $g_{\phi\phi
O}^{2}=g_{*}^{2}$. From (\ref{4.2.6}) this implies that to leading order in
$1/N$
\begin{eqnarray}
g_{\phi\phi O}^{2} & = & -\frac{2}{N}C(2)C_{\phi}^{2}C_{\tilde{O}}
\nonumber \\
 &  = &
\frac{2}{N}\frac{\Gamma(2\mu-2)}{\Gamma(3-\mu)\Gamma^{3}(\mu-1)}
C_{\phi}^{2}C_{\tilde{O}}.\label{4.2.7}
\end{eqnarray}

For simplicity we henceforth use the fact that in CFT one can
arbitrarily adjust  the normalisations of the two-point functions, (except of
those of the
conserved currents), by rescaling the relevant fields, and we set
$C_{\phi}=C_{O}=C_{\tilde{O}}=1$ (in the skeleton expansion these are
not modified). The discussion of (\ref{4.2.4}) to $O(1/N)$ and
(\ref{4.2.1}) to $O(1/N^{2})$ is achieved assuming a $1/N$
expansion for the dynamical parameters of the theory as follows
\begin{eqnarray}
\eta & = & \mu-1+\frac{1}{N}\eta_{1}, \label{4.2.8}\\
\tilde{\eta}_{o}=\eta_{o} & = & 2+\frac{1}{N}\eta_{o,1}, \label{4.2.9}\\
g_{*}^{2} = g_{\phi\phi O}^{2} & = &
\frac{2}{N}\frac{\Gamma(2\mu-2)}{\Gamma(3-\mu)\Gamma^{3}(\mu-1)}
\Bigg(1+\frac{1}{N}g_{*,1}\Bigg),\label{4.2.10}
\\
C_{J} & = &
\frac{2}{(d-2)S_{d}^{2}}\Bigg(1+\frac{1}{N}C_{J,1}\Bigg),\label{4.2.11} \\
C_{T} & = &
N\frac{d}{(d-1)S_{d}^{2}}\Bigg(1+\frac{1}{N}C_{T,1}\Bigg).\label{4.2.12}
\end{eqnarray}

It is simpler to first consider
${\cal{F}}_{V}^{(\tilde{\eta}_{o})}(Y,W)$ in (\ref{4.2.4}). Using the results
in Appendix \ref{apdx1} we can expand
$g_{*}^{2}{\cal{F}}_{V,1}^{(\tilde{\eta}_{o})}(Y,W)$ in powers
of $Y$ and then, by virtue of the required agreement of
${\cal{F}}_{V}^{(\tilde{\eta}_{o})}(Y,W)$ in (\ref{4.2.4}) with
${\cal{F}}_{V}(Y,W)$ in (\ref{2.2.20}), we obtain to $O(1/N)$ using
(\ref{2.1.22})
\begin{eqnarray}
\frac{1}{2}\frac{1}{S_{d}^{2}C_{J}} & & W^{\mu-1-\frac{2}{3}\eta}\,Y  \,= \,
\eta
\,W^{\frac{1}{3}\eta}\,Y \nonumber \\
 &  &
{}+g_{*}^{2}\,\frac{\Gamma^{2}(\mu-\frac{1}{2}\tilde{\eta}_{o})
\Gamma(\tilde{\eta}_{o})}{\Gamma^{2}(\frac{1}{2}\tilde{\eta}_{o})
\Gamma(\mu-\tilde{\eta}_{o})\Gamma(\mu)}\,W^{\frac{1}{3}\eta}
\,Y\,\Bigg[\left(\eta-\frac{\tilde{\eta}_{o}
(d-\tilde{\eta}_{o})}{d}\right)(b_{00}-\mbox{ln}W)\nonumber
\\
 &  &
\hspace{7cm}{}+1-4\frac{\tilde{\eta}_{o}(d-\tilde{\eta}_{o})}
{d^{2}}\Bigg],\label{4.2.13}
\end{eqnarray}
where
\begin{equation}
b_{00}=2\Bigl(\psi(1)+\psi(\mu)-\psi({\textstyle{\frac{1}{2}}}
\tilde{\eta}_{o})-\psi(\mu-{\textstyle{\frac{1}{2}}}
\tilde{\eta}_{o})\Bigl).\label{4.2.14}
\end{equation}
Note that (\ref{4.2.13}) is manifestly symmetric in
$\tilde{\eta}_{o}\leftrightarrow d-\tilde{\eta}_{o}$. In the first line of
(\ref{4.2.13}) we expand $\eta$ and $C_{J}$ as in (\ref{4.2.8}) and
(\ref{4.2.11}) respectively. From the results for free fields in section 3
the $O(1)$ terms  are in agreement in both
sides of (\ref{4.2.13}). To $O(1/N)$ we have also to consider the last two
lines of (\ref{4.2.13}) using  (\ref{4.2.7}) and setting $\eta=\mu-1$,
$\tilde{\eta}_{o}=2$ when $b_{00}=2/(\mu-1)$. Consistency of the
$O(1/N)$ terms in both sides of (\ref{4.2.13})  requires
\begin{eqnarray}
\eta_{1} & = & \frac{2\Gamma(2\mu
-2)}{\Gamma(1-\mu)\Gamma(\mu)\Gamma(\mu +1)\Gamma(\mu -2)},\label{4.2.15} \\
C_{J,1} & = & -\frac{2(2\mu-1)}{\mu(\mu-1)}\eta_{1}.\label{4.2.16}
\end{eqnarray}

Next, agreement of
${\cal{F}}^{(\tilde{\eta}_{o})}_{S}(Y,W)$ in (\ref{4.2.1}) with
${\cal{F}}_{S}(Y,W)$ in (\ref{2.2.19}) to $O(1/N^{2})$ requires, by virtue of
(\ref{2.1.21})
\begin{eqnarray}
 &  &
\left(\frac{d\eta}{d-1}\right)^{2}\frac{1}{S_{d}^{2}C_{T}}
W^{\mu-1-\frac{2}{3}\eta}\Bigg(\frac{1}{4}Y^{2} - \frac{1}{d}W\Bigg)
\, =\,
\frac{1}{N}W^{\frac{1}{3}\eta}\Bigg(2+\frac{1}{4}\eta^{4}Y^{2}\Bigg)
\nonumber \\
 & +& g_{\phi\phi
O}^{2}C(d-\eta_{o})W^{\frac{1}{2}(d-\eta_{o})-\frac{2}{3}\eta}\Bigg(1-
\frac{(d-\eta_{o})^{2}}{32(d-\eta_{o}+1)}Y^{2} +
\frac{(d-\eta_{o})^{3}}{16(d-\eta_{o}+1)(\mu-\eta_{o}+1)}W\Bigg)
\nonumber \\
  & + & \frac{1}{N}g_{\phi\phi
O}^{2}\,{\cal{F}}_{S,1}^{(\eta_{o})}(Y,W) + g_{\phi\phi
O}^{4}\,{\cal{F}}_{S,2}^{(\eta_{o})}(Y,W) .\label{4.2.17}
\end{eqnarray}
In obtaining (\ref{4.2.17}) we have used the result (\ref{2.2.12}) for
$C^{\eta_{o}}(x_{12},\pa_{2})C^{\eta_{o}}(x_{34},\pa_{4})
(1/x_{24}^{2\eta_{o}})$ in
${\cal{F}}_{S}(Y,W)$ to cancel  the term
$g_{*}^{2}\,W^{-\frac{2}{3}\eta}\,{\cal{H}}_{\tilde{\eta}_{o}}(u,v)$
in ${\cal{F}}^{(\tilde{\eta}_{o})}_{S}(Y,W)$ when $\eta_{o}=\tilde{\eta}_{o}$
and $g_{\phi\phi
O}^{2}=g_{*}^{2}$. We have also expanded
${\cal{H}}_{d-\tilde{\eta}_{o}}(u,v)$ as in (\ref{2.2.19}). From the
results in Appendices \ref{apdx2} and \ref{apdx3} using (\ref{4.2.7}) and
setting $\eta=\mu-1$
and $\tilde{\eta}_{o}=2$ we also obtain after some algebra
\begin{eqnarray}
 & & \frac{1}{N}  g_{\phi\phi
O}^{2}\,{\cal{F}}_{S,1}^{(\eta_{o})}(Y,W)+g_{\phi\phi
O}^{4}\,{\cal{F}}_{S,2}^{(\eta_{o})}(Y,W) \nonumber \\
 &  & =
\frac{1}{N^{2}}\frac{2\mu(4\mu-5)}{\mu-2}\,\eta_{1}\,W^{\frac{1}{3}(\mu-1)}
\nonumber  \\
& &
{}\times\Bigg[\Bigl(A_{1}-\mbox{ln}W\Bigl)+\frac{(\mu-1)^{2}}
{\mu(4\mu-5)}\Bigl(B_{1}-\mbox{ln}W\Bigl)W-\frac{(\mu-1)^{3}}
{8(4\mu-5)}\Bigl(C_{1}-\mbox{ln}W\Bigl)Y^{2}\Bigg]+\cdots\label{4.2.18}
\end{eqnarray}
where the dots stand for terms $O(Y^{3},W^{2})$ and
\begin{eqnarray}
A_{1} & = &
\frac{8\mu-11}{(\mu-1)(4\mu-5)}+\frac{2(2\mu-3)}{4\mu-5}
{\cal{C}}(\mu),\label{4.2.19}
\\
B_{1} & = &
\frac{2\mu^{2}+\mu-2}{\mu(\mu-1)(\mu+1)}+\frac{\mu^{2}-2}
{(\mu-1)(\mu+1)}{\cal{C}}(\mu),\label{4.2.20} \\
C_{1} & = &
\frac{\mu^{2}-3\mu+4}{\mu(\mu-1)(\mu+1)}+\frac{2(\mu-1)}
{\mu+1}{\cal{C}}(\mu),\label{4.2.21}
\end{eqnarray}
with
\begin{equation}
{\cal{C}}(\mu)=\psi(3-\mu)+\psi(2\mu-1)-\psi(1)-\psi(\mu)
\,\,\,\,\,,\,\,\,\,\,\psi(x)=\Gamma'(x)/\Gamma(x).\label{4.2.22}
\end{equation}
In the first two lines of (\ref{4.2.17}) we expand  $\eta$, $\eta_{o}$,
$g_{\phi\phi
O}^{2}$ and $C_{T}$ as in (\ref{4.2.8})-(\ref{4.2.10}) and
(\ref{4.2.12}) respectively. The leading order terms\footnote{These  terms are
actually $ O(1/N)$ here.} are in agreement in both sides
of (\ref{4.2.17}) as expected from the results for  free fields in section 3.
Agreement of the $O(1/N^{2})$ terms in both
sides of (\ref{4.2.17}) requires (\ref{4.2.15}) again and also
\begin{eqnarray}
\eta_{o,1} & = & 4\frac{(2\mu-1)(\mu-1)}{\mu-2}\eta_{1}, \label{4.2.23}\\
g_{*,1} & = &
-2\left(\frac{2\mu^{2}-3\mu+2}{\mu-2}{\cal{C}}(\mu)
+\frac{8\mu^{3}-24\mu^{2}+21\mu-2}{2(\mu-1)(\mu-2)}
\right)\eta_{1},\label{4.2.24}
\\
C_{T,1} & = &
-\Bigg(\frac{2}{\mu+1}{\cal{C}}(\mu)+\frac{\mu^{2}+3\mu-2}
{\mu(\mu-1)(\mu+1)}\Bigg)\eta_{1}.\label{4.2.25}
\end{eqnarray}

The discussion of the consistency relations presented above, between the
amplitudes
constructed according to the graphical expansion in Fig. \ref{fg4} and the ones
obtained from  straightforward application of  OPE (\ref{2.2.4}) on the
four-point function, is not unique. To the
order in $1/N$ considered in the present work and by virtue of the
{\it{shadow symmetry}} properties (B.3) and (C.7) proved in Appendices
B and C respectively, we could have chosen instead
$\tilde{\eta}_{o}\equiv\tilde{\eta}_{o}'=d-\eta_{o}$ and
$g_{*}^{2}\equiv\lambda^{2}_{*}=g_{\phi\phi O}^{2}C(\tilde{\eta}_{o}')$ in the
discussion of (\ref{4.2.1}). This
would have been equivalent to  identifying    the field $O(x)$ in the OPE
(\ref{2.2.4}) with the
{\it{shadow field}}\footnote{This is an
$O(d)$ scalar field associated with $O(x)$ and having dimension
$d-\tilde{\eta}_{o}$.
It is crucial to avoid having both fields and their shadows in the
OPE. For the notion of {\it{shadow fields}} in CFT see \cite{Ferrara2} and
references therein.} of  $\tilde{O}(x)$ in the graphical
expansion or vise versa. Such a choice would lead to
the same results (\ref{4.2.15}), (\ref{4.2.16}) and (\ref{4.2.23}) for the
dynamical
parameters $\eta$, $C_{J}$ and $C_{T}$ respectively of the
theory. In this case  we also   obtain
\begin{eqnarray}
\tilde{\eta}_{o}' & = &
d-2-\frac{1}{N}\frac{4(2\mu-1)(\mu-1)}{\mu-2}\eta_{1}, \label{4.2.26}\\
\lambda^{2}_{*} & = & -\frac{2}{N}\Bigg[
1+\frac{1}{N}\left(\frac{2\mu(2\mu-3)}{\mu-2}{\cal{C}}(\mu)
+\frac{\mu(8\mu-11)}{(\mu-1)(\mu-2)}\right)\eta_{1}\Bigg]. \label{4.2.27}
\end{eqnarray}
Note that, at least for $N$ large enough, $\lambda_{*}$ is purely
imaginary  when $2<d<4$ and therefore the underlying field theory is
non-unitary. However, it is perhaps interesting
to remark  that such a  non-unitary theory may be related, at least to
leading order in $1/N$,  to the free
theory of section \ref{sec3} through the correspondence
$\tilde{O}(x)\rightarrow \lambda_{*}\,\phi^{2}(x)/2$ where
\begin{equation}
\phi^{2}(x)=:\phi^{\a}(x)\phi^{\a}(x):.\label{4.2.28}
\end{equation}
Therefore, on account of
our conjecture at the end of section \ref{sbsec41} we  arrive at a
possible surprising
duality property of the $O(N)$ invariant theory. Namely,
the  field
theory underlying   the graphical expansion in Fig. \ref{fg4}, which
corresponds to
the interaction of the field $\phi^{\a}(x)$ with a scalar field
$\tilde{O}(x)$ of dimension $0<\tilde{\eta}_{o}<d$ via a unique vertex
having coupling constant $g_{*}$, may be  equivalent to a field
theory underlying an identical  graphical expansion but  corresponding to the
interaction of $\phi^{\a}(x)$ with a scalar field $\tilde{O}_{s}(x)$ of
dimension
$\tilde{\eta}_{o}'=d-\tilde{\eta}_{o}$ via a unique vertex having an
imaginary coupling coupling constant
$\lambda_{*}=\left[C(\tilde{\eta}_{o}')\right]^{\frac{1}{2}}g_{*}$. The
equivalence holds presumably in $2<d<4$. From (\ref{4.2.26}), (\ref{4.2.27})
and
(\ref{4.2.28}) one may view the theory dual to
the $O(N)$ vector model  as a possible non-unitary $1/N$
deformation of the free field theory of section 3.

\section{Four-Point Function Involving the Field $O(x)$} \label{sec5}
\setcounter{equation}{0}

In the previous section we argued for the possibility of constructing
the amplitude for the four-point function $\langle\phi\phi\phi\phi\rangle$ in
terms of skeleton graphs with internal
lines corresponding to the full two-point functions of
the fields $\phi^{\alpha}(x)$ and $\tilde{O}(x)$, the latter being
identified either with the field $O(x)$
appearing in the OPE  (\ref{2.2.4}) or with its {\it{shadow field}}. However,
up to the order  considered only graphs involving the vertex formed by two
$\phi$'s and one $\tilde{O}$ were necessary. In higher
orders, graphs involving the vertex formed by
three $\tilde{O}(x)$ fields will also appear in any  consistent graphical
treatment of $n$-point functions for  $\phi^{\a}(x)$ with $n\geq 4$,
based on a skeleton expansion with no self-energy
 or vertex insertions. This vertex  introduces a
new coupling constant $g_{\tilde{O}}$ which has to be determined in
our approach   by
requiring consistency of the algebraic and the graphical treatments of
$n$-point functions  involving  $O(x)$ or its {\it{shadow field}}. To this end,
consider the following four-point function
\begin{equation}
\langle\phi^{\alpha}(x_{1})\phi^{\beta}(x_{2})O(x_{3})
O(x_{4})\rangle=Y(x_{1},x_{2},x_{3},x_{4})\delta^{\alpha\beta},
\label{add1}
\end{equation}
which is symmetric in $x_{1}\leftrightarrow x_{2}$ and
$x_{3}\leftrightarrow x_{4}$. By virtue of conformal invariance we can write
\begin{equation}
Y(x_{1},x_{2},x_{3},x_{4})=\frac{1}{x_{12}^{2\eta}x_{34}^{2\eta_{o}}}\,Y(u,v),
\label{add01}
\end{equation}
where $Y(u,v)=Y(v,u)$ is
an arbitrary function of the usual invariant ratios $u$, $v$ given in
(\ref{2.1.6}).

There are basically two independent algebraic treatments of
(\ref{add1}) based on OPE's. One is to consider the OPE\footnote{In this
section we set to one the normalisations of the two-point
functions of all  fields unless explicitly stated otherwise.} for
$\phi^{\a}(x_{1})\phi^{\b}(x_{2})$ in (\ref{2.2.4}) together with
\begin{equation}
O(x_{1})O(x_{2})=\frac{1}{x_{12}^{2\eta_{o}}}
+g_{O}\frac{1}{(x_{12}^{2})^{\frac{1}{2}\eta_{o}}}
C^{\eta_{o}}(x_{12},\partial_{2})O(x_{2})+\cdots ,
\label{add2}
\end{equation}
where the dots stand for other possible quasiprimary fields. Following
the procedure of subsection \ref{sbsec22} it is easily seen that the
ansatz (\ref{add2}) is consistent with the conformally invariant three point
function
\begin{equation}
\langle
O(x_{1})O(x_{2})O(x_{3})\rangle=g_{O}\frac{1}
{(x_{12}^{2}x_{13}^{2}x_{23}^{2})^{\frac{1}{2}\eta_{o}}},
\label{add3}
\end{equation}
for $C^{\eta_{o}}(x_{12},\pa_{2})$ as in (\ref{2.2.11}). Substituting
(\ref{2.2.4}) and (\ref{add2}) into
(\ref{add1}) we obtain, using the results in Appendix \ref{apdx1},
\begin{equation}
Y(u,v)=1+g_{\phi\phi
O}\,g_{O}{\cal{H}}_{\eta_{o}}(u,v)+\cdots.
\label{add4}
\end{equation}
Clearly, (\ref{add4}) is appropriate in considering the limit of $Y(u,v)$ as
$u$,$v$
$\rightarrow 0 $ and the dots stand for less singular terms.

The other way of algebraically evaluating  the four-point function (\ref{add1})
is to
consider the OPE ansatz for $\phi^{\a}(x_{1})O(x_{3})$ which has a
leading term
\begin{equation}
\phi^{\alpha}(x_{1})O(x_{3})=g_{\phi\phi
O}\frac{1}{(x_{13}^{2})^{\frac{1}{2}\eta_{o}}}
C^{\eta,\eta_{o}-\eta}(x_{31},\partial_{1})\,\phi^{\alpha}(x_{1})
+\cdots , \label{add5}
\end{equation}
where,  for compatibility with (\ref{2.1.13}), $C^{a,b}(y,\pa)$
generalising (\ref{2.2.6}) is defined by the requirement
\begin{equation}
C^{a,-b}(x_{21},\pa_{1})\frac{1}{x_{13}^{2a}}
=C^{a,b}(x_{12},\pa_{2})\frac{1}{x_{23}^{2a}}
=\frac{1}{x_{13}^{2a_{+}}x_{23}^{2a_{-}}}\,\,\,\,\,
,\,\,\,\,\,a_{\pm}={\textstyle{\frac{1}{2}}}(a\pm b),
\label{add02}
\end{equation}
and, following a similar argument to the one in subsection \ref{sbsec22},
explicitly given as
\begin{eqnarray}
C^{a,b}(y,\partial) & = & \frac{1}{B(a_{+},a_{-})}\int_{0}^{1}dt
\,t^{a_{+}-1}(1-t)^{a_{-}-1} \nonumber
\\
 &  &
\times{}\sum_{m=0}^{\infty}\frac{1}{m!}\frac{1}
{(a+1-\mu)_{m}}\Bigl[-\frac{1}{4}y^{2}t(1-t)\Bigl]
^{m}\partial^{2m}e^{t\,y\,\partial}.
\label{add6}
\end{eqnarray}
The OPE for  two scalar fields $A(x_{1})$, $B(x_{2})$ with arbitrary
dimensions $\eta_{A}$, $\eta_{B}$ then has the form
\begin{equation}
A(x_{1})B(x_{2})=\cdots
+g_{ABC}\,\frac{1}{(x_{12}^{2})^{\frac{1}{2}(\eta_{A}+\eta_{B}
-\eta_{C})}}C^{\eta_{C},\eta_{A}-\eta_{B}}(x_{12}
,\partial_{2})C(x_{1})+\cdots
{}.
\end{equation}
The corresponding generalisation of
(\ref{2.2.12}) is then
\begin{equation}
C^{a,b}(x_{12},\partial_{2})\,C^{a,b}(x_{34},\partial_{4})
\frac{1}{x_{24}^{2a}}=\frac{1}{x_{13}^{2a_{+}}x_{24}^{2a_{-}}}
\,{\cal{E}}_{a,b}(u,v),
\label{add7}
\end{equation}
where
\begin{eqnarray}
 &  & {\cal{E}}_{a,b}(u,v) ={\cal{E}}_{a,-b}(u,v)\nonumber \\
 &  &
=\left(\frac{v}{u}\right)^{a_{+}}\sum_{n=0}^{\infty}
\frac{v^{n}}{n!}\frac{(a_{+})_{n}^{2}(a_{-})_{n}^{2}}
{(a)_{2n}(a+1-\mu)_{n}}{}_{2}F_{1}(a_{+}+n,a_{+}+n;a+2n;1-\frac{v}{u}).
\nonumber \\
\label{add8}
\end{eqnarray}
Note that for $a=\eta_{o}$ and $b=0$, (\ref{add8}) coincides with
(\ref{2.2.11}). By virtue of (\ref{add5}) and (\ref{add8}) we then find
\begin{equation}
Y(u,v)=g_{\phi\phi
O}^{2}\,u^{\frac{1}{2}\eta_{o}}\,{\cal{E}}_{\eta,\eta-\eta_{o}}
(\frac{1}{u},\frac{v}{u})+\cdots.
\label{add9}
\end{equation}
Clearly, the result (\ref{add9}) is appropriate in  considering the limit of
$Y(u,v)$ as $1/u$,
$v/u$ $\rightarrow 0$ or equivalently ${v/u \rightarrow 0}$ and $v$
$\rightarrow 1$ and the dots stand for less singular terms. We assume
that no other fields having dimensions $<d$ contribute in this limit
to  the OPE of $\phi^{\a}(x_{1})O(x_{3})$.

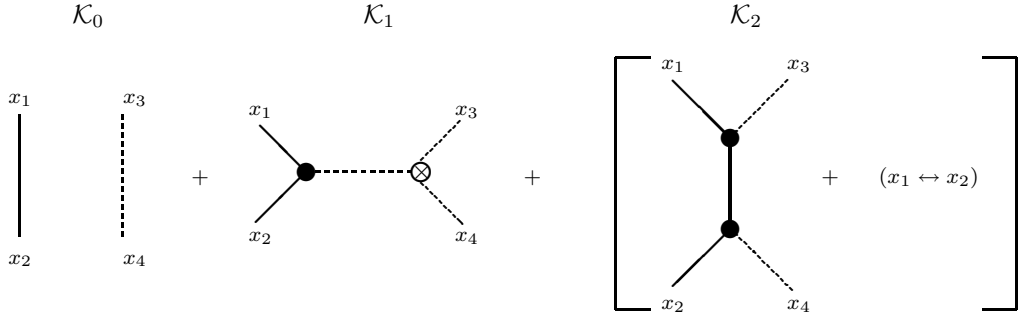
\begin{figure}
\setlength{\unitlength}{0.0120in}%
\begingroup\makeatletter
\def\x#1#2#3#4#5#6#7\relax{\def\x{#1#2#3#4#5#6}}%
\expandafter\x\fmtname xxxxxx\relax \def\y{splain}%
\ifx\x\y   
\gdef\SetFigFont#1#2#3{%
  \ifnum #1<17\tiny\else \ifnum #1<20\small\else
  \ifnum #1<24\normalsize\else \ifnum #1<29\large\else
  \ifnum #1<34\Large\else \ifnum #1<41\LARGE\else
     \huge\fi\fi\fi\fi\fi\fi
  \csname #3\endcsname}%
\else
\gdef\SetFigFont#1#2#3{\begingroup
  \count@#1\relax \ifnum 25<\count@\count@25\fi
  \def\x{\endgroup\@setsize\SetFigFont{#2pt}}%
  \expandafter\x
    \csname \romannumeral\the\count@ pt\expandafter\endcsname
    \csname @\romannumeral\the\count@ pt\endcsname
  \csname #3\endcsname}%
\fi
\endgroup
\begin{picture}(300,111)(0,675)
\thicklines
\multiput(110,760)(0.0,-3.9285714){14}{\line( 0,-1){  2.2}}
\put( 65,760){\line( 0,-1){ 53}}
\put(170,755){\line( 1,-1){ 20}}
\put(190,735){\line(-1,-1){ 22}}

\multiput(190,735)(4.0,0.0){12}{\line( 1, 0){  2}}
\put(190,735){\circle*{8}}
\put(240,735){\circle{8}}
\put(236,732){\makebox(0,0)[lb]{\smash{\SetFigFont{10}{14.4}{rm}$\times$}}}

\multiput(240,740)(0.41667,0.41667){3}{\makebox(0.4444,0.6667)
{\SetFigFont{7}{8.4}{rm}.}}
\multiput(242,742)(0.41667,0.41667){3}{\makebox(0.4444,0.6667)
{\SetFigFont{7}{8.4}{rm}.}}

\multiput(244,744)(0.41667,0.41667){3}{\makebox(0.4444,0.6667)
{\SetFigFont{7}{8.4}{rm}.}}
\multiput(246,746)(0.41667,0.41667){3}{\makebox(0.4444,0.6667)
{\SetFigFont{7}{8.4}{rm}.}}
\multiput(248,748)(0.41667,0.41667){3}{\makebox(0.4444,0.6667)
{\SetFigFont{7}{8.4}{rm}.}}
\multiput(250,750)(0.41667,0.41667){3}{\makebox(0.4444,0.6667)
{\SetFigFont{7}{8.4}{rm}.}}
\multiput(252,752)(0.41667,0.41667){3}{\makebox(0.4444,0.6667)
{\SetFigFont{7}{8.4}{rm}.}}
\multiput(254,754)(0.41667,0.41667){3}{\makebox(0.4444,0.6667)
{\SetFigFont{7}{8.4}{rm}.}}
\multiput(256,756)(0.41667,0.41667){3}{\makebox(0.4444,0.6667)
{\SetFigFont{7}{8.4}{rm}.}}

\multiput(240,730)(0.41667,-0.41667){3}{\makebox(0.4444,0.6667)
{\SetFigFont{7}{8.4}{rm}.}}
\multiput(242,728)(0.41667,-0.41667){3}{\makebox(0.4444,0.6667)
{\SetFigFont{7}{8.4}{rm}.}}
\multiput(244,726)(0.41667,-0.41667){3}{\makebox(0.4444,0.6667)
{\SetFigFont{7}{8.4}{rm}.}}
\multiput(246,724)(0.41667,-0.41667){3}{\makebox(0.4444,0.6667)
{\SetFigFont{7}{8.4}{rm}.}}
\multiput(248,722)(0.41667,-0.41667){3}{\makebox(0.4444,0.6667)
{\SetFigFont{7}{8.4}{rm}.}}
\multiput(250,720)(0.41667,-0.41667){3}{\makebox(0.4444,0.6667)
{\SetFigFont{7}{8.4}{rm}.}}
\multiput(252,718)(0.41667,-0.41667){3}{\makebox(0.4444,0.6667)
{\SetFigFont{7}{8.4}{rm}.}}
\multiput(254,716)(0.41667,-0.41667){3}{\makebox(0.4444,0.6667)
{\SetFigFont{7}{8.4}{rm}.}}
\multiput(256,714)(0.41667,-0.41667){6}{\makebox(0.4444,0.6667)
{\SetFigFont{7}{8.4}{rm}.}}

\thinlines
\put(340,785){\line(-1, 0){ 15}}
\put(325,785){\line( 0,-1){100}}
\put(325,685){\line( 0,-1){ 10}}
\put(325,675){\line( 1, 0){ 15}}
\thicklines
\put(350,775){\line( 1,-1){ 25}}
\put(375,750){\line( 0,-1){ 40}}
\put(375,710){\line(-1,-1){ 25}}
\put(375,750){\circle*{8}}
\put(375,710){\circle*{8}}
\multiput(375,750)(0.41667,0.41667){3}{\makebox(0.4444,0.6667)
{\SetFigFont{7}{8.4}{rm}.}}
\multiput(377,752)(0.41667,0.41667){3}{\makebox(0.4444,0.6667)
{\SetFigFont{7}{8.4}{rm}.}}
\multiput(379,754)(0.41667,0.41667){3}{\makebox(0.4444,0.6667)
{\SetFigFont{7}{8.4}{rm}.}}
\multiput(381,756)(0.41667,0.41667){3}{\makebox(0.4444,0.6667)
{\SetFigFont{7}{8.4}{rm}.}}
\multiput(383,758)(0.41667,0.41667){3}{\makebox(0.4444,0.6667)
{\SetFigFont{7}{8.4}{rm}.}}
\multiput(385,760)(0.41667,0.41667){3}{\makebox(0.4444,0.6667)
{\SetFigFont{7}{8.4}{rm}.}}
\multiput(387,762)(0.41667,0.41667){3}{\makebox(0.4444,0.6667)
{\SetFigFont{7}{8.4}{rm}.}}
\multiput(389,764)(0.41667,0.41667){3}{\makebox(0.4444,0.6667)
{\SetFigFont{7}{8.4}{rm}.}}
\multiput(391,766)(0.41667,0.41667){3}{\makebox(0.4444,0.6667)
{\SetFigFont{7}{8.4}{rm}.}}
\multiput(393,768)(0.41667,0.41667){3}{\makebox(0.4444,0.6667)
{\SetFigFont{7}{8.4}{rm}.}}
\multiput(395,770)(0.41667,0.41667){3}{\makebox(0.4444,0.6667)
{\SetFigFont{7}{8.4}{rm}.}}
\multiput(397,772)(0.41667,0.41667){3}{\makebox(0.4444,0.6667)
{\SetFigFont{7}{8.4}{rm}.}}
\multiput(399,774)(0.41667,0.41667){3}{\makebox(0.4444,0.6667)
{\SetFigFont{7}{8.4}{rm}.}}
\multiput(375,710)(0.41667,-0.41667){3}{\makebox(0.4444,0.6667)
{\SetFigFont{7}{8.4}{rm}.}}
\multiput(377,708)(0.41667,-0.41667){3}{\makebox(0.4444,0.6667)
{\SetFigFont{7}{8.4}{rm}.}}
\multiput(379,706)(0.41667,-0.41667){3}{\makebox(0.4444,0.6667)
{\SetFigFont{7}{8.4}{rm}.}}

\put(385,700){\makebox(0.4444,0.6667){\SetFigFont{10}{12}{rm}.}}
\multiput(381,704)(0.41667,-0.41667){3}{\makebox(0.4444,0.6667)
{\SetFigFont{7}{8.4}{rm}.}}
\multiput(383,702)(0.41667,-0.41667){3}{\makebox(0.4444,0.6667)
{\SetFigFont{7}{8.4}{rm}.}}
\multiput(385,700)(0.41667,-0.41667){3}{\makebox(0.4444,0.6667)
{\SetFigFont{7}{8.4}{rm}.}}
\multiput(387,698)(0.41667,-0.41667){3}{\makebox(0.4444,0.6667)
{\SetFigFont{7}{8.4}{rm}.}}
\multiput(389,696)(0.41667,-0.41667){3}{\makebox(0.4444,0.6667)
{\SetFigFont{7}{8.4}{rm}.}}
\multiput(391,694)(0.41667,-0.41667){3}{\makebox(0.4444,0.6667)
{\SetFigFont{7}{8.4}{rm}.}}
\multiput(393,692)(0.41667,-0.41667){3}{\makebox(0.4444,0.6667)
{\SetFigFont{7}{8.4}{rm}.}}
\multiput(395,690)(0.41667,-0.41667){3}{\makebox(0.4444,0.6667)
{\SetFigFont{7}{8.4}{rm}.}}
\multiput(397,688)(0.41667,-0.41667){3}{\makebox(0.4444,0.6667)
{\SetFigFont{7}{8.4}{rm}.}}
\multiput(399,686)(0.41667,-0.41667){3}{\makebox(0.4444,0.6667)
{\SetFigFont{7}{8.4}{rm}.}}
\multiput(401,684)(0.41667,-0.41667){3}{\makebox(0.4444,0.6667)
{\SetFigFont{7}{8.4}{rm}.}}

\thinlines
\put(485,785){\line( 1, 0){ 15}}
\put(500,785){\line( 0,-1){110}}
\put(500,675){\line(-1, 0){ 15}}
\thicklines
\put(140,730){\makebox(0,0)[lb]{\smash{\SetFigFont{8}{14.4}{rm}$+$}}}
\put(285,730){\makebox(0,0)[lb]{\smash{\SetFigFont{8}{14.4}{rm}$+$}}}
\put(415,730){\makebox(0,0)[lb]{\smash{\SetFigFont{8}{14.4}{rm}$+$}}}
\put(440,730){\makebox(0,0)[lb]{\smash{\SetFigFont{8}{14.4}{rm}$
(x_{1}\leftrightarrow x_{2})$}}}
\put( 60,765){\makebox(0,0)[lb]{\smash{\SetFigFont{8}{14.4}{rm}$x_{1}$}}}
\put(110,765){\makebox(0,0)[lb]{\smash{\SetFigFont{8}{14.4}{rm}$x_{3}$}}}
\put(110,695){\makebox(0,0)[lb]{\smash{\SetFigFont{8}{14.4}{rm}$x_{4}$}}}
\put( 60,695){\makebox(0,0)[lb]{\smash{\SetFigFont{8}{14.4}{rm}$x_{2}$}}}
\put(165,760){\makebox(0,0)[lb]{\smash{\SetFigFont{8}{14.4}{rm}$x_{1}$}}}
\put(165,705){\makebox(0,0)[lb]{\smash{\SetFigFont{8}{14.4}{rm}$x_{2}$}}}
\put(255,705){\makebox(0,0)[lb]{\smash{\SetFigFont{8}{14.4}{rm}$x_{4}$}}}
\put(255,760){\makebox(0,0)[lb]{\smash{\SetFigFont{8}{14.4}{rm}$x_{3}$}}}
\put(345,780){\makebox(0,0)[lb]{\smash{\SetFigFont{8}{14.4}{rm}$x_{1}$}}}
\put(400,780){\makebox(0,0)[lb]{\smash{\SetFigFont{8}{14.4}{rm}$x_{3}$}}}
\put(400,675){\makebox(0,0)[lb]{\smash{\SetFigFont{8}{14.4}{rm}$x_{4}$
}}}
\put(345,675){\makebox(0,0)[lb]{\smash{\SetFigFont{8}{14.4}{rm}$x_{2}$}}}

\put(88,800){\makebox(0,0)[lb]{\smash{\SetFigFont{10}{14.4}{rm}$
{\cal{K}}_{0}$}}}
\put(215,800){\makebox(0,0)[lb]{\smash{\SetFigFont{10}{14.4}{rm}$
{\cal{K}}_{1}$}}}
\put(375,800){\makebox(0,0)[lb]{\smash{\SetFigFont{10}{14.4}{rm}$
{\cal{K}}_{2}$}}}
\end{picture}

\caption{The Graphical Expansion corresponding to $\langle\phi\phi OO\rangle$.}
\label{fig?}
\end{figure}

In the spirit of the present work we will require consistency of
(\ref{add4}) and (\ref{add9}) with an evaluation of (\ref{add1}) based
on a graphical expansion. It is crucial that both  consistency
relations obtained should give the  same results for the couplings and
the field dimensions of the theory. As previously, we assume that a graphical
expansion for the four-point function
(\ref{add1}) can be constructed using graphs with internal lines
corresponding  to the full propagators of
$\phi^{\alpha}(x)$ and $O(x)$, the latter being the field
appearing in the OPE (\ref{2.2.4}),  which are glued together using the
vertex discussed in subsection \ref{sbsec41} and also the new vertex
obtained from the full leg amputation of the three-point function (\ref{add3}).
The first few graphs in increasing order according to the
total number of vertices are shown in Fig. \ref{fig?}
where the new vertex in represented by a crossed circle. Without further input
the magnitude of the
new coupling $g_{\tilde{O}}\equiv g_{O}$ is undetermined. Nevertheless, we
assume  that $g^{2}_{O}\sim
O(1/N)$ so that the graphs displayed in Fig. \ref{fig?} give the full
four-point function (\ref{add1}) up to $O(1/N)$. This assumption will
be justified in what follows by the consistency of our approach.

The evaluation of the
amplitudes corresponding to the graphs in Fig. \ref{fig?} is
straightforward if one uses the results of  Appendix \ref{apdx2}.
However, as it should be clear by now, one must conduct the
calculations of these amplitudes having in mind the corresponding
short-distance limit of (\ref{add1}) with which the resulting
expressions  have  to be compared. All our calculations are based on
formula (\ref{b10}) with the basic amplitudes corresponding to the
graphs given in Appendix \ref{apdx4}. Thus to $O(1/N)$
\begin{eqnarray}
Y(u,v) & = & 1+g_{\phi\phi O}\,g_{O}\,{\cal{K}}_{1}(u,v) \nonumber \\
 &  & {}+g_{\phi\phi
O}^{2}\Bigl({\cal{K}}_{2}(u,v)+{\cal{K}}_{2}(v,u)\Bigl){}+\cdots,\label{add10}
\end{eqnarray}
where
\begin{eqnarray}
{\cal{K}}_{1}(u,v) & = &
u^{\frac{1}{2}\eta_{o}}\,{\cal{H}}_{\eta_{o}}(u,v)
+C(d-\eta_{o})\,u^{\mu-\frac{1}{2}\eta_{o}}\,{\cal{H}}_{d-\eta_{o}}(u,v)
\label{add20} \\
{\cal{K}}_{2}(u,v) & = &
\frac{\Gamma(\eta)\Gamma(\eta_{o}-\eta)\Gamma^{2}
(\mu-\frac{1}{2}\eta_{o})}{\Gamma(\mu-\eta)
\Gamma(\mu+\eta-\eta_{o})\Gamma^{2}(\frac{1}{2}\eta_{o})}
\label{add21}
 \\
 &  & {}\times
\Bigl(\,u^{\eta}\,{\cal{E}}_{\mu+\eta-\eta_{o},\mu-\eta}(u,v)
+C(\mu+\eta-\eta_{o},2\mu-\eta_{o})\,u^{\eta_{o}}
\,{\cal{E}}_{\mu-\eta+\eta_{o},\mu-\eta}(u,v)\Bigl),
\nonumber  \\
C(\eta,\eta_{o}) & = &
\frac{\Gamma(\eta)\Gamma(\eta-\mu)\Gamma^{2}
(\mu-\eta+\frac{1}{2}\eta_{o})\Gamma^{2}
(\mu-\frac{1}{2}\eta_{o})}{\Gamma(\mu-\eta)
\Gamma(2\mu-\eta)\Gamma^{2}(\eta-\frac{1}{2}\eta_{o})
\Gamma^{2}(\frac{1}{2}\eta_{o})},
\label{add22}
\end{eqnarray}
with the same ${\cal{E}}_{a,b}(u,v)$ as in (\ref{add8}).
${\cal{K}}_{2}(u,v)$ corresponds to the third graph in Fig. \ref{fig?}
and ${\cal{K}}_{2}(v,u)$ to the fourth (not drawn) graph.

In the limit $u$, $v$ $\rightarrow 0$ with $v/u\rightarrow 1$, which
can be found using (\ref{add20}), (\ref{add21}), it is necessary to
cancel the {\it{shadow singularities}} coming from the
${\cal{H}}_{d-\eta_{o}}(u,v)$ term in (\ref{add20}). This cancellation
is achieved by the first term in (\ref{add21}) if one requires requires
$\mu-{\textstyle{\frac{1}{2}}}\eta_{o}=\eta$ which is satisfied for
$\eta=\mu-1$, $\eta_{o}=2$, and then using the expression (\ref{4.1.9})
for $C(d-\eta_{o})$ as well as (\ref{add22}) above we easily find
\begin{equation}
g_{O}=2\,(2\mu-3)\,g_{\phi\phi O},  \label{add13}
\end{equation}
where $g_{\phi\phi O}^{2}$ is given by (\ref{4.2.7}). Therefore, our
assumption  $g_{O}^{2}\sim O(1/N)$ is justified. Only the terms $\propto
u^{\eta}$ in   ${\cal{K}}_{2}(u,v)$ in (\ref{add21}) were used in  obtaining
(\ref{add13}). The terms $\propto u^{\eta_{o}}$  in ${\cal{K}}_{2}(u,v)$
correspond to contributions in the four-point function from a scalar
field appearing in the OPE's (\ref{2.2.4}) and (\ref{add2}), having
dimension 4 to leading
order in $1/N$.

Using (\ref{add13}) then we find from (\ref{add10})
\begin{equation}
Y(u,v) =1+g_{\phi\phi
O}\,g_{O}\,{\cal{H}}_{2}(u,v)+\frac{1}{N}\frac{4\mu(\mu-1)}
{(2\mu-1)}\Bigg(\frac{1}{4}(1-\frac{v}{u})^{2}
-\frac{1}{2\mu}(uv)^{\frac{1}{2}}\Bigg)+\cdots.
\label{add50}
\end{equation}
The last term in (\ref{add50})  corresponds exactly to the contribution of the
energy momentum tensor
in the OPE's (\ref{2.2.4}) and (\ref{add3}) with $C_{T}$ given to
leading order by (\ref{3.1.7}) or (\ref{4.2.12}) again,  which is another
consistency check for our approach.

To discuss  $Y(u,v)$ in the limit as  $x_{13}^{2}$, $x_{24}^{2}$
$\rightarrow 0$ or ${v/u
\rightarrow 0}$ and $v$
 $\rightarrow 1$, it is necessary to find an alternative expansion  for
${\cal{K}}_{1}(u,v)$ and ${\cal{K}}_{2}(u,v)$. This is achieved in much the
same
way as the one used in Appendix \ref{apdx2} to obtain different
expansions for the one-particle exchange amplitudes
${\cal{G}}^{(\tilde{\eta}_{o})}_{1}$.
Of course, the resulting expressions should  be equal to the corresponding
ones given in (\ref{add20}) and (\ref{add21}), but they are appropriate
in considering the limit as $v/u\rightarrow 0$ and $v\rightarrow 1$.
We obtain
\begin{eqnarray}
{\cal{K}}_{1}(u,v) & = & g_{\phi\phi
O}\,g_{O}\,v^{\frac{1}{2}\eta_{o}}\sum_{n,m=0}^{\infty}
\frac{\left(\frac{v}{u}\right)^{n}(1-v)^{m}}{n!m!}
a_{nm}[-\mbox{ln}\frac{v}{u}+b_{nm}],
\label{add23}  \\
{\cal{K}}_{2}(u,v) & = & g_{\phi\phi
O}^{2}\,\Bigg[u^{\frac{1}{2}\eta_{o}}\,{\cal{E}}_
{\eta,\eta-\eta_{o}}(\frac{1}{u},\frac{v}{u})
+C(\eta,\eta_{o})\,u^{\eta+\frac{1}{2}\eta_{o}-\mu}
\,{\cal{E}}_{d-\eta,\eta-\eta_{o}}(\frac{1}{u},\frac{v}{u})\Bigg],
\label{add24} \\
{\cal{K}}_{2}(v,u) & = & g_{\phi\phi
O}^{2}\,v^{\eta}\sum_{n,m=0}^{\infty}\frac{\left(
\frac{v}{u}\right)^{n}(1-v)^{m}}{n!m!}C_{nm}[-\mbox{ln}
\frac{v}{u}+D_{nm}],
\label{add25}
\end {eqnarray}
with the same $C(\eta,\eta_{o})$ as in (\ref{add22}). The coefficients
$a_{nm}$ and  $b_{nm}$ are given by (\ref{b13}) and (\ref{b14}) while $C_{nm}$
and
$D_{nm}$ are given by (\ref{d13}) and (\ref{d14}) respectively.

Discussion of this limit requires more care. The important point is
that $C(\eta,\eta_{o})$ in (\ref{add24}) as given by (\ref{add22})
is  $ O(N)$ since it
contains the factor $\Gamma(\eta-\mu)$ in the numerator. To
leading order
\begin{equation}
C(\eta,\eta_{o})\rightarrow
-\frac{N}{\eta_{1}}\frac{(\mu-2)^{2}}{\mu(\mu-1)}=-\frac{1}{g_{\phi\phi
O}^{2}}.
\label{add14}
\end{equation}
Therefore, $g_{\phi\phi O}^{2}C(\eta,\eta_{o})=-1+O(1/N)$ and this  will
generate
terms which are $ O(1)$. These are essential to ensure the cancellation
of terms corresponding to the {\it{shadow field}} of $\phi^{\a}(x)$
which has dimension $d-\eta$, and hence obtain the  correct
structure for
the OPE
(\ref{add5}). Writing in (\ref{add21})
\begin{equation}
u^{\eta+\frac{1}{2}\eta_{o}-\mu}=1+\frac{1}{N}(\eta_{1}
+{\textstyle{\frac{1}{2}}}\eta_{o,1})\mbox{ln}u+\cdots,
\label{add60}
\end{equation}
then in this limit ${\cal{K}}_{2}(u,v)\rightarrow -1+O(1/N)$ which ensures the
cancellation of the first term on the r.h.s. of  (\ref{add10}). In
order to cancel the $O(1/N)$ $\mbox{ln}u$ terms present in
(\ref{add23})-(\ref{add25}) it is necessary that\footnote{We only consider the
$n=m=0$ contributions in the series in (\ref{add23})-(\ref{add25}) which
suffice for
our purposes. Discussion of the $n,m\neq 1$ contributions requires the
introduction of more quasiprimary fields in the OPE (\ref{add5}).}
\begin{equation}
-g_{\phi\phi
O}^{2}\frac{(\mu-2)^{2}}{\mu(\mu-1)}\frac{\eta_{1}
+{\textstyle{\frac{1}{2}}}\eta_{o,1}}{\eta_{1}}
+g_{\phi\phi O}\,g_{O}\,a_{00}+g_{\phi\phi
O}^{2}\,C_{00}{}=0 . \label{add15}
\end{equation}
It is easily seen that, by virtue of (\ref{4.2.7}), (\ref{4.2.15}),
(\ref{4.2.23}) and the results in Appendix \ref{apdx4},
(\ref{add15}) requires (\ref{add13}) again. Agreement of the two
approaches of  evaluating  the four-point function (\ref{add1})
is essentially a  check on  the associativity of the OPE. It is perhaps
interesting to note that $g_{O}$ in (\ref{add13}) vanishes for $d=3$, at least
to
leading order in $1/N$, which implies the possible existence of a discrete
symmetry
$O(x)\leftrightarrow -O(x)$ in this dimension in the sector generated
by $O(x)$.

For completeness we briefly discuss the free field theory of section
\ref{sec3}. In this case we take
\begin{equation}
O(x)=\frac{1}{\sqrt{2N}}:\phi^{2}(x):, \label{add26}
\end{equation}
since we set to one the normalisation of the two-point function of
$O(x)$. Therefore, from Wick's theorem we easily find
\begin{equation}
Y(u,v)_{f}=1+\frac{2}{N}(u^{\eta}+v^{\eta}). \label{add27}
\end{equation}
In the limit $u$, $v$ $\rightarrow 0$ with $u/v\rightarrow 1$ we find
\begin{equation}
Y(u,v)_{f}=1+\frac{2}{N}v^{\eta}\Bigl(2+\eta(1-\frac{v}{u})+\cdots\Bigl).
\label{add28}
\end{equation}
Requiring agreement of (\ref{add28}) with (\ref{add4}) we obtain
\begin{equation}
\eta_{o}=2\eta\,\,\,\,\,,\,\,\,\,\,g_{O}=2g_{\phi\phi O}=2\sqrt{\frac{2}{N}},
\label{add29}
\end{equation}
which is consistent with free field theory and $g_{\phi\phi O}^{2}$ in
(\ref{3.1.6}).

To discuss the limit $v/u\rightarrow 0$, $v\rightarrow 1$ of
(\ref{add27}) one needs to introduce an additional quasiprimary field
in the free field theory version of the OPE (\ref{add5}), namely
\begin{equation}
\phi^{\a}(x_{1})O(x_{3})=g_{\phi\phi
O}\frac{1}{x_{13}^{2\eta}}\phi^{\a}(x_{1})+g_{\phi
OF}\,F^{\a}(x_{1})+\cdots. \label{add30}
\end{equation}
The new $O(N)$ vector $O(d)$ scalar
field $F^{\a}(x)$ having dimension $3\eta$ may be taken
to be
\begin{equation}
F^{\a}(x)=C\,:\phi^{\a}(x)\phi^{2}(x):.  \label{add31}
\end{equation}
Normalising to one its two-point function  requires $2(N+2)C^{2}=1$
which in turn implies that $g_{\phi O F}=\sqrt{(N+2)/N}$. Then, by
virtue of (\ref{add30}), the relevant terms of $Y(u,v)_{f}$ in the
limit as
$v/u\rightarrow 0$, $v\rightarrow 1$ are
\begin{equation}
Y(u,v)_{f}=1+\frac{2}{N}+\frac{2}{N}u^{\eta}+\cdots, \label{add40}
\end{equation}
which agrees with the corresponding limit of (\ref{add27}).

Regarding the possible duality property of the $O(N)$ vector model we remark
that one  could have built the non-trivial graphs in the skeleton
expansion in
Fig. \ref{fig?} for
$Y(x_{1},x_{2},x_{3},x_{4})$ using dashed lines corresponding to the
{\it{shadow field}} of $O(x)$. Then, the dark blobs would correspond
to the coupling $\lambda_{*}$ and the crossed circle to a coupling
$\lambda_{O}$ while $\eta_{o}\equiv\eta'_{o}=d-2$ to leading order in
$1/N$. In this case one can show that $\lambda_{O}=2\lambda_{*}$ at
least to leading order in $1/N$. But this is exactly  the relationship
(\ref{add29}) between the couplings $g_{\phi\phi O}$ and $g_{O}$ in the
free field theory of section \ref{sec3}. This is in accord with our conjecture
that
the theory dual to the $O(N)$ vector model  is somehow related to the
theory of $N$ massless free scalars.

Clearly, one can discuss  many more consistency checks regarding the
non-trivial
$O(N)$ invariant CFT in $2<d<4$. For example, one may try to
reproduce our next-to-leading order in $1/N$ result (\ref{4.2.25}) for $C_{T}$
by
introducing the energy momentum tensor into the OPE (\ref{add2}). Our
formalism can also be applied to discussing other  four-point
functions such as  $\langle
OOOO\rangle$ or  those involving conserved
currents. We leave
these investigations for future developments of our approach.

\section{Discussion and Concluding Remarks}\label{sec6}
\setcounter{equation}{0}

The results of the present work may have possible applications to
other conformally invariant field theories in $d>2$. Starting
from the OPE ansatz (\ref{2.2.4}) in a $O(N)$ invariant CFT we were able to
find explicit expressions for the four-point function of the
fundamental field $\phi^{\a}(x)$ in a  suitably chosen short-distance limit.
However, these expressions  contain undetermined dynamical parameters
of the theory. Motivated by the form of the ansatz (\ref{2.2.4}), we  assumed
that the four-point function of
$\phi^{\a}(x)$ can alternatively be calculated using a skeleton graph
expansion with internal lines corresponding to the two-point functions
of the fields $\phi^{\a}(x)$ and $\tilde{O}(x)$ where the latter is a scalar
field of dimension $\tilde{\eta}_{o}$ with
$0<\tilde{\eta}_{o}<d$. To low orders only the interaction of
$\phi^{\a}(x)$ with $\tilde{O}(x)$ via a unique vertex having coupling
constant $g_{*}$ is essential and our
assumption was shown to be valid when  $\tilde{O}(x)$ is identified with
the field $O(x)$ in the OPE (\ref{2.2.4}) and
$g_{*}^{2}=g_{\phi\phi O}^{2}$. Since a consistent skeleton graphical
expansion for $n$-point functions with $n\geq 4$ requires the
introduction of the triple $O(x)$ vertex, we have also discussed  a four-point
function involving the scalar field $O(x)$ and we showed that our
approach of requiring consistency of algebraically and graphically
based evaluations of correlation functions is consistent in this case as well.
This last case provided a non-trivial check for the associativity
property of OPE's in $d>2$. Our treatment of the
four-point functions may be considered as an application  of the
well known bootstrap program of CFT \cite{Parisi,Vasiliev,Gracey} and as
such we expect that it is  also applicable  \cite{newanastasios} in the case of
 other known
examples of  non-trivial CFT's in $2<d<4$, like the Four-Fermi or the
Gross-Neveu model in the $1/N$ expansion \cite{Gracey,Semenoff} at its critical
point.

Requiring consistency
of the graphical expansions with the  algebraic treatments of
four-point functions resulted  in a set of consistency relations which
determine
the dynamical parameters of the theory at least within the context of
a $1/N$ expansion. More specifically, our results (\ref{4.2.15}) and
(\ref{4.2.23}) for
$\eta$ and $\eta_{o}$ agree with the well known results (\eg see
\cite{Ruhl1} and references therein) for the
anomalous dimensions of the fundamental and auxiliary fields
respectively in the $O(N)$ sigma model at its critical point which
exists in the context of a $1/N$ expansion for $2<d<4$. We
conclude that the $O(N)$ sigma model  and the non-trivial theory considered in
section \ref{sec4} coincide since they  seem to have the same field
algebra. Therefore, we may identify the auxiliary field $\sigma(x)$ which
enforces the constraint $\phi^{2}(x)=1$ in the $O(N)$ sigma model, with
the field $O(x)$ which appears in the OPE (\ref{2.2.4}), \eg{} both these
fields have dimension close to $2$ for large $N$. Note that for large $N$ the
field algebras of the trivial and
the non-trivial $O(N)$ invariant CFT's are quite different since in the
former the low dimension $<d$ scalar field which appears in the OPE
(\ref{2.2.4}) has dimension close to $d-2$. Consequently, these two theories
correspond to different universality classes in $2<d<4$ although
strong evidence indicates that they coincide for $d=4$. The results
(\ref{4.2.24})
for the coupling $g_{\phi\phi O}$ and (\ref{4.2.16}) for $C_{J}$ agree with
previously obtained
results in \cite{Ruhl2}.

The duality property of the graphical
expansion mentioned at the end of section \ref{sbsec42} is another
novel result of the present work. It seems  that the non-unitary
field theory which underlies  the graphical expansion corresponding to
the interaction of $\phi^{\a}(x)$ with $\tilde{O}_{s}(x)$ via a unique
vertex having coupling constant $\lambda_{*}$ as given in (\ref{4.2.27}), is
related to the free field theory of section 3.

Apart from these general considerations it is of interest to
discuss our next-to-leading
order  in $1/N$ results for the important quantities $C_{J}$ and
$C_{T}$. The normalisation of the energy momentum tensor two-point
function $C_{T}$ has
been considered \cite{Cappelli,Shore} as one possible generalisation
of the two-dimensional Virasoro central charge in
higher-dimensional CFT's.
Similarly, (\eg see \cite{Cardona}) one may view $C_{J}$
as one possible generalisation in higher dimensions of the Kac-Moody algebra
level
of a two-dimensional conformal WZW model. These generalisations refer to the
property of both the central
charge and the algebra level  to be  the fixed point values of important
quantities which are
monotonously decreasing  along the renormalisation group flow from
UV to  IR fixed points of  unitary
two-dimensional quantum field theories \cite{Cardona}. Such a property of the
corresponding $d>2$ quantities $C_{J}$ and $C_{T}$ is no longer true
in general in higher dimensions. However, it is well known
\cite{Zinn-Justin} that each order in the $1/N$ expansion includes
contributions from all orders in the usual weak coupling perturbation
expansion. Hence, our next-to-leading order in $1/N$ results
(\ref{4.2.25}) and
(\ref{4.2.16}) give non-perturbative information for the renormalisation
group flow from   the UV fixed point (free theory of section 3) to the
IR one (non-trivial theory of section \ref{sec4}), when looked at from the
point
of view of weak coupling expansions.

\begin{figure}[t]
\epsfxsize=18cm
\epsfbox{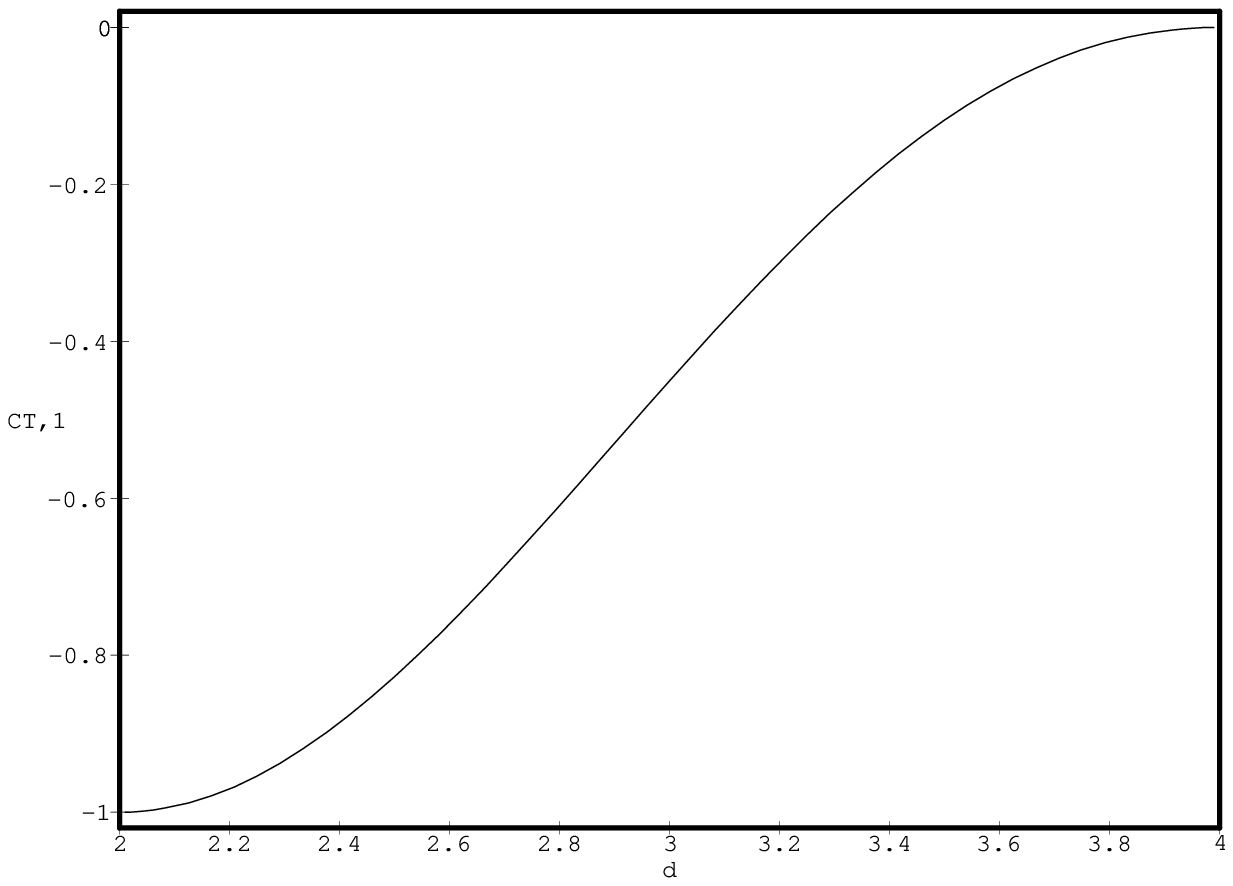}
\caption{$C_{T,1}$ for $2<d<4$.}\label{fg5}
\end{figure}

\begin{figure}[t]
\epsfxsize=18cm
\epsfbox{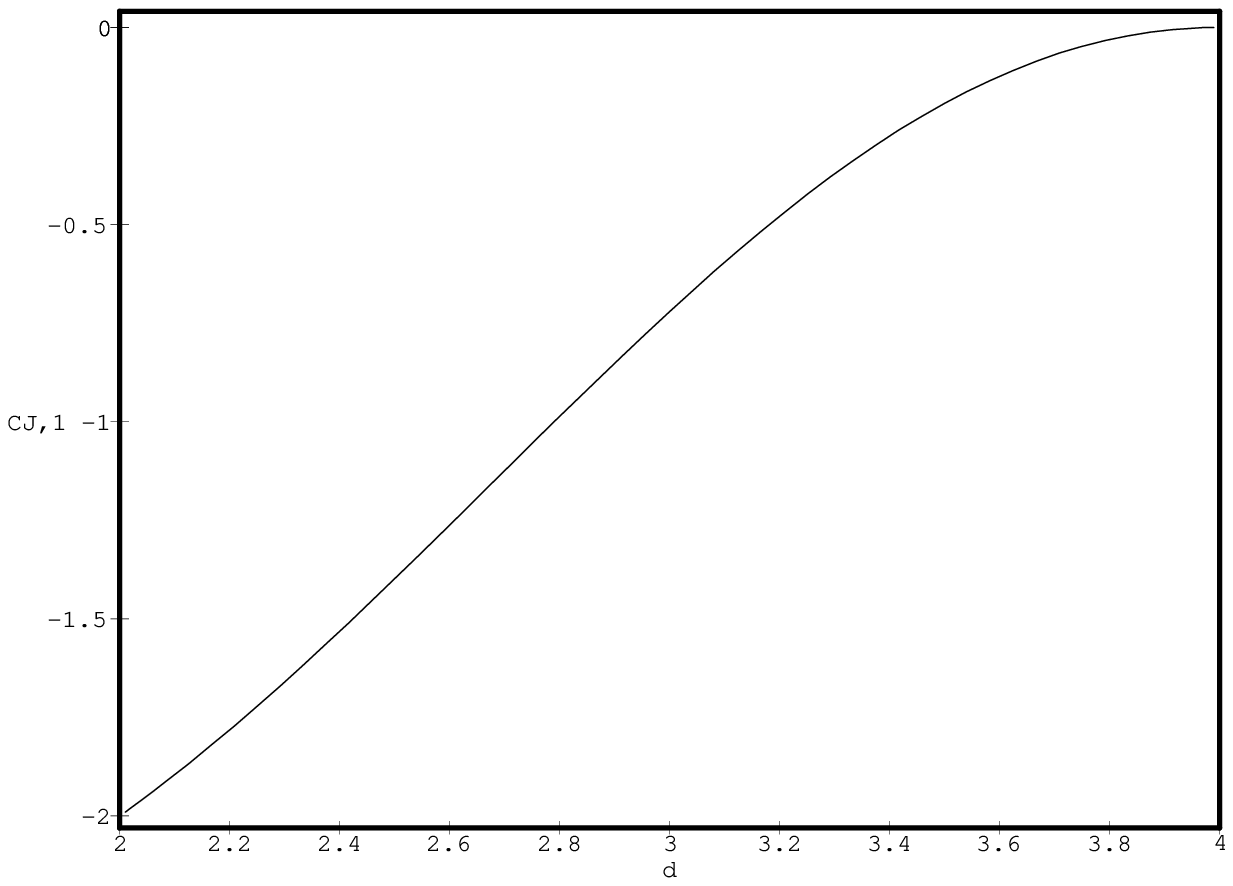}
\caption{$C_{J,1}$ for $2<d<4$.}\label{fg6}
\end{figure}

In Fig. \ref{fg5} we plot $C_{T,1}$ for $2<d<4$
and by virtue of  (\ref{4.2.12}) we  see that for   this range of
dimensions $C_{T}$ at the UV fixed point (gaussian theory) is always
greater than $C_{T}$ at  the IR fixed point (non-trivial
theory) at least to next-to-leading  order in a $1/N$ expansion. A
similar conclusion can be drawn for $C_{J}$ based on the plot of
$C_{J,1}$ for $2<d<4$ in Fig. \ref{fg6}.

It is also interesting to consider the case when $d\rightarrow 4$ and
to compare our results for $C_{T}$ and $C_{J}$ with known results in
the context of $\e$-expansion when $\e=4-d>0$. In four dimensions, if the
theory is
defined for a background metric $g_{\mu\nu}$ and has a gauge field
$A_{\mu}^{\a}$ coupled to the conserved vector current, even for a
conformal theory there is a trace anomaly \cite{anastasios}
\begin{equation}
g^{\mu\nu}\la
T_{\mu\nu}\ra=-\b_{a}F-\kappa{\textstyle{\frac{1}{4}}}
F_{\mu\nu}^{\a}F^{\a,\mu\nu}+\cdots,\label{5.1.1}
\end{equation}
where $F$ is the square of the Weyl tensor and terms which are
irrelevant here are neglected. The quantities $\b_{a}$
and $\kappa$ can be perturbatively calculated and for a  $O(N)$
invariant
renormalisable  field
theory with ${\textstyle{\frac{1}{24}}}g\left(\phi^{2}\right)^{2}$
interaction, a three loop
calculation yields \cite{Osborn}
\begin{eqnarray}
\b_{a} & =  &
-\frac{1}{16\pi^{2}}\frac{N}{120}\Bigl(1-\frac{5}{108}
(N+2)u^{2}\Bigl),\label{5.1.2} \\
\kappa & = &
\frac{1}{3}\frac{1}{16\pi^{2}}\,R\,\Bigl(1-\frac{1}{12}
(N+2)u^{2}\Bigl),\label{5.1.3}
\end{eqnarray}
where $u=g/16\pi^{2}$ with $g$ the renormalised coupling
and
$\mbox{tr}(t^{\a}t^{\b})=-\delta^{\a\b}R$. For
the adjoint representation of $O(N)$ we have
$(t^{\a\b})^{\c\delta}=-(\delta^{\a\c}\delta^{\b\delta}
-\delta^{\a\delta}\delta^{\b\c})$
and $R=2$. The results of our previous work \cite{anastasios} show
that   for a conformal theory when  $d=4$
\begin{equation}
C_{T}=-\frac{640}{\pi^{2}}\b_{a}\,\,\,\,\,,\,\,\,\,\,C_{J}
=\frac{6}{\pi^{2}}\kappa.\label{5.1.4}
\end{equation}
In general we suppose that we may
write $C_{T}(\e,u_{*})$, $C_{J}(\e,u_{*})$ where $u_{*}$ is the
critical coupling. The free or Gaussian field theory results
(\ref{3.1.7}),  (\ref{3.1.9}) correspond to $C_{T,f}=C_{T}(\e,0)$ and
$C_{J,f}=C_{J}(\e,0)$
while (\ref{5.1.4}) gives  $C_{T}(0,u)$ and $C_{J}(0,u)$. Using then
(\ref{5.1.2}) and (\ref{5.1.3}) with $u_{*}=3\e/(N+8)+O(\e^{2})$
\cite{Zinn-Justin} gives the leading corrections in the $\e$-expansion
\footnote{The result (\ref{5.1.5}) for $N=1$ was found in \cite{Cappelli}.}
\begin{eqnarray}
C_{T} & = &
C_{T,f}\Bigl(1-\frac{5}{12}\frac{N+2}{(N+8)^{2}}\e^{2}
+O(\e^{3})\Bigl),\label{5.1.5}
\\
C_{J} & = &
C_{J,f}\Bigl(1-\frac{3}{4}\frac{N+2}{(N+8)^{2}}\e^{2}
+O(\e^{3})\Bigl).\label{5.1.6}
\end{eqnarray}
As $d\rightarrow 4$ we see from (\ref{4.2.15}) that
$\eta_{1}\sim\e^{2}/4$ and then we can easily show  that
out results (\ref{4.2.25}) and (\ref{4.2.16}) agree correspondingly with
(\ref{5.1.5})
and (\ref{5.1.6}), something  which is a remarkable independent
check for their validity at least up to the order considered here.

Finally, it was shown in  \cite{Cardy2}  that $C_{T}$
parametrises universal finite size effects of statistical systems at their
critical points in two dimensions which provides a natural method for
its measurement both numerically and experimentally. For $d>2$, although Cardy
\cite{Cardy3} has pointed out that $C_{T}$ may be in principle
measurable, the finite scaling of the
free energy is parametrised \cite{Cardy3,Sachdev} by a universal number
$\tilde{c}$ whose relation with $C_{T}$ is not  clear. Sachdev
\cite{Sachdev} has calculated $\tilde{c}$ for the $O(N)$ vector model
to leading order in $1/N$ for $d=3$ and found it to be a rational number
however
different from the leading order in $1/N$ value for $C_{T}$
\footnote{It is interesting to point out  that the leading order in $1/N$ value
for $C_{T}$
coincides with the Gaussian theory value $C_{T,f}$ in any dimension
whereas as shown in \cite{Sachdev} the leading order in $1/N$ and the
Gaussian values for $\tilde{c}$ differ in $2<d<4$.}. Using our results
(\ref{4.2.25}) and (\ref{4.2.16}) we obtain for $d=3$
\begin{eqnarray}
C_{T}|_{d=3} & = &
N\frac{3}{2S_{3}^{2}}\Bigl(1-\frac{1}{N}\frac{40}{9\pi^{2}}\Bigl),\label{5.1.7}
\\
C_{J}|_{d=3} & = &
\frac{2}{S_{3}^{2}}\Bigl(1-\frac{1}{N}\frac{32}{9\pi^{2}}\Bigl).\label{5.1.8}
\end{eqnarray}
Noting from  (\ref{1.1.3}) and (\ref{2.2.11}) that with our normalisation
$C_{T}$
has to be multiplied by $S_{d}^{2}/2$ to agree with  the corresponding
quantity in  \cite{Sachdev}, we can answer by virtue of (\ref{5.1.7}) part
of the question addressed in that reference: $C_{T}$ does not seem to be a
rational number for finite $N$ in three dimensions.

\section*{Acknowledgements}

I would like to thank my  supervisor Hugh Osborn
for much guidance and helpful suggestions during the preparation of this
manuscript and
also for providing me with a copy of \cite{Lang}.

\section*{{\underline{Appendix}}}

\begin{appendix}

\section{Scalar Field Contribution to the Operator Product
Expansion}\label{apdx1}

\setcounter{equation}{0}

We describe here details of the calculation leading to (\ref{2.2.12}). Using
(\ref{2.2.7}) we readily find
\begin{eqnarray}
 &  &
C^{\eta_{o}}(x_{12},\pa_{2})C^{\eta_{o}}(x_{34},\pa_{4})
\frac{1}{x_{24}^{2\eta_{o}}} \nonumber \\
 &  &
=\frac{1}{\Bigl[B(\frac{1}{2}\eta_{o},\frac{1}{2}\eta_{o})\Bigl]^{2}}
\nonumber \\
 &  &
\times\int_{0}^{1}dtds[t(1-t)s(1-s)]^{\frac{1}{2}\eta_{o}-1}
\left(\sum_{k=0}^{\infty}\frac{1}{k!}\frac{1}{(\eta_{o}
+1-\mu)_{k}}[-{\textstyle{\frac{1}{4}}}x_{12}^{2}t(1-t)
\pa^{2}_{2}]^{k}\right)
\nonumber \\
&  &
\times\left(\sum_{l=0}^{\infty}\frac{1}{l!}\frac{1}
{(\eta_{o}+1-\mu)_{l}}[-{\textstyle{\frac{1}{4}}}
x_{34}^{2}t(1-t)\pa^{2}_{4}]^{l}\right)\frac{1}{[tx_{1}
+(1-t)x_{2}-sx_{3}-(1-s)x_{4}]^{2\eta_{o}}}.\label{a1}
\end{eqnarray}
Next we write
\begin{equation}
[tx_{1}+(1-t)x_{2}-sx_{3}-(1-s)x_{4}]^{2}=\Lambda^{2}-A^{2}-B^{2},\label{a2}
\end{equation}
where
\begin{eqnarray}
\Lambda^{2} & =
& tsx_{13}^{2}+t(1-s)x_{14}^{2}+s(1-t)x_{23}^{2}+(1-t)(1-s)x_{24}^{2},
\label{a3}\\
A^{2} & = & x_{12}^{2}t(1-t), \label{a4}\\
B^{2} & = & x_{34}^{2}s(1-s).\label{a5}
\end{eqnarray}
Using (\ref{2.2.9}) and the notation above we may cast (\ref{a1}) into the
following
form
\begin{eqnarray}
&  &
\frac{1}{\Bigl[B(\frac{1}{2}\eta_{o},\frac{1}{2}\eta_{o})
\Bigl]^{2}}\int_{0}^{1}dtds[t(1-t)s(1-s)]^{\frac{1}{2}
\eta_{o}-1}\left[\frac{1}{\Lambda^{2}-A^{2}-B^{2}}\right]^{\eta_{o}}
\nonumber \\
&  &
{}\times\sum_{k,l=0}^{\infty}\frac{1}{k!l!}\frac{(\eta_{o})
_{k+l}(\eta_{o}+1-\mu)_{k+l}}{(\eta_{o}+1-\mu)_{k}
(\eta_{o}+1-\mu)_{l}}\left[\frac{-A^{2}}{\Lambda^{2}
-A^{2}-B^{2}}\right]^{k}\left[\frac{-B^{2}}{\Lambda^{2}
-A^{2}-B^{2}}\right]^{l}
\nonumber \\
& =  &
\frac{1}{\Bigl[B(\frac{1}{2}\eta_{o},\frac{1}{2}\eta_{o})
\Bigl]^{2}}\int_{0}^{1}dtds[t(1-t)s(1-s)]^{\frac{1}{2}
\eta_{o}-1}\left[\frac{1}{\Lambda^{2}-A^{2}-B^{2}}\right]^{\eta_{o}}
\nonumber \\
&  & {}\times
F_{4}\left(\eta_{o},\eta_{o}+1-\mu;\eta_{o}+1-\mu,\eta_{o}
+1-\mu;\frac{-A^{2}}{\Lambda^{2}-A^{2}-B^{2}},\frac{-B^{2}}
{\Lambda^{2}-A^{2}-B^{2}}\right),\label{a6}
\end{eqnarray}
where
\begin{equation}
F_{4}(a,b;c,d;x,y)=\sum_{m=0}^{\infty}\sum_{n=0}^
{\infty}\frac{1}{m!n!}\frac{(a)_{m+n}(b)_{m+n}}
{(c)_{m}(d)_{n}}x^{m}y^{n},\label{a7}
\end{equation}
is one of the Appell functions (hypergeometric functions of two
variables). Then, by virtue of the following property \cite{integrals}
\begin{equation}
F_{4}(a,b;b,b;x,y)=(1-x-y)^{-a}{}_{2}F_{1}
({\textstyle{\frac{1}{2}}}a,{\textstyle{\frac{1}{2}}}(a+1)
;b;\frac{4xy}{(1-x-y)^{2}}),\label{a8}
\end{equation}
(\ref{a6}) becomes
\begin{eqnarray}
&  &
\frac{1}{\Bigl[B(\frac{1}{2}\eta_{o},\frac{1}{2}\eta_{o})\Bigl]^{2}} \nonumber
\\
&  &
{}\times\int_{0}^{1}dtds[t(1-t)s(1-s)]^{\frac{1}{2}\eta_{o}-1}
\frac{1}{\Lambda^{2\eta_{o}}}{}_{2}F_{1}\left({\textstyle
{\frac{1}{2}}}\eta_{o},{\textstyle{\frac{1}{2}}}(\eta_{o}+1);\eta_{o}+1
-\mu;\frac{4A^{2}B^{2}}{\Lambda^{4}}\right).\label{a9}
\end{eqnarray}
Next, we use the following representation for the hypergeometric
function
\begin{equation}
{}_{2}F_{1}(a,b;c;z)=\frac{1}{2\pi
i}\frac{\Gamma(c)}{\Gamma(a)\Gamma(b)}\int_{-i\infty}^
{i\infty}\!dx\,\Gamma(-x)\frac{\Gamma(a+x)\Gamma(b+x)}
{\Gamma(c+x)}(-z)^{x},\label{a10}
\end{equation}
obtaining
\begin{eqnarray}
&  & \frac{1}{B^{2}(\frac{1}{2}\eta_{o},\frac{1}{2}\eta_{o})}
\nonumber \\
 & & {}\times\frac{1}{2\pi
i}\int_{-i\infty}^{i\infty}\!dx\,\Gamma(-x)\Bigg(\frac{
\Gamma({\textstyle{\frac{1}{2}}}\eta_{o}+x)
\Gamma({\textstyle{\frac{1}{2}}}(\eta_{o}+1)+x)
\Gamma(\eta_{o}+1-\mu)}{\Gamma(\eta_{o}+1-\mu+x)
\Gamma({\textstyle{\frac{1}{2}}}\eta_{o})
\Gamma({\textstyle{\frac{1}{2}}}\eta_{o}
+{\textstyle{\frac{1}{2}}})}[-4x_{12}^{2}x_{34}^{2}]^{x}
\nonumber \\
 &  &
{}\times\int_{0}^{1}dtds\frac{[t(1-t)s(1-s)]^{\frac{1}{2}
\eta_{o}+x-1}}{[tsx_{13}^{2}+t(1-s)x_{14}^{2}+s(1-t)x_{23}
^{2}+(1-s)(1-t)x_{24}^{2}]^{\eta_{o}+2x}}\Bigg).\label{a11}
\end{eqnarray}
We may then successively integrate over $t$ and $s$ using
\begin{eqnarray}
\int_{0}^{1}dt\,t^{a-1}(1-t)^{b-1}[tx+(1-t)y]^{-a-b} & = & x^{-a}y^{-b}B(a,b),
\label{a12}\\
\int_{0}^{1}ds\,s^{a-1}\,(1-s)^{b-1}(1-sx)^{-\rho}\,(1-sy)^{-\sigma}
 & = &  B(a,b)\,F_{1}(a,\rho,\sigma;a+b;x,y),\label{a13}
\end{eqnarray}
and also
\begin{eqnarray}
F_{1}(a,b,c;b+c;x,y) & = &
(1-y)^{-a}{}_{2}F_{1}(a,b;b+c;\frac{x-y}{1-y}), \label{a14}\\
4^{n}({\textstyle{\frac{1}{2}}}\eta_{o})_{n}({\textstyle{
\frac{1}{2}}}\eta_{o}+{\textstyle{\frac{1}{2}}})_{n}
& = & (\eta_{o})_{2n},\label{a15}
\end{eqnarray}
to finally obtain
\begin{equation}
C^{\eta_{o}}(x_{12},\pa_{2})\,C^{\eta_{o}}(x_{34},\pa_{4})
\frac{1}{x_{24}^{2\eta_{o}}}=\frac{1}{(x_{13}^{2}x_{24}^{2})
^{\frac{1}{2}\eta_{o}}}{\cal{H}}_{\eta_{o}}(u,v),
\label{a16}
\end{equation}
where
\begin{equation}
{\cal{H}}_{\eta_{o}}(u,v)=\left(\frac{v}{u}\right)^{\frac{1}{2}
\eta_{o}}\sum_{n=0}^{\infty}\frac{v^{n}}{n!}\frac{({\textstyle
{\frac{1}{2}}}\eta_{o})_{n}^{4}}{(\eta_{o})_{2n}(\eta_{o}
+1-\mu)_{n}}{}_{2}F_{1}({\textstyle{\frac{1}{2}}}\eta_{o}
+n,{\textstyle{\frac{1}{2}}}\eta_{o}+n;\eta_{o}+2n;
1-\frac{v}{u}),\label{a17}
\end{equation}
which is the result used in the text.

\section{The One-Particle Exchange Graphs
${\cal{G}}_{1}^{(\tilde{\eta_{o}})}$}\label{apdx2}
\setcounter{equation}{0}

After simple manipulations as described in the text we obtain the amplitude
\begin{eqnarray}
{\cal{G}}_{1}^{(\tilde{\eta}_{o})}(x_{1},x_{2},x_{3},x_{4}) & = & \int
d^{d}xd^{d}y\la\phi(x_{1})\phi(x_{2})\tilde{O}(x)\ra\,\Phi^{-1}
_{\tilde{O}}(x-y)\,\la\phi(x_{3})\phi(x_{4})\tilde{O}(y)\ra
\nonumber \\
 & =  &
\frac{g_{*}^{2}}{C_{\tilde{O}}}\rho(\tilde{\eta}_{o})\,
U({\textstyle{\frac{1}{2}}}\tilde{\eta}_{o},{\textstyle{
\frac{1}{2}}}\tilde{\eta}_{o},d-\tilde{\eta}_{o})\frac{1}{
(x_{12}^{2})^{\eta-\frac{1}{2}\tilde{\eta}_{o}}(x_{34}^{2})^{\eta
-\mu +\frac{1}{2}\tilde{\eta}_{o}}} \nonumber \\
 &  &
\times
S_{4}(x_{1},{\textstyle{\frac{1}{2}}}\tilde{\eta}_{o};x_{2},
{\textstyle{\frac{1}{2}}}\tilde{\eta}_{o};x_{3},\mu
-{\textstyle{\frac{1}{2}}}\tilde{\eta}_{o};x_{4},\mu
-{\textstyle{\frac{1}{2}}}\tilde{\eta}_{o}),\label{b1}
\end{eqnarray}
where we have used once the D'EPP formula (\ref{4.1.6}) and have  defined
\begin{eqnarray}
 &  & S_{4}(x_{1},a_{1};x_{2},a_{2};x_{3},a_{3};x_{4},a_{4})
\nonumber \\
 &  & {}=\int
d^{d}x\frac{1}{(x_{1}-x)^{2a_{1}}(x_{2}-x)^{2a_{2}}(x_{3}-x)
^{2a_{3}}(x_{4}-x)^{2a_{4}}},\label{b2}
\end{eqnarray}
which is conformally invariant if $a_{1}+a_{2}+a_{3}+a_{4}=d$. In
(\ref{b1}), from (\ref{4.1.3}) and (\ref{4.1.7})
\begin{equation}
\pi^{\mu}\,\rho(\tilde{\eta}_{o})\,U({\textstyle{\frac{1}{2}}}
\tilde{\eta}_{o},{\textstyle{\frac{1}{2}}}\tilde{\eta}_{o},
d-\tilde{\eta}_{o})=\frac{\Gamma(\tilde{\eta}_{o})
\Gamma(\mu-\frac{1}{2}\tilde{\eta}_{o})}{\Gamma(\mu-\tilde{\eta}_{o})
\Gamma(\frac{1}{2}\tilde{\eta}_{o})}.
\label{.}
\end{equation}
Formula (\ref{b1}) clearly demonstrates the ``shadow symmetry''
property  for the graphs ${\cal{G}}_{1}^{(\tilde{\eta}_{o})}$ since we easily
see that
\begin{equation}
{\cal{G}}_{1}^{(d-\tilde{\eta}_{o})}(x_{1},x_{2},x_{3},x_{4})
=C(\tilde{\eta}_{o}){\cal{G}}_{1}^{(\tilde{\eta}_{o})}
(x_{1},x_{2},x_{3},x_{4}),\label{b3}
\end{equation}
where the quantity $C(\tilde{\eta}_{o})$ was defined in (\ref{4.1.9}).
The integral (\ref{b2})   can be performed
in various ways when $a_{1}+a_{2}+a_{3}+a_{4}=d$. Here we apply
Symanzik's \cite{Symanzik} method for its evaluation.
Using standard procedures we obtain
\begin{eqnarray}
&  & S_{4}(x_{1},a_{1};x_{2},a_{2};x_{3},a_{3};x_{4},a_{4}) \nonumber \\
&
&{}=\frac{\pi^{\mu}}{\left[\prod_{i=1}^{4}\Gamma(a_{i})\right]}
\int_{0}^{\infty}d\lambda_{1}..d\lambda_{4}\prod_{i=1}^{4}
\left[\lambda_{i}^{a_{i}-1}\right]\left[S_{\lambda}\right]
^{-\mu}\mbox{exp}\left[-\frac{1}{S_{\lambda}}\sum_{j>i=1}
^{4}\Bigl(\lambda_{i}\lambda_{j}x_{ij}^{2}\Bigl)\right], \label{b4}
\end{eqnarray}
where
\begin{equation}
S_{\lambda}=\sum_{i=1}^{4}\lambda_{i}.\label{b5}
\end{equation}
The conformal invariance of (\ref{b4}) manifests itself in the fact
that on the r.h.s one may replace $S_{\lambda}$ with
$\sum_{i=1}^{4}c^{2}_{i}\lambda_{i}$ and obtain the same result for
arbitrary positive $c^{2}_{i}$, $i=1,..4$. Following Symanzik we
 then choose $c^{2}_{1}=1$ and $c^{2}_{i}=0$, for
$i=2,3,4.$, or $S_{\lambda}=\lambda_{1}$.  Next, using the representation
\begin{equation}
\mbox{exp}[-z]=\frac{1}{2\pi
i}\int_{c-i\infty}^{c+i\infty}ds\,\Gamma(-s)z^{s}\,\,\,\,\, ,\,\,\,\,\,
(c<0,\,|\mbox{arg}\,z|<{\textstyle{\frac{1}{2}}}\pi),\label{b6}
\end{equation}
for the two exponentials involving $x_{24}^{2}$ and $x_{34}^{2}$, we
can perform the $\lambda$-integrations obtaining
\begin{eqnarray}
 &  & S^{4} (x_{1},a_{1};x_{2},a_{2};x_{3},a_{3};x_{4},a_{4})
\nonumber \\
 &  &
{}=\frac{\pi^{\mu}}{\left[\prod_{i=1}^{4}\Gamma(a_{i})\right]}
\frac{1}{(x_{23}^{2})^{\mu-a_{4}}(x_{14}^{2})^{a_{1}}
(x_{34}^{2})^{a_{3}+a_{4}-\mu}(x_{24}^{2})^{a_{2}+a_{4}-\mu}}
\nonumber \\
 &  & {}\times\left(\frac{1}{2\pi
i}\right)^{2}\int_{c-i\infty}^{c+i\infty}\!\!\!dtds\,
\Bigg(\Gamma(-s)\Gamma(-t)\Gamma(a_{3}+a_{4}-\mu
-t) \nonumber \\
 &  & {}\times\Gamma(a_{2}+a_{4}-\mu -s)\Gamma(\mu
-a_{4}+s+t)\Gamma(a_{1}+s+t)\left(\frac{v}{u}\right)^{s}v^{t}\Bigg),\label{b7}
\end{eqnarray}
where
\begin{equation}
v=\frac{x_{12}^{2}x_{34}^{2}}{x_{14}^{2}x_{23}^{2}} \,\,\,\,\,
,\,\,\,\,\,
\frac{v}{u}=\frac{x_{13}^{2}x_{24}^{2}}{x_{14}^{2}x_{23}^{2}}.\label{b8}
\end{equation}
The $s$-integration can be performed using
\begin{eqnarray}
\frac{1}{2\pi
i} &  &
\!\int_{c-i\infty}^{c+i\infty}ds\Gamma(a+s)\Gamma(b+s)
\Gamma(c-a-b-s)\Gamma(-s)(1-z)^{s}
\nonumber \\
 & & {}=
\Gamma(c-a)\Gamma(b-a)\frac{\Gamma(a)\Gamma(b)}{\Gamma(c)}
{}_{2}F_{1}(a,b;c;z),\label{b9}
\end{eqnarray}
and the remaining $t$-integration results in the sum of two infinite
series of poles of the Gamma functions $\Gamma(-t)$ and
{$\Gamma(a_{3}+a_{4}-\mu -t)$} situated at the points $t=n$ and
$t=a_{3}+a_{4}-\mu +n$ respectively with $n=0,1,2..$. The general
result, valid for $a_{1}+a_{2}+a_{3}+a_{4}=d$, reads
\begin{eqnarray}
 &  & S_{4}(x_{1},a_{1};x_{2},a_{2};x_{3},a_{3};x_{4},a_{4}) \, = \,
\frac{\pi^{\mu}}{\left[\prod_{i=1}^{4}\Gamma(a_{i})\right]}
\frac{1}{(x_{23}^{2})^{\mu
-a_{4}}(x_{14}^{2})^{a_{1}}(x_{34}^{2})^{a_{3}+a_{4}-\mu}
(x_{24}^{2})^{a_{2}+a_{4}-\mu}} \nonumber \\
\times &  & \!\Bigg(\Gamma(\mu
-a_{1}-a_{2})\sum_{n=0}^{\infty}\frac{v^{n}}{n!}
\frac{\Gamma(1-\mu+a_{1}+a_{2})\Gamma(\mu
-a_{3}+n)\Gamma(a_{2}+n)}{\Gamma(1-\mu
+a_{1}+a_{2}+n)} \nonumber \\
 &  &
{}\times\frac{\Gamma(\mu-a_{4}+n)\Gamma(a_{1}+n)}
{\Gamma(a_{1}+a_{2}+2n)}{}\,_{2}F_{1}(\mu-a_{4}+n,a_{1}
+n;a_{1}+a_{2}+2n;1-\frac{v}{u})
\nonumber \\
+ &  & \!v^{a_{3}+a_{4}-\mu}\Gamma(\mu
-a_{3}-a_{4})\sum_{n=0}^{\infty}\frac{v^{n}}{n!}
\frac{\Gamma(1-\mu+a_{3}+a_{4})\Gamma(a_{4}+n)
\Gamma(\mu-a_{1}+n)\Gamma(a_{3}+n)}{\Gamma(1-\mu
+a_{3}+a_{4}+n)\Gamma(a_{3}+a_{4}+2n)}
\nonumber \\
 & &
{}\times\Gamma(\mu-a_{2}+n){}\,_{2}F_{1}(a_{3}+n,\mu-a_{2}+n;
a_{3}+a_{4}+2n;1-\frac{v}{u})\Bigg).\label{b10}
\end{eqnarray}

Note that we have conducted the calculation leading to (\ref{b10}) so
that one can unambiguously consider the limits as $uv\rightarrow 0$,
$1-v/u\rightarrow 0$ {\it{independently}}. The amplitude for the graph
${\cal{G}}_{1}^{(\tilde{\eta}_{o})}(x_{1},x_{2},x_{3},x_{4})$ is
obtained using  (\ref{b10}) when $a_{1}=a_{2}=\frac{1}{2}\tilde{\eta}_{o}$ and
$a_{3}=a_{4}=\mu -\frac{1}{2}\tilde{\eta}_{o}$. The result is
\begin{equation}
{\cal{G}}_{1}^{(\tilde{\eta}_{o})}(x_{1},x_{2},x_{3},x_{4})
=\frac{g_{*}^{2}}{C_{\tilde{O}}}\frac{1}{(x_{12}^{2}x_{34}^{2})
^{\eta}}\Bigl[u^{\frac{1}{2}\tilde{\eta}_{o}}\,{\cal{H}}
_{\tilde{\eta}_{o}}(u,v)+C(d-\tilde{\eta}_{o})\,u^{\mu-
\frac{1}{2}\tilde{\eta}_{o}}\,{\cal{H}}_{d-\tilde{\eta}_{o}}
(u,v)\Bigl],\label{b11}
\end{equation}
with ${\cal{H}}(u,v)$ defined in (\ref{2.2.13}).

The amplitude for the graph
${\cal{G}}_{1}^{(\tilde{\eta}_{o})}(x_{1},x_{4},x_{3},x_{2})$ is
obtained using   (\ref{b10}) when $a_{1}+a_{2}-\mu
=\Delta=\mu-a_{3}-a_{4}$ in the limit $\Delta\rightarrow 0$. In this
case, the Gamma functions $\Gamma(\mu-a_{1}-a_{2})$ and $\Gamma(\mu
-a_{3}-a_{4})$ have poles $\propto \mp\frac{1}{\Delta}$ which cancel
between the two terms in (\ref{b10}). The finite result reads
\begin{equation}
{\cal{G}}_{1}^{(\tilde{\eta}_{o})}(x_{1},x_{4},x_{3},x_{2}) =
\frac{g_{*}^{2}}{C_{\tilde{O}}}\,\frac{1}{(x_{12}^{2}x_{34}^{2})^{\eta}}
v^{\eta}\sum_{n,m=0}^{\infty}\frac{v^{n}(1-\frac{v}{u})
^{m}}{n!m!}a_{nm}\left[-\mbox{ln}v+b_{nm}\right],\label{b12}
\end{equation}
where the overall coefficient is given in (\ref{.}) and
\begin{eqnarray}
a_{nm} & = &
\pi^{\mu}\,\rho(\tilde{\eta}_{o})\,U({\textstyle{\frac{1}{2}}}
\tilde{\eta}_{o},{\textstyle{\frac{1}{2}}}\tilde{\eta}_{o},d-
\tilde{\eta}_{o})\frac{({\textstyle{\frac{1}{2}}}\tilde{\eta}_{o})
_{n}(\mu-{\textstyle{\frac{1}{2}}}\tilde{\eta}_{o})_{n}
({\textstyle{\frac{1}{2}}}\tilde{\eta}_{o})_{n+m}(\mu
-{\textstyle{\frac{1}{2}}}\tilde{\eta}_{o})_{n+m}}{\Gamma(1+n)
\Gamma(\mu+2n+m)}, \label{b13}\\
b_{nm} & = & 2\psi(1+n)+2\psi(\mu
+m+2n)-\psi({\textstyle{\frac{1}{2}}}\tilde{\eta}_{o}+n)-\psi(\mu
-{\textstyle{\frac{1}{2}}}\tilde{\eta}_{o}+n) \nonumber \\
 &  & -\psi({\textstyle{\frac{1}{2}}}\tilde{\eta}_{o}+n+m)-\psi(\mu
-{\textstyle{\frac{1}{2}}}\tilde{\eta}_{o}+n+m).\label{b14}
\end{eqnarray}
The amplitude for the graph
${\cal{G}}^{(\tilde{\eta}_{o})}_{1}(x_{1},x_{3},x_{2},x_{4})$ is
simply obtained from (\ref{b12}) by the interchange  $x_{3}\leftrightarrow
x_{4}$ or $ u\leftrightarrow v$. The results in this Appendix are in accord
with similar results given
in \cite{Ruhl1}.

\section{The Box-Graphs ${\cal{G}}_{2}^{(\tilde{\eta}_{o})}$}\label{apdx3}

\setcounter{equation}{0}

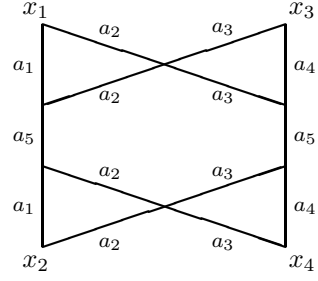
\begin{figure}[t]

\setlength{\unitlength}{0.005300in}%
\begingroup\makeatletter
\def\x#1#2#3#4#5#6#7\relax{\def\x{#1#2#3#4#5#6}}%
\expandafter\x\fmtname xxxxxx\relax \def\y{splain}%
\ifx\x\y   
\gdef\SetFigFont#1#2#3{%
  \ifnum #1<17\tiny\else \ifnum #1<20\small\else
  \ifnum #1<24\normalsize\else \ifnum #1<29\large\else
  \ifnum #1<34\Large\else \ifnum #1<41\LARGE\else
     \huge\fi\fi\fi\fi\fi\fi
  \csname #3\endcsname}%
\else
\gdef\SetFigFont#1#2#3{\begingroup
  \count@#1\relax \ifnum 25<\count@\count@25\fi
  \def\x{\endgroup\@setsize\SetFigFont{#2pt}}%
  \expandafter\x
    \csname \romannumeral\the\count@ pt\expandafter\endcsname
    \csname @\romannumeral\the\count@ pt\endcsname
  \csname #3\endcsname}%
\fi
\endgroup
\begin{picture}(200,280)(-650,400)
\thicklines
\put(220,660){\line( 3,-1){240}}
\put(460,580){\line( 0, 1){  0}}
\put(460,580){\line( 0, 1){  0}}
\put(460,580){\line( 0, 1){  0}}
\put(460,660){\line(-3,-1){240}}
\put(220,660){\line( 0,-1){220}}
\put(220,440){\line( 3, 1){240}}
\put(460,520){\line( 0,-1){ 80}}
\put(220,520){\line( 3,-1){240}}
\put(460,660){\line( 0,-1){140}}
\put(-620,540){\makebox(0,0)[lb]{\smash{\SetFigFont{10}{14.4}{rm}$
{\cal{G}}_{2}^{(\tilde{\eta}_{o})}(x_{1},x_{2},x_{3},x_{4}) \,\,\,\,\,=
\,\,\,\,\,\left(\frac{g_{*}^{2}}{C_{\tilde{O}}C_{\phi}}\right)^{2}
\rho^{2}(\eta)\rho^{2}(\tilde{\eta}_{o})U^{2}({\textstyle
{\frac{1}{2}}}\tilde{\eta}_{o},{\textstyle{\frac{1}{2}}}\tilde{\eta}_{o},d-\tilde{\eta}_{o})
  $}}}

\put(200,670){\makebox(0,0)[lb]{\smash{\SetFigFont{10}{14.4}{rm}$x_{1}$}}}
\put(463,670){\makebox(0,0)[lb]{\smash{\SetFigFont{10}{14.4}{rm}$x_{3}$}}}
\put(200,420){\makebox(0,0)[lb]{\smash{\SetFigFont{10}{14.4}{rm}$x_{2}$}}}
\put(463,420){\makebox(0,0)[lb]{\smash{\SetFigFont{10}{14.4}{rm}$x_{4}$}}}
\put(190,615){\makebox(0,0)[lb]{\smash{\SetFigFont{8}{14.4}{rm}$a_{1}$}}}
\put(275,650){\makebox(0,0)[lb]{\smash{\SetFigFont{8}{14.4}{rm}$a_{2}$}}}
\put(387,652){\makebox(0,0)[lb]{\smash{\SetFigFont{8}{14.4}{rm}$a_{3}$}}}
\put(275,585){\makebox(0,0)[lb]{\smash{\SetFigFont{8}{14.4}{rm}$a_{2}$}}}
\put(387,585){\makebox(0,0)[lb]{\smash{\SetFigFont{8}{14.4}{rm}$a_{3}$}}}
\put(468,615){\makebox(0,0)[lb]{\smash{\SetFigFont{8}{14.4}{rm}$a_{4}$}}}
\put(468,545){\makebox(0,0)[lb]{\smash{\SetFigFont{8}{14.4}{rm}$a_{5}$}}}
\put(190,545){\makebox(0,0)[lb]{\smash{\SetFigFont{8}{14.4}{rm}$a_{5}$}}}
\put(190,475){\makebox(0,0)[lb]{\smash{\SetFigFont{8}{14.4}{rm}$a_{1}$}}}
\put(275,510){\makebox(0,0)[lb]{\smash{\SetFigFont{8}{14.4}{rm}$a_{2}$}}}
\put(387,510){\makebox(0,0)[lb]{\smash{\SetFigFont{8}{14.4}{rm}$a_{3}$}}}
\put(387,440){\makebox(0,0)[lb]{\smash{\SetFigFont{8}{14.4}{rm}$a_{3}$}}}
\put(275,440){\makebox(0,0)[lb]{\smash{\SetFigFont{8}{14.4}{rm}$a_{2}$}}}
\put(468,475){\makebox(0,0)[lb]{\smash{\SetFigFont{8}{14.4}{rm}$a_{4}$}}}
\put(155,540){\makebox(0,0)[lb]{\smash{\SetFigFont{10}{14.4}{rm}$\times$}}}
\end{picture}

\caption{The amplitude for
${\cal{G}}_{2}^{(\tilde{\eta}_{o})}(x_{1},x_{2},x_{3},x_{4})$.}\label{fg7}
\end{figure}

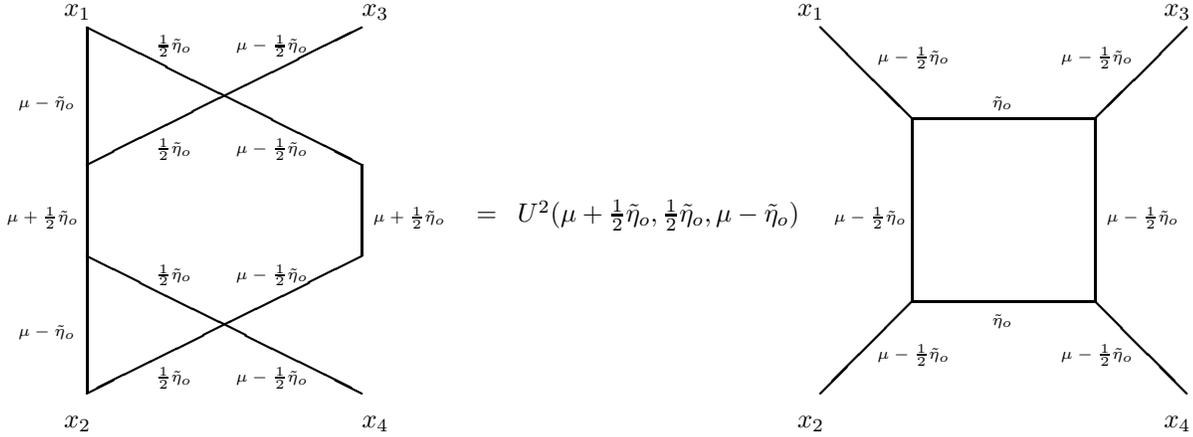
\begin{figure}[t]

\setlength{\unitlength}{0.01200in}%
\begingroup\makeatletter
\def\x#1#2#3#4#5#6#7\relax{\def\x{#1#2#3#4#5#6}}%
\expandafter\x\fmtname xxxxxx\relax \def\y{splain}%
\ifx\x\y   
\gdef\SetFigFont#1#2#3{%
  \ifnum #1<17\tiny\else \ifnum #1<20\small\else
  \ifnum #1<24\normalsize\else \ifnum #1<29\large\else
  \ifnum #1<34\Large\else \ifnum #1<41\LARGE\else
     \huge\fi\fi\fi\fi\fi\fi
  \csname #3\endcsname}%
\else
\gdef\SetFigFont#1#2#3{\begingroup
  \count@#1\relax \ifnum 25<\count@\count@25\fi
  \def\x{\endgroup\@setsize\SetFigFont{#2pt}}%
  \expandafter\x
    \csname \romannumeral\the\count@ pt\expandafter\endcsname
    \csname @\romannumeral\the\count@ pt\endcsname
  \csname #3\endcsname}%
\fi
\endgroup
\begin{picture}(480,192)(25,605)
\thicklines
\put( 60,780){\line( 2,-1){120}}
\put(180,720){\line( 0,-1){ 40}}
\put(180,680){\line(-2,-1){120}}
\put( 60,620){\line( 0, 1){ 60}}
\put( 60,680){\line( 2,-1){120}}
\put( 60,780){\line( 0,-1){100}}
\put( 60,720){\line( 2, 1){120}}
\put(380,780){\line( 1,-1){ 40}}
\put(420,740){\line( 1, 0){ 80}}
\put(500,740){\line( 0,-1){ 80}}
\put(420,660){\line( 1, 0){ 80}}
\put(420,740){\line( 0,-1){ 80}}
\put(420,660){\line(-1,-1){ 40}}
\put(500,660){\line( 1,-1){ 40}}
\put(540,780){\line(-1,-1){ 40}}
\put(230,695){\makebox(0,0)[lb]{\smash{\SetFigFont{10}{14.4}{rm}$
=\,\,\,U^{2}(\mu+{\textstyle{\frac{1}{2}}}\tilde{\eta}_{o},
{\textstyle{\frac{1}{2}}}\tilde{\eta}_{o},\mu-\tilde{\eta}_{o})$}}}
\put( 50,785){\makebox(0,0)[lb]{\smash{\SetFigFont{10}{14.4}{rm}$x_{1}$}}}
\put( 50,605){\makebox(0,0)[lb]{\smash{\SetFigFont{10}{14.4}{rm}$x_{2}$}}}
\put(180,785){\makebox(0,0)[lb]{\smash{\SetFigFont{10}{14.4}{rm}$x_{3}$}}}
\put(180,605){\makebox(0,0)[lb]{\smash{\SetFigFont{10}{14.4}{rm}$x_{4}$}}}
\put(
30,745){\makebox(0,0)[lb]{\smash{\SetFigFont{6}{14.4}{rm}$
\mu-\tilde{\eta}_{o}$}}}
\put(30,645){\makebox(0,0)[lb]{\smash{\SetFigFont{6}{14.4}{rm}$
\mu-\tilde{\eta}_{o}$}}}
\put(25,695){\makebox(0,0)[lb]{\smash{\SetFigFont{6}{14.4}{rm}$
\mu+\frac{1}{2}\tilde{\eta}_{o}$}}}
\put(185,695){\makebox(0,0)[lb]{\smash{\SetFigFont{6}{14.4}{rm}$
\mu+\frac{1}{2}\tilde{\eta}_{o}$}}}
\put(90,770){\makebox(0,0)[lb]{\smash{\SetFigFont{6}{14.4}{rm}$
\frac{1}{2}\tilde{\eta}_{o}$}}}
\put(90,725){\makebox(0,0)[lb]{\smash{\SetFigFont{6}{14.4}{rm}$
\frac{1}{2}\tilde{\eta}_{o}$}}}
\put(125,770){\makebox(0,0)[lb]{\smash{\SetFigFont{6}{14.4}{rm}$
\mu-\frac{1}{2}\tilde{\eta}_{o}$}}}
\put(125,725){\makebox(0,0)[lb]{\smash{\SetFigFont{6}{14.4}{rm}$
\mu-\frac{1}{2}\tilde{\eta}_{o}$}}}
\put(90,670){\makebox(0,0)[lb]{\smash{\SetFigFont{6}{14.4}{rm}$
\frac{1}{2}\tilde{\eta}_{o}$}}}
\put(90,625){\makebox(0,0)[lb]{\smash{\SetFigFont{6}{14.4}{rm}$
\frac{1}{2}\tilde{\eta}_{o}$}}}
\put(125,670){\makebox(0,0)[lb]{\smash{\SetFigFont{6}{14.4}{rm}$
\mu-\frac{1}{2}\tilde{\eta}_{o}$}}}
\put(125,625){\makebox(0,0)[lb]{\smash{\SetFigFont{6}{14.4}{rm}$
\mu-\frac{1}{2}\tilde{\eta}_{o}$}}}
\put(370,785){\makebox(0,0)[lb]{\smash{\SetFigFont{10}{14.4}{rm}$x_{1}$}}}
\put(530,785){\makebox(0,0)[lb]{\smash{\SetFigFont{10}{14.4}{rm}$x_{3}$}}}
\put(370,605){\makebox(0,0)[lb]{\smash{\SetFigFont{10}{14.4}{rm}$x_{2}$}}}
\put(530,605){\makebox(0,0)[lb]{\smash{\SetFigFont{10}{14.4}{rm}$x_{4}$}}}
\put(405,765){\makebox(0,0)[lb]{\smash{\SetFigFont{6}{14.4}{rm}$
\mu-\frac{1}{2}\tilde{\eta}_{o}$}}}
\put(485,765){\makebox(0,0)[lb]{\smash{\SetFigFont{6}{14.4}{rm}$
\mu-\frac{1}{2}\tilde{\eta}_{o}$}}}
\put(386,695){\makebox(0,0)[lb]{\smash{\SetFigFont{6}{14.4}{rm}$
\mu-\frac{1}{2}\tilde{\eta}_{o}$}}}
\put(505,695){\makebox(0,0)[lb]{\smash{\SetFigFont{6}{14.4}{rm}$
\mu-\frac{1}{2}\tilde{\eta}_{o}$}}}
\put(455,745){\makebox(0,0)[lb]{\smash{\SetFigFont{6}{14.4}{rm}$
\tilde{\eta}_{o}$}}}
\put(455,650){\makebox(0,0)[lb]{\smash{\SetFigFont{6}{14.4}{rm}$
\tilde{\eta}_{o}$}}}
\put(405,635){\makebox(0,0)[lb]{\smash{\SetFigFont{6}{14.4}{rm}$
\mu-\frac{1}{2}\tilde{\eta}_{o}$}}}
\put(485,635){\makebox(0,0)[lb]{\smash{\SetFigFont{6}{14.4}{rm}$
\mu-\frac{1}{2}\tilde{\eta}_{o}$}}}
\end{picture}

\caption{The Reduction to a Box-Graph.}\label{fg8}
\end{figure}

Using the diagrammatic representations in Fig. \ref{fg3} for the
amputation of the three-point function $\la\phi\phi \tilde{O}\ra$ in the
$\tilde{O}$ leg we obtain the amplitude shown in Fig. \ref{fg7}
with
\begin{eqnarray}
a_{1} & = & \eta-{\textstyle{\frac{1}{2}}}\tilde{\eta}_{o}, \label{c1}\\
a_{2} & = & {\textstyle{\frac{1}{2}}}\tilde{\eta}_{o}, \label{c2}\\
a_{3} & = & \mu -{\textstyle{\frac{1}{2}}}\tilde{\eta}_{o}, \label{c3}\\
a_{4} & = & \eta +{\textstyle{\frac{1}{2}}}\tilde{\eta}_{o}-\mu, \label{c4}\\
a_{5} & = & d-\eta.\label{c5}
\end{eqnarray}
The lines in Fig. \ref{fg7} represent   powers of the separation  with exponent
twice the
attached numbers given in (\ref{c1})-(\ref{c5}). The internal vertices are
simple integration vertices. It is then clear that the {\it{shadow symmetry}}
property
\begin{equation}
{\cal{G}}_{2}^{(d-\tilde{\eta}_{o})}(x_{1},x_{2},x_{3},x_{4})
=\left[C(\tilde{\eta}_{o})\right]^{2}{\cal{G}}_{2}
^{(\tilde{\eta}_{o})}(x_{1},x_{2},x_{3},x_{4}),\label{c6}
\end{equation}
can be easily proved in general on account of the  symmetry property
\begin{equation}
{\cal{G}}_{2}^{(\tilde{\eta}_{o})}(x_{1},x_{2},x_{3},x_{4})
={\cal{G}}_{2}^{(\tilde{\eta}_{o})}(x_{3},x_{4},x_{1},x_{2}).\label{c7}
\end{equation}
The graph in Fig. \ref{fg7} has not been evaluated for general values of the
exponents $a_{1},\cdots a_{5}$. However, in certain cases it reduces to a
simpler box-graph
for which an analytic expression in closed form has been found using
techniques similar to the ones applied  for the
evaluation of the one-particle exchange graphs in Appendix B. Such is
the case when
\begin{equation}
\tilde{\eta}_{o}=d-2\eta\,\,\,\,\,\Rightarrow \,\,\,\,\,a_{4}=0.\label{c8}
\end{equation}
Then, using only D'EPP formula (\ref{4.1.6}) we find the equation shown in
Fig. \ref{fg8}.
Moreover, it can be seen either from the {\it{shadow symmetry}}
property (\ref{c6}) or by direct calculation that the graph in Fig. \ref{fg7}
reduces
to a simple box-graph with unique \footnote{An integration vertex made out of
lines which are
powers of the separation is called {\cal{unique}} if the sum of the exponents
of these
lines is $2d$. One can easily prove that scalar amplitudes which
correspond to graphs made out of power lines and unique vertices
behave like conformal scalars (\ie see (\ref{1.1.1})) under conformal
transformations of the external coordinates.}
vertices  when
\begin{equation}
\tilde{\eta}_{o}=2\eta\,\,\,\,\,\Rightarrow\,\,\,\,\, a_{1}=0\label{c9}
\end{equation}
The box-graph in Fig. \ref{fg8} has been evaluated by Lang \cite{Lang}. In the
context of the present work and to the order in $1/N$ we are
interested, we just need the value of this box-graph when
$\tilde{\eta}_{o}=2$. Putting it all  together and using
the general result of \cite{Lang} we find
\begin{eqnarray}
{\cal{G}}_{2}^{(\tilde{\eta}_{o})}(x_{1},x_{2},x_{3},x_{4}) & = &
\left(\frac{g_{*}^{2}}{C_{\phi}C_{\tilde{O}}}\right)^{2}\,
{\cal{A}}(d)\,\frac{1}{(x_{12}^{2}x_{34}^{2})^{\mu
-1}}\nonumber \\
 &  & {}\times\,
v^{\mu-1}\sum_{n,m=0}^{\infty}\frac{v^{n}(1-\frac{v}{u})^{m}}{n!m!}[
-c_{nm}\mbox{ln}v+d_{nm}],\label{c10}
\end{eqnarray}
with
\begin{eqnarray}
c_{nm} & = & \pi^{4\mu}\frac{n!\Gamma(3-\mu)\Gamma^{2}(\mu -1+n+m)}{(\mu
-2)^{2}\Gamma(\mu -2)\Gamma^{3}(\mu
-1)\Gamma(\mu+m+2n)} \nonumber \\
 &  & {}\times\sum_{s=0}^{n
}\frac{1}{s!}\frac{\Gamma(\mu
-2+s)\Gamma(\mu -1+n+m+s)}{\Gamma(2\mu -3+n+m+s)}, \label{c11}\\
d_{nm} & = & -\pi^{4\mu}\frac{\Gamma(\mu -2)\Gamma^{2}(\mu
-1+n+m)}{\Gamma^{6}(\mu -1)} \nonumber \\
 &  &
{}\times\sum_{r=0}^{n}\sum_{s=0}^{n-r}\Bigg(\frac{1}{r!s!(n-r-s)!}
\frac{\Gamma(1+n-r)\Gamma(\mu
-2+r)\Gamma(\mu -1+n+m+s)}{\Gamma(1-r)\Gamma(\mu+m+2n-r)} \nonumber \\
 &  & {}\times\frac{\Gamma(\mu -2+r+s)\Gamma(1+n-r-s)\Gamma(3-\mu
-r)}{\Gamma(2\mu-3+n+m+r+s)} \nonumber \\
 & & {}\times\Bigl[\psi(\mu
-1+n+m+s)+\psi(1+n-r)+\psi(1-r)  \nonumber \\
 &  & {}+\psi(1+n-r-s)+2\psi(\mu -1+n+m)-\psi(3-\mu
-r)-2\psi(1+n)\nonumber \\
&  &  {}-\psi(2\mu
-3+n+m+r+s)-2\psi(\mu+m+2n-r)\Bigl]\Bigg),\label{c12}
\end{eqnarray}
where
\begin{eqnarray}
{\cal{A}}(d) &  = &
\Bigl[\rho(\mu-1)\,\rho(2)\,U(1,1,2\mu-2)\,U(\mu+1,1,\mu-2)\Bigl]^{2}
\nonumber \\
 & = & \frac{1}{\pi^{4\mu}}\Bigl[(\mu-2)\Gamma(\mu-1)\Bigl]^{4}.\label{c13}
\end{eqnarray}
The amplitude for the graph
${\cal{G}}_{2}^{(\tilde{\eta}_{o})}(x_{1},x_{2},x_{4},x_{3})$ is obtained from
${\cal{G}}_{2}^{(\tilde{\eta}_{o})}(x_{1},x_{2},x_{3},x_{4})$ by the
interchange
$x_{3}\leftrightarrow x_{4}\Leftrightarrow u\leftrightarrow v$. Note that in
evaluating
$d_{10}$ the $r=1$ contribution does not vanish
despite the presence of  $\Gamma(1-r)$  in the
denominator of (\ref{c12}), by virtue of
\begin{equation}
\frac{1}{\Gamma(1-r)}\psi(1-r)=\frac{1-r}{\Gamma(2-r)}\psi(2-r)
-\frac{1}{\Gamma(2-r)},\label{c14}
\end{equation}
and we  obtain
\begin{eqnarray}
d_{10} & = &
\pi^{4\mu}\frac{\Gamma(3-\mu)}{\Gamma(\mu)\Gamma(2\mu-2)}
\frac{(\mu-1)^{2}(\mu^{2}-2)}{2\mu(\mu
+1)(\mu
-2)^{3}}\Bigg(\frac{2\mu^{3}+3\mu^{2}-5\mu-4}{\mu(\mu+1)
(\mu^{2}-2)}+{\cal{C}}(\mu)\Bigg),\label{c15}
\end{eqnarray}
where ${\cal{C}}(\mu)$ is defined in (\ref{4.2.22}).

\section{The Graphs in the Four-Point Function $\langle\phi\phi OO\rangle$}
\label{apdx4}

\setcounter{equation}{0}

For the graphs shown in Fig. \ref{fig?}, after simple manipulation as
described in the text,  we obtain
\begin{eqnarray}
{\cal{K}}_{1}(x_{1},x_{2},x_{3},x_{4}) & = & g_{\phi\phi
O}\,g_{O}\,\rho(\eta_{o})\,U({\textstyle{\frac{1}{2}}}\eta_{o},
{\textstyle{\frac{1}{2}}}\eta_{o},d-\eta_{o})
\frac{1}{(x_{12}^{2})^{\mu-\frac{1}{2}\eta_{o}}(x_{34}^{2})
^{\frac{3}{2}\eta_{o}-\mu}}\nonumber \\
 &  & {}\times
S_{4}(x_{1},{\textstyle{\frac{1}{2}}}\eta_{o};x_{2},
{\textstyle{\frac{1}{2}}}\eta_{o};x_{3},\mu-{\textstyle
{\frac{1}{2}}}\eta_{o};x_{4},\mu-{\textstyle{\frac{1}{2}}}\eta_{o}).
\label{d1}
\end{eqnarray}
The limit $x_{12}^{2}$, $x_{34}^{2}$ $\rightarrow 0$ leads to
(\ref{add20}). The limit $x_{13}^{2}$, $x_{24}^{2}$ $ \rightarrow
0$ corresponds to the case explained just before (\ref{b12}) when
poles appear in the evaluation of $S_{4}$ which however cancel giving
as finite result (\ref{add23}).

Similarly
\begin{eqnarray}
{\cal{K}}_{2}(x_{1},x_{2},x_{3},x_{4}) & = & g_{\phi\phi
O}^{2}\,\rho(\eta)\,U(\eta-{\textstyle{\frac{1}{2}}}\eta_{o},
{\textstyle{\frac{1}{2}}}\eta_{o},d-\eta)\frac{1}{(x_{13}^{2})
^{\frac{1}{2}\eta_{o}}(x_{24}^{2})^{\eta-\mu-\frac{1}{2}\eta_{o}}}
\nonumber \\
 &  & {}\times
S_{4}(x_{1},\eta-{\textstyle{\frac{1}{2}}}\eta_{o};x_{2},
\mu-{\textstyle{\frac{1}{2}}}\eta_{o};x_{3},{\textstyle
{\frac{1}{2}}}\eta_{o};x_{4},\mu-\eta+{\textstyle{\frac{1}{2}}}\eta_{o}).
\label{d2}
\end{eqnarray}
Three different limits are relevant here. When $x_{12}^{2}$, $x_{34}^{2}$
$\rightarrow 0$ we obtain (\ref{add21}). When $x_{13}^{2}$,
$x_{24}^{2}$, $\rightarrow 0$ we obtain (\ref{add24}) while the case
when $x_{14}^{2}$, $x_{23}^{2}$ $\rightarrow 0$ corresponds to a
cancellation of poles in the integral as before and leads to the
finite result (\ref{add25}) with
\begin{eqnarray}
C_{nm} & =
&\pi^{\mu}\rho(\eta)\,U(\eta-{\textstyle{\frac{1}{2}}}\eta_{o},
{\textstyle{\frac{1}{2}}}\eta_{o},d-\eta)
\frac{(\mu-\eta+\frac{1}{2}\eta_{o})_{n}(\frac{1}{2}\eta_{o})_{n}
(\eta-\frac{1}{2}\eta_{o})_{n+m}(\mu-\frac{1}{2}\eta_{o})_{n+m}}
{\Gamma(1+n)\Gamma(\mu+2n+m)},
\label{d13} \nonumber \\
  &  & \\
D_{nm} & = &
2\psi(1+n)+2\psi(\mu+2n+m)-\psi(\mu-\eta+{\textstyle{
\frac{1}{2}}}+n)-\psi({\textstyle{\frac{1}{2}}}\eta_{o}+n)
\nonumber \\
 &  &
-\psi(\eta-{\textstyle{\frac{1}{2}}}\eta_{o}+n+m)-\psi(\mu
-{\textstyle{\frac{1}{2}}}\eta_{o}+n+m)
. \label{d14}
\end{eqnarray}

\end{appendix}

\end{document}